\documentclass[fleqn,usenatbib]{rasti}

\usepackage{newtxtext,newtxmath}

\usepackage[T1]{fontenc}

\usepackage{graphics,graphicx}
\hypersetup{urlcolor=blue}
\usepackage{natbib}
\usepackage[outdir=./]{epstopdf}
\usepackage{textcomp,gensymb}
\usepackage{xfrac}
\usepackage{xcolor}

\usepackage{booktabs} 
\usepackage{longtable} 
\usepackage{array} 
\usepackage{threeparttable}
\usepackage{caption}

\newcommand{\istar}{$i_{\star}$}

\newcommand{\cosi}{$\cos i_{\star}$}

\newcommand{\rstar}{$R_{\star}$}
\newcommand{\mstar}{$M_{\star}$}

\newcommand{\rhostar}{$\rho_{\star}$}
\newcommand{\vsini}{$v\sin i_{\star}$}
\newcommand{\prot}{$P_{\textnormal{rot}}$}
\newcommand{\idisk}{$i_{\textnormal{d}}$}

\newcommand{\av}{$A_V$}
\newcommand{\veq}{$v_{\textnormal{eq}}$}

\gdef\teff{effective temperature ($T_{\mathrm{eff}}$)\gdef\teff{$T_{\mathrm{eff}}$}}

\newcommand{\fnu}{$F_{\nu}$}

\newcommand{\Alambda}{$A_{\lambda}$}
\newcommand{\lambdapiv}{$\lambda_{\textnormal{piv}}$}

\newcommand{\kms}{km\,s$^{-1}$}
\renewcommand{\micron}{$\mu$m}
\newcommand{\msun}{$M_{\odot}$}
\newcommand{\rsun}{$R_{\odot}$}
\newcommand{\rhosun}{$\rho_{\odot}$}

\newcommand{\flamunits}{erg\,s$^{-1}$\,cm$^{-2}$\,\AA$^{-1}$}
\newcommand{\flam}{$F_{\lambda}$}
\newcommand{\flamalt}{$F(\lambda)$}

\gdef\tess{the Transiting Exoplanet Survey Satellite (\textit{TESS})\gdef\tess{\textit{TESS}}}
\gdef\mcmc{Markov chain Monte Carlo (MCMC)\gdef\mcmc{MCMC}}
\gdef\alma{the Atacama Large Millimeter/submillimeter Array (ALMA)\gdef\alma{ALMA}}
\gdef\hbm{hierarchical Bayesian model (HBM)\gdef\hbm{HBM}}

\gdef\twomass{the Two Micron All Sky Survey \citep[2MASS;][]{skrutskie_two_2006}\gdef\twomass{2MASS}}
\gdef\gaia{the \textit{Gaia} mission \citep{prusti_gaia_2016}\gdef\gaia{\textit{Gaia}}}
\gdef\apass{the AAVSO Photometric All-Sky Survey \citep[APASS;][]{henden_aavso_2009}\gdef\apass{APASS}}
\gdef\hipparcos{the \textit{Hipparcos} mission \citep{esa_hipparcos_1997}\gdef\hipparcos{\textit{Hipparcos}}}
\gdef\tycho{the \hipparcos{} TYCHO-2 catalog \citep{hog_tycho-2_2000}\gdef\tycho{TYCHO}}
\gdef\ps1{the Pan-STARRS1 survey \citep[PS1;][]{chambers_pan-starrs1_2018}\gdef\ps1{PS1}}
\gdef\sdss{the Sloan Digital Sky Survey (SDSS)\gdef\sdss{SDSS}}
\gdef\soar{the Southern Astrophysical Research (SOAR) Telescope\gdef\soar{SOAR}}
\gdef\goodman{the Goodman High Throughput Spectrograph (Goodman)\gdef\goodman{Goodman}}
\gdef\igrins{the Immersion Grating Infrared Spectrograph \citep[IGRINS;][]{park_design_2014}\gdef\igrins{IGRINS}}
\gdef\mast{the Barbara A. Mikulski Archive for Space Telescopes (MAST)\gdef\mast{MAST}}
\gdef\bpmg{the $\beta$\,Pictoris Moving Group ($\beta$PMG)\gdef\bpmg{$\beta$PMG}}
\gdef\tuchor{the Tucana-Horologium Moving Group (Tuc-Hor)\gdef\tuchor{Tuc-Hor}}
\gdef\carext{the Carina-Extended Association \citep[Car-Ext;][]{luhman_census_2024}\gdef\carext{Car-Ext}}
\gdef\ktwo{the Kepler-\textit{K2} Mission\gdef\ktwo{\textit{K2}}}
\gdef\pms{pre-main-sequence (PMS)\gdef\pms{PMS}}
\gdef\hbm{hierarchical Bayesian model (HBM)\gdef\hbm{HBM}}
\gdef\dsep{the Dartmouth Stellar Evolution Program \citep[DSEP;][]{chaboyer_heavy-element_2001,dotter_dartmouth_2008}\gdef\dsep{DSEP}}
\gdef\parsec{the PAdova and TRieste Stellar Evolution
Code \citep[PARSEC;][]{bressan_parsec_2012}\gdef\parsec{PARSEC}}
\gdef\ruwe{Renormalized Unit Weight Error (RUWE)\gdef\ruwe{RUWE}}

\newcommand{\rstarcode}{\texttt{stelpar}}

\newcommand{\BT}{{B$_{\textnormal{T}}$}}
\newcommand{\VT}{{V$_{\textnormal{T}}$}}
\newcommand{\Hp}{{H$_{\textnormal{p}}$}}

\graphicspath{{./}{}}

\title[Method for Estimating Disk-Star Alignment]{Disk-Star Alignment I: Pre-Main-Sequence Stellar Parameters and the Statistical Alignment Between Disks and Stellar Rotation}

\author[Fields et al.]{
Matthew J. Fields,$^{1}$\thanks{Contact e-mail: mjfields@live.unc.edu} 
Andrew W. Mann,$^{1}$ 
Aurora Kesseli,$^{2}$ 
and Andrew W. Boyle$^{1}$ 
\\
$^{1}$Department of Physics and Astronomy, The University of North Carolina at Chapel Hill, Chapel Hill, NC 27599, USA\\
$^{2}$IPAC, Mail Code 100-22, Caltech, 1200 E. California Boulevard, Pasadena, CA 91125, USA
}

\date{Accepted XXX. Received YYY; in original form ZZZ}

\pubyear{\the\year{}}

\begin{document}
\label{firstpage}
\pagerange{\pageref{firstpage}--\pageref{lastpage}}
\maketitle

\begin{abstract}
Astronomers generally assume planet-forming disks are aligned with the rotation of their host star. However, recent observations have shown evidence of warping in protoplanetary disks. One can measure the statistical alignment between the inclination angles of the disk and stellar spin using the projected rotational velocity, radius, and rotation period of the star and interferometric measurements of the protoplanetary disk. Such work is  challenging due to the difficulty in measuring the properties of young stars and biases in methods to combine them for population studies. Here, we provide an overview of the required observables, realistic uncertainties, and complications when using them to constrain the orientation of the system. 
We show in several tests that we are able to constrain the uncertainties on the necessary stellar parameters to better than 5\% in most cases. We show that by using a hierarchical Bayesian model, we can account for many of the systematic effects (e.g., biases in measured stellar and disk orientations) by fitting for the alignments of each system simultaneously. We demonstrate our hierarchical model on a realistic synthetic sample and verify that we can recover our input alignment distribution to $\lesssim$5$^\circ$ with a modest ($\simeq$30\,star) sample. As the sample of systems with disk inclinations grows, future studies can improve upon our approach with a three-dimensional treatment of misalignment and better handling of non-Gaussian errors.
\end{abstract}


\section{Introduction} \label{sec:intro}

Disks of gas and dust surrounding newly-formed stars (called `protoplanetary' or `planet-forming' disks) are natural consequences of angular momentum conservation as molecular clouds collapse to form protostars. Young stars are initially embedded in gas-rich envelopes, but after $\simeq$1\,Myr they become optically visible as the dust and gas settle into a disk and evolve through viscous accretion. The protoplanetary disks around such \pms{} stars are thought to form planets within 5--10\,Myr, after which their gaseous disk material dissipates, first into dusty debris disks, then into mature planetary systems \citep{williams_protoplanetary_2011,morbidelli_building_2012}.

The existence of protoplanetary disks was implied by the shape of the Solar System; the planets appeared to occupy a flat plane, orbiting in-line with the rotation of the Sun. In fact, the rotational axis of the Sun is only inclined $6^{\circ}$ relative to the average angular momentum of the eight major planets \citep{2005ApJ...621L.153B}, and the spread in inclinations is no more than $\sim$7$^{\circ}$. The observed geometry of the Solar System inspired the Kant-Laplace nebular hypothesis, which led to the widely accepted idea that protoplanetary disks should be aligned with the rotation of their newly-formed host stars.

Observations of the Rossiter–McLaughlin effect during planetary transits have revealed a significant population of planets with orbits misaligned from their host star's spin axis \citep[e.g.,][]{winn_hot_2010, albrecht_stellar_2022}. This seemingly contradicts the basic planet-formation model, so early explanations often focused on later-stage many-body interactions, such as Kozai-Lidov \citep{wu_planet_2003}, planet–planet scattering \citep{ford_origins_2008, naoz_hot_2011}, or angular momentum transport within the host star \citep{rogers_internal_2012}. Other work has shown that planetary misalignment can arise from early interactions with the protoplanetary disk \citep[e.g.,][]{petrovich_disk-driven_2020}, particularly if the disk is misaligned \citep[e.g.,][]{batygin_primordial_2012}.

A protoplanetary disk can be quickly misaligned from its host due to the influence of a wide binary companion, a dense stellar environment (stellar flybys), or interactions with the surrounding star-forming cloud \citep[e.g.,][]{bate_chaotic_2010, fielding_turbulent_2015, takaishi_stardisc_2020, kuffmeier_misaligned_2021}. The inner disk may stay bound to the orientation of the host star, but hydro-dynamical simulations suggest the outer disk ($\gtrsim$10\,au) can end up with a near-random orientation  \citep[e.g.,][]{bate_diversity_2018}. 

\begin{figure}
    \centering
    \includegraphics[width=0.5\textwidth]{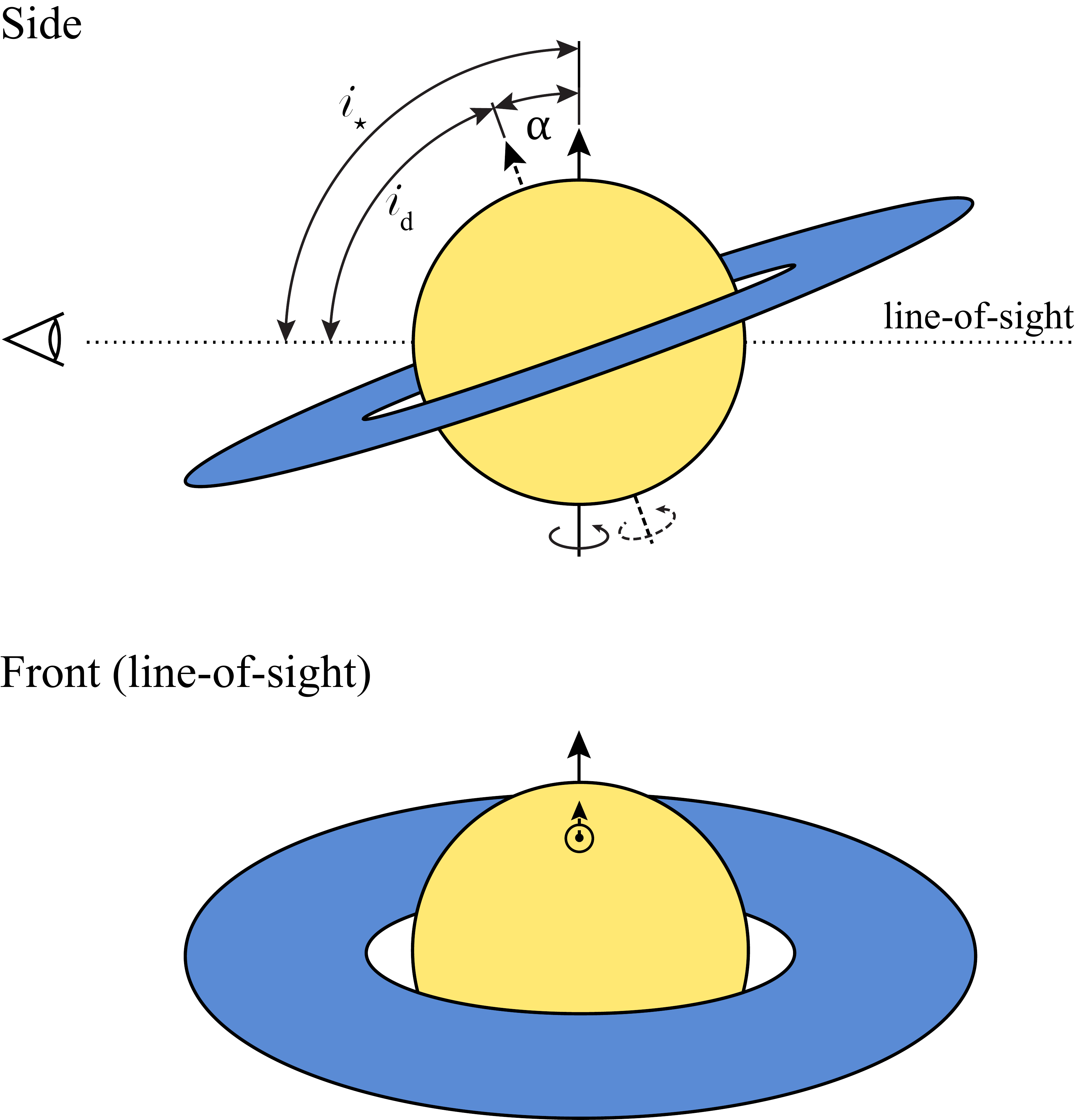}
    \caption{Cartoon of a disk-star system and the relevant coordinates. Top: side-on profile of the system where the line-of-sight of the observer enters from the left. Bottom: front profile of the system along the line-of-sight (i.e., the perspective of the observer where the view from the top panel is rotated 90$^{\circ}$ with respect to the stellar spin axis). The stellar and disk axes of rotation are shown as solid and dashed vectors, respectively, and make angles \istar{} and \idisk{} with the line-of-sight, respectively. The angular difference between the inclinations is defined as $\alpha \equiv i_{\star} - i_{\mathrm{d}}$.}
    \label{fig:alignment_diagram}
\end{figure}

We define the disk-star alignment angle ($\alpha$) as the difference between the inclination angles of the disk orbital axis and the stellar spin axis. This is not to be confused with the sky-projected spin-orbit angle in transiting exoplanet studies \citep[i.e., the sky-projected obliquity $\lambda$; see, e.g.,][]{fabrycky_exoplanetary_2009, dong_hierarchical_2023} which is perpendicular to $\alpha$.
We show a diagram of a misaligned protoplanetary disk in Figure~\ref{fig:alignment_diagram}.

Recent studies using observations from \alma{} have found misaligned protoplanetary disks that are pole-on \citep[$\sim$0$^{\circ}$; e.g.,][]{kennedy_circumbinary_2019, ansdell_are_2020} or have potentially retrograde \citep[e.g.,][]{kraus_gw_2020} orbits. High-resolution imaging has also revealed shadows in a significant fraction of so-called transition disks that can be best explained by a misalignment between inner and outer disks \citep[e.g.,][]{avenhaus_structures_2014, casassus_inner_2018}. How often such misalignments occur is unclear. 

There have only been a few statistical surveys of disk-star alignment, primarily focusing on small (15--20) samples of debris \citep{2011MNRAS.413L..71W, greaves_alignment_2014} or protoplanetary \citep{davies_stardisc_2019} disks. These have generally found that most systems are aligned or weakly misaligned ($<$30$^{\circ}$). One more recent study found 6 of 31 debris disks were significantly misaligned \citep{hurt_evidence_2023}.

These studies often used previously published stellar parameters which can be unreliable for cool \citep[e.g.,][]{mann_how_2015, newton_rotation_2016} and \pms{} stars \citep[e.g.,][]{kraus_massradius_2015, rizzuto_dynamical_2016}. They were also limited in sample size and stuck to a basic comparison between the inferred stellar and disk inclinations without accounting for the statistical biases and co-dependence of these inputs. While such studies are valuable, these limitations make it difficult to draw statistical conclusions.

This work generally relies on computing the stellar inclination (\istar) from the combined rotation period (\prot), stellar radius (\rstar), and rotational spectral broadening (\vsini). This is effectively comparing the line-of-sight broadening to the equatorial velocity (\veq) as in \citet{campbell_inclination_1985}:
\begin{equation}
    \label{eqn:sini}
    \sin i_{\star} = \frac{v\sin i_{\star}}{v_\mathrm{eq}} = \frac{P_{\textnormal{rot}} \, v\sin i_{\star}}{2\pi R_{\star}}
    \,\textnormal{.}
\end{equation}

This method has been similarly used for studying the alignment between the orbits of (transiting) planets and the rotation of their stellar hosts \citep[e.g.][]{mann_tess_2022,wood_tess_2023}. However, the approach is far more complicated than implied by Equation~\ref{eqn:sini} due to the dependent and non-Gaussian probability distributions and (large) measurement uncertainties \citep[see][for further discussion]{morton_obliquities_2014, masuda_inference_2020}. Further, young disk-bearing stars tend to have faster rotation (and hence more easily measured \vsini{} and \prot{}), but the presence of the protoplanetary disk and the difficulties obtaining accurate radii of \pms{} stars make this more challenging than studying mature planet hosts. 

In this paper, we explore the challenges associated with estimating precise stellar parameters and a means of combining them with the disk inclination (\idisk{}) to study the disk-star alignment distribution. Our goal is to provide an approach to measure the statistical projected alignment between stars and their protoplanetary disks with realistic uncertainties. In Section~\ref{sec:methods}, we outline our methodology for estimating the disk-star alignment angle.
We test part of our basic methodology to measure stellar parameters for a sample of \pms{} stars lacking protoplanetary disks, which we discuss in Section~\ref{sec:betapic}.
In Section~\ref{sec:synthetic_sample}, we test the full method on a synthetic population of disk-bearing stars.
Finally, in Section~\ref{sec:conclusion}, we summarize the results of the various tests and offer concluding remarks on how our methods can be applied to a real disk-bearing star population.

\section{Methods} \label{sec:methods}

Most simply, we need \istar{} and \idisk{} for a sample of targets. For \istar{}, we need \vsini{}, \rstar{}, and \prot{}, and realistic uncertainties, as well as a framework to handle complications when applying Equation~\ref{eqn:sini}. We describe each of these parameters in the subsections below, followed by
a description on how one can combine the sample within a hierarchical Bayesian framework, as well as a discussion of complications from missing spatial information and multi-star systems.

\subsection{Projected Stellar Rotation Velocity} \label{sec:vsini}

\vsini{} is most commonly derived from the broadening of spectral lines in high-resolution spectra. \vsini{} is observationally derived as a single parameter despite containing information on both the rotation and inclination of the star. A particular problem for this program, \vsini{} can be biased in the youngest stars due to degeneracies with pressure and magnetic broadening. 

\begin{figure*}
    \centering
    \includegraphics[width=\textwidth]{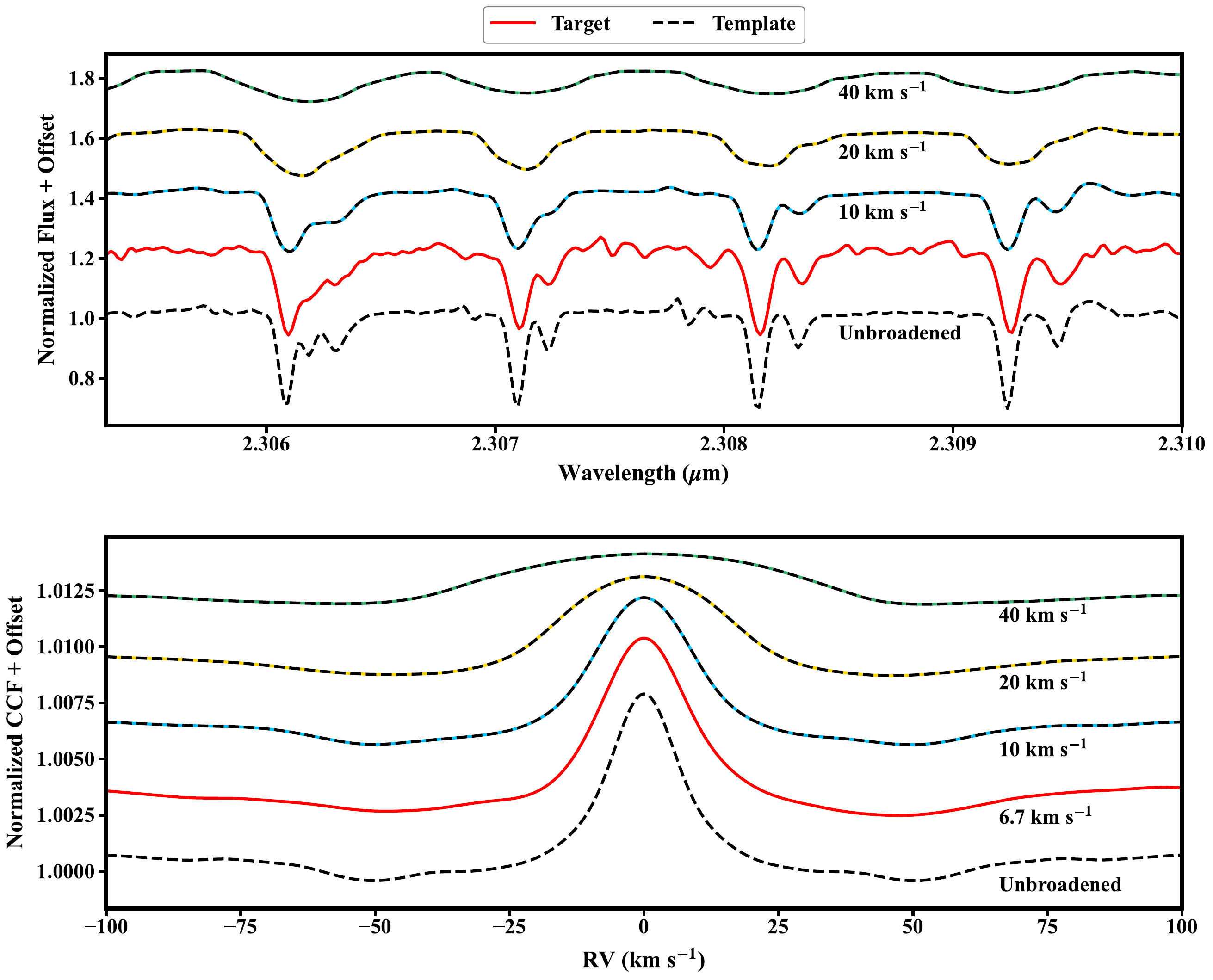}
    \caption{An application of the \vsini{} method to an arbitrary \pms{} star. Top: the normalized spectrum of the target (red solid line), the unbroadened spectral template (black and white dashed line), and the same spectral template broadened at 10\,\kms{}, 20\,\kms{}, and 40\,\kms{} (black and {blue/yellow/green} dashed lines, respectively). Bottom: cross-correlation of each spectrum from the top panel with the unbroadened template (matching colors). The target CCF is broader than the unbroadened CCF and narrower than the 10\,\kms{} broadened CCF, hence the estimated \vsini{} lies somewhere in between (6.7\,\kms{}).}
    \label{fig:vsini_method}
\end{figure*}

We adopted the method used by \citet{kesseli_magnetic_2018}, which is similar to methods used in some previous studies \citep[e.g.,][]{west_first_2009, muirhead_characterizing_2013, reiners_carmenes_2018}. In brief, we compared a given target spectrum to a slowly-rotating template spectrum of a star with the same spectral type as the target. The templates have negligible \vsini{} ($\ll$2\,\kms{}) compared to the target spectrum and instrumental broadening. We artificially broadened the template with a grid of \vsini{} values (from 2--62\,\kms{}), and cross-correlated each broadened template with the original, unbroadened template. Then, we cross-correlated the target spectrum with the unbroadened template spectrum. Finally, we measured the full-width-at-half-maximum (FWHM) of the target's cross-correlation function (CCF), and we linearly interpolated the FWHM onto the grid of FWHM values from the broadened template CCFs to determine the \vsini{} value. In Figure~\ref{fig:vsini_method}, we show an example of this method.

Artificial broadening requires a limb-darkening coefficient, which we calculated using the Python Limb-Darkening Toolkit \citep[LDTk;][]{parviainen_ldtk_2015}, which uses models from \citet{husser_new_2013}, and stellar parameters which we describe in Section~\ref{sec:rstar}. Varying limb-darkening for a range of stars suggests uncertainties in limb-darkening parameters have a negligible impact on the final \vsini{} when compared to other sources of uncertainty. 

For our tests, we opted to use K-band spectra from \igrins{}, as were used in \citet{kesseli_magnetic_2018}. \igrins{} spectra are high-resolution ($R\simeq45,000$), more than sufficient for the expected \vsini{} seen in disk-bearing stars (10's of \kms). \igrins{} also covers the near-infrared (NIR) CO bands, which have relatively equally-spaced and well-separated lines which are resistant to pressure and magnetic broadening that can be degenerate with \vsini{} \citep[see also][]{lavail_characterising_2019, lopez-valdivia_igrins_2021}. NIR spectrographs are also favorable because of high extinction in many star-forming regions.

A downside of using the CO bands is that the lines are weak in warmer ($\gtrsim$5000\,K) stars. At ages where disks are present \citep[0--5\,Myr][]{mamajek_initial_2009}, only stars $\gtrsim$1.5\,\mstar{} are this warm. For a typical initial mass function, this is only $\simeq$6\% of stars; these targets are also more challenging in terms of assigning stellar parameters. 

\igrins{} has archival spectra from which we can draw slow-rotating templates.\footnote{\url{https://igrinscontact.github.io/}.} We also tested using PHOENIX \citep{allard_progress_2013} model spectra, and found they often yield lower \vsini{} values (by 0.1--1\,\kms{}) and larger order-to-order uncertainties. This may be due to inaccuracies in the instrumental broadening and/or effects of missing astrophysics in the models. While the offset is small, because it is systematic, we opted for empirical templates. 

For each target, we estimated \vsini{} from echelle orders [4,~5,~6,~11,~12,~13,~14] (2.09--2.36\,\micron{}), sometimes excluding an order due to a strong emission line and/or no significant absorption lines. We combined the resulting \vsini{} values using a weighted mean where the weights were chosen as the maximum values of the normalized CCF peaks for each order and the errors were given as the standard error of the weighted mean. 
Typical uncertainties for this method are $\lesssim$1\,\kms{}, regardless of \vsini{} or spectral type. When one or multiple orders give $v\sin i_{\star} \leq 2$\,\kms{}, we treat the final combined \vsini{} value as an upper limit.

\subsection{Stellar Radius} \label{sec:rstar}

Stellar radii of young (single) stars are commonly estimated using the Stefan-Boltzmann relation \citep[e.g.,][]{mann_zodiacal_2016,davies_stardisc_2019}, the scale factor between a template or model spectrum and an absolutely calibrated spectrum of the star \citep[the infrared-flux method;][]{blackwell_stellar_1977,casagrande_absolutely_2010,newton_tess_2019}, and/or interpolation from a grid of stellar models \citep[e.g.,][]{muirhead_characterizing_2012,mayo_275_2018,loaiza-tacuri_spectroscopic_2023}. The first two methods tend to be applied when handling a single star or a set of similar systems (e.g., just field M dwarfs). The reason is because they depend on the choice of template, model, and available photometry, which varies significantly with age, \teff{}, surface gravity, and extinction. When working with a more diverse set of stars, it is preferable to use evolutionary models, where one can achieve results across the mass function at scale. 

The challenge for evolutionary models is they tend to under-predict the radii of cold \citep[e.g.,][]{mann_how_2015} and/or young stars \citep[e.g.][]{rizzuto_dynamical_2016}. However, models from \parsec{} contain an empirical correction for this. Models from \dsep{} also reproduce observed properties of young stars \citep[e.g.,][]{kraus_massradius_2015,david_age_2019} after adding in effects of magnetic fields \citep[e.g.,][]{feiden_magnetic_2016}. 

For this work, we focus on estimating \rstar{} from a stellar evolution model given multi-band photometry. To this end, we have developed \rstarcode{}, a Python-based pipeline and analysis tool for estimating \rstar{} and other stellar parameters from a general set of observational photometry (and a parallax) and an input model grid.\footnote{The stellar parameter estimation code is available at \url{https://github.com/mjfields/stelpar}.} The code is similar to others \citep[e.g.,][]{koposov_segasaiminimint_2023}, but ensures a homogeneous treatment across stellar types.

\subsubsection{Stellar Evolutionary Models} \label{sec:evolution_models}

We consider three evolutionary model grids: non-magnetic \dsep{}, a \dsep{}-based grid with magnetic enhancement \citep{feiden_self-consistent_2012}, and a \parsec{} grid. Both \dsep{}-derived models cover ages across the entire pre-main- and main-sequence (1\,Myr to 10\,Gyr) and stellar masses (\mstar{}) from 0.09--2.45\,\msun{}, depending on age. The \parsec{} grid has ages ranging from 1--500\,Myr.\footnote{We downloaded our custom \parsec{} isochrone model grid from \url{http://stev.oapd.inaf.it/cgi-bin/cmd}.}

Prior work on young stars often used the MESA Isochrones \& Stellar Tracks \citep[MIST;][]{choi_mesa_2016}, SPOTS models \citep{somers_spots_2020}, or BHAC15 \citep{baraffe_new_2015}. The first does not include the effects of magnetic fields, spots, or other activity. As such, they tend to give ages significantly lower than other methods, like lithium depletion \citep{malo_banyan_2014, wood_lithium_2023}. The SPOTS models only go to 1.3\,\msun{}, while our sample would go to at least 1.5\,\msun{} (age dependent cutoff). The BHAC15 models grid is more coarse and has a more limited set of photometry, which severely limits the precision of the output parameters. 

Studies of star-forming regions near the Sun have found most associations are Solar or slightly sub-Solar metallicity \citep{spina_gaia-eso_2017}. We only consider Solar metallicity for these models. Tests on slightly sub-Solar metallicities did not significant impact the final radii. 

To lessen the computational cost of interpolating the model grid at every iteration, we used \texttt{DFInterpolator} from the \texttt{isochrones} Python package \citep{morton_isochrones_2015} to pre-interpolate each \dsep{} model bilinearly in age and mass. This results in a grid spacing of $\leq$10\% in age (e.g., 0.1\,Myr from 1--10\,Myr, 1\,Myr from 10--100\,Myr, etc.), and 0.005\,\msun{} for $M_{\star} < 0.1\,M_{\odot}$ and 0.01\,\msun{} for $M_{\star} \geq 0.1\,M_{\odot}$ in mass. We did not repeat this for the \parsec{} model since we could choose the age spacing when we downloaded the grid. As a result, the mass spacings are not the same between all the models, but spacings were still smaller than the measurement uncertainties. 

\subsubsection{Input Photometry} \label{sec:meas_phot}

For all fits, we download (when available) photometry from \twomass{}; \gaia{} Data Release 3 \citep[DR3;][]{vallenari_gaia_2023}; \sdss{} Data Release 16 \citep[DR16;][]{ahumada_16th_2020}; $B$ and $V$ from \apass{} Data Release 9 \citep[DR9;][]{henden_apass_2015}; \Hp{} from \hipparcos{}; and \BT{} and \VT{} from \tycho{}. We did not include \ps1{} photometry for now, as our test samples are too bright (beyond PAN-STARRS saturation limit) and due to evidence of offsets for cooler stars \citep{kado-fong_m_2016}. We also found that most fits are not limited by the photometry; Pan-STARRS and similar surveys \citep[e.g., SkyMapper;][]{wan_galactic_2018} can be included following the same method if needed. We also excluded photometry from NASA's Wide-field Infrared Survey Explorer \citep[WISE;][]{wright_wide-field_2010}, which would likely be contaminated by emission from the disk.

\subsubsection{Photometry Model Construction} \label{sec:phot_model}

We compared the model to observed photometry within a \mcmc{} framework with the \texttt{emcee} Python package \citep{foreman-mackey_emcee_2013}. At each iteration, we interpolate the evolutionary model grid by age and \mstar{}, neglecting metallicity (with Solar metallicity the model point is uniquely determined by only age and \mstar{}). 

We used models with close grid spacings (either from pre-interpolation or custom download; see Section~\ref{sec:evolution_models}), so there is no advantage to bilinear interpolation \citep[e.g., with \texttt{DFInterpolator} from \texttt{isochrones};][]{morton_isochrones_2015} which comes with a significant decrease in computation speed. Therefore, we choose to interpolate by searching for the nearest-neighbor value in age, and linearly interpolating in \mstar{}. This interpolation method can create a bias toward grid points, since we are choosing discrete age values. However, this does not change any of our results because the age and \mstar{} uncertainties are much larger than the grid spacing.

In total, \rstarcode{} contains four free parameters: age, \mstar{}, extinction (\av{}), and $f$. The first two are set by the models. For \av{}, we use the \texttt{synphot} Python package \citep{stsci_development_team_synphot_2018} following the extinction model of \citet{cardelli_relationship_1989}. The final parameter, $f$ is a factor describing the underestimation of the errors on the measured photometry and/or uncertainties in the model, measured in magnitudes. In practice, $f$ would be more accurately modeled as a vector in wavelength and age, which would account for the fact that models tend to struggle more at optical wavelengths due to missing opacities in cool stars \citep{mann_spectro-thermometry_2013}, stronger variability in the blue \citep{gully-santiago_placing_2017, mori_characterization_2024}, and accretion \citep{herczeg_uv_2008}. Until such external constraints are available, we kept this as a single number applied to all photometry for a given star.

When comparing the model to photometry, \rstarcode{} assumes Gaussian uncertainties. By default, the code includes uniform priors within a set of user-configurable bounds for each fit parameter. However, the user may optionally choose to define Gaussian priors for any of the fit parameters and/or \teff{}, the latter of which acts as a prior on both age and \mstar{}. Since the parameters above uniquely determine the model selection, the full posterior distributions for age and \mstar{} can be turned into posteriors on \rstar{}, \teff{}, log-surface gravity ($\log g$), log-luminosity ($\log L$), and stellar density (\rhostar{}).

\subsubsection{Comparison to Empirical Values} \label{sec:density_results}

To test our method and check for additional systematic uncertainties, we compared the inferred \rhostar{} to those for seven young transiting systems. In cases of low eccentricity, the transit duration can yield a strong constraint on the stellar density, \rhostar{} \citep{seager_unique_2003}. The transiting sample encompasses a reasonable range of stellar masses (M dwarfs to F stars) and includes stars with transitional and debris disks (3--20\,Myr). 

A comparison in \rhostar{} can be misleading because inaccurate \rstar{} can be masked by matching inaccurate \mstar{} values. However, since \rhostar{} is scaled steeply with \rstar{} (cubic), even (modest) 10--15\% precision on \rhostar{} would only lead to 3--5\% errors in \rstar{}.

The systems were selected based on their age and planet multiplicity. The youngest systems are expected to have low eccentricity \citep[due to dampening from the protoplanetary disk;][]{papaloizou_orbital_2000}. Multi-planet systems tend to have low eccentricities and yield stronger constraints on stellar density \citep{van_eylen_eccentricity_2015}.

\begin{figure*}
    \centering
    \includegraphics[width=\textwidth]{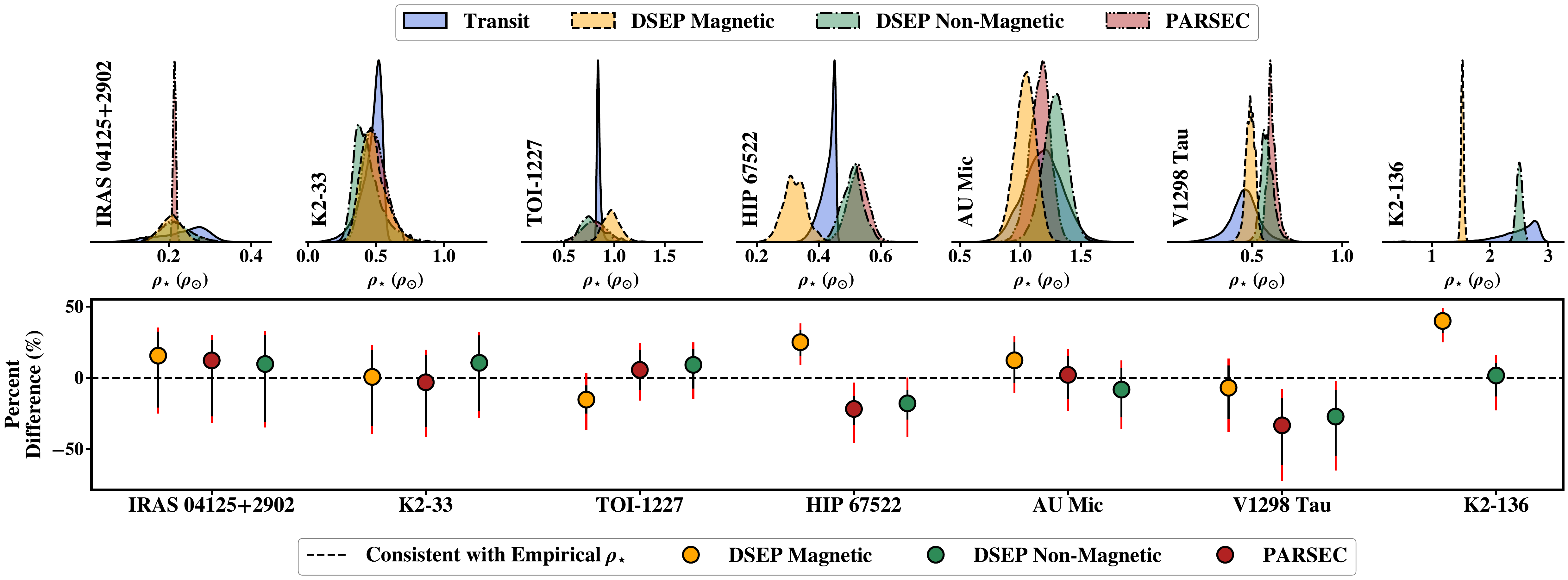}
    \caption{Comparison of \rhostar{} posteriors for seven young planet-hosts (top panels), and percent differences between the model-derived \rhostar{} posteriors and the empirically-derived posteriors (bottom panel). Transit-based \rhostar{} measurements are in blue, with those from the \dsep{}-derived magnetic in orange, \dsep{} non-magnetic in green, and \parsec{} models in red. Most of our estimates agreed with empirical \rhostar{} values, independent of the model used. The two outliers were K2-136 and HIP 67522. K2-136 is the oldest system ($\simeq$800\,Myr), and was meant primarily to ensure the non-magnetic \dsep{} model was preferred for less active/older systems. The empirical \rhostar{} of HIP 67522 is not captured by any single model (3$\sigma$ agreement at best), suggesting that multiple models may be needed to flag such cases and provide an aggregated estimate. Red errorbars in the bottom panel are the results of a 5\% perturbation in \rstar{}, and their agreement affirms we can estimate \rstar{} to better than 5\% for the sample.}
    \label{fig:rstar_density}
\end{figure*}

We show the comparison in Figure~\ref{fig:rstar_density} with supplementary information in Table~\ref{tab:young_planet_hosts}. Most of our results agreed with observations regardless of the model. The \dsep{}-magnetic model result for K2-136 shows a significant offset, but this is as expected. K2-136 is the oldest target in the sample ($\simeq$800\,Myr) and is relatively quiet; magnetic enhancement is probably not required. Indeed, the non-magnetic \dsep{} model gave better agreement. We did not use the \parsec{} model for K2-136 because our custom \parsec{} grid only goes to 500\,Myr in age.

Results for HIP~67522 also show disagreement. This is unlikely to be due to our assumption that the planetary orbits are circular; a second planet has since been identified in that system \citep{barber_tess_2024} and the lack of dynamical interactions between the planets (transit timing variations) suggest the masses and eccentricities of both planets are low \citep{thao_featherweight_2024}. However, a large part of this disagreement is because the models yield exceptionally small uncertainties, sometimes better than 2\% on radius. If we adopt a flat 5\% uncertainty on radius \citep{tayar_guide_2022}, there is agreement across the sample.

One surprising piece is that the DSEP non-magnetic model is consistent too. On further inspection, this is because we had to use \rhostar{} rather than a direct comparison in \mstar{}. Non-magnetic \dsep{} is yielding radii that are, on average, smaller than the values from the other two grids and smaller when compared to radii measurements from direct spectral analysis in the discovery papers. However, the masses are also smaller, which masks some of the difference. Another effect is that we placed priors on the age, derived from the parent population. Upon removing this prior, the DSEP non-magnetic model tended to yield discrepant ages, as seen in more extensive studies \citep{rizzuto_dynamical_2016}. 

Overall, we find that our model-based radii are reliable, although the uncertainties are underestimated. This holds down to the 3\,Myr system with a protoplanetary disk \citep[IRAS 04125+2902;][]{barber_giant_2024}. In absolute terms, the agreement suggests we can achieve a $\simeq$5\% uncertainty on \rstar{}. There is also no evidence of a systematic offset. The \dsep{}-magnetic model appears to perform the best, although \parsec{} does similarly well. While the sample is small, they are broadly consistent with prior comparisons in mass-radius space using young eclipsing binary (EB) systems \citep[e.g.,][]{kraus_massradius_2015, gillen_new_2017, david_age_2019}. We note that such studies often find issues with the \teff{}-luminosity scale, but our method appears less impacted by this offset.

\begin{table*}
    \footnotesize
    \centering
    \rotatebox{90}{
        \begin{threeparttable}
            \caption{Literature information and estimated stellar parameters for the sample of young planet hosts.}
            \label{tab:young_planet_hosts}
            
            \begin{tabular}{p{1.75cm} c c c c c c c}
                \toprule
                Parameter & IRAS 04125+2902 & K2-33 & TOI-1227 & HIP 67522 & AU Mic & V1298 Tau & K2-136 \\
                \midrule
                2MASS & J04154278$+$2909597 & J16101473$-$1919095 & J12270432$-$7227064 & J13500627$-$4050090 & J20450949$-$3120266 & J04051959$+$2009256 & J04293897$+$2252579 \\
                RA (J2000)\tnote{a} (h:m:s) & 04:15:42.80 & 16:10:14.73 & 12:27:04.16 & 13:50:06.24 & 20:45:09.88 & 04:05:19.60 & 04:29:39.09 \\
                DEC (J2000) (d:m:s) & +29:09:59.54 & $-$19:19:09.79 & $-$72:27:06.67 & $-$40:50:09.24 & $-$31:20:33.00 & +20:09:25.31 & +22:52:57.22 \\
                Age (Myr) & $3.3^{+0.6}_{-0.5}$ & $9.3^{+1.1}_{-1.3}$ & $11 \pm 2$ & $17 \pm 2$ & $22 \pm 3$ & $23 \pm 4$ & $\simeq$800 \\
                \midrule
                \multicolumn{8}{c}{Prior} \\
                \midrule
                Age (Myr) & $3 \pm 1$ & $11 \pm 3$ & $11 \pm 2$ & $17 \pm 2$ & $22 \pm 3$ & $23 \pm 4$ & $725 \pm 100$ \\
                \teff{} (K) & $3922 \pm 111$ & $3450 \pm 70$ &  & $5675 \pm 75$ & $3700 \pm 100$ & $4970 \pm 120$ & $4499 \pm 50$ \\
                \av{} (mag) & $2.253 \pm 0.128$ & $0.64 \pm 0.08$ & $0.21 \pm 0.1$ &  &  &  & \phantom{} \\
                \midrule
                \multicolumn{8}{c}{DSEP Magnetic} \\
                \midrule
                \mstar{} (\msun{}) & $0.90^{+0.06}_{-0.06}$ & $0.47^{+0.04}_{-0.04}$ & $0.16^{+0.01}_{-0.01}$ & $1.24^{+0.05}_{-0.04}$ & $0.65^{+0.02}_{-0.02}$ & $1.07^{+0.02}_{-0.02}$ & $0.79^{+0.01}_{-0.01}$ \\
                \addlinespace[1ex]
                \rstar{} (\rsun{}) & $1.64^{+0.07}_{-0.05}$ & $1.00^{+0.04}_{-0.04}$ & $0.55^{+0.01}_{-0.01}$ & $1.57^{+0.08}_{-0.06}$ & $0.85^{+0.02}_{-0.02}$ & $1.30^{+0.03}_{-0.03}$ & $0.80^{+0.01}_{-0.01}$ \\
                \addlinespace[1ex]
                \rhostar{} (\rhosun{}) & $0.21^{+0.03}_{-0.03}$ & $0.47^{+0.10}_{-0.07}$ & $0.97^{+0.08}_{-0.08}$ & $0.32^{+0.03}_{-0.03}$ & $1.04^{+0.08}_{-0.09}$ & $0.49^{+0.03}_{-0.03}$ & $1.52^{+0.02}_{-0.02}$ \\
                \midrule
                \multicolumn{8}{c}{DSEP Non-Magnetic} \\
                \midrule
                \mstar{} (\msun{}) & $0.63^{+0.05}_{-0.04}$ & $0.37^{+0.03}_{-0.03}$ & $0.12^{+0.01}_{-0.01}$ & $1.19^{+0.03}_{-0.03}$ & $0.64^{+0.03}_{-0.03}$ & $1.16^{+0.02}_{-0.02}$ & $0.72^{+0.01}_{-0.01}$ \\
                \addlinespace[1ex]
                \rstar{} (\rsun{}) & $1.43^{+0.05}_{-0.05}$ & $0.96^{+0.04}_{-0.05}$ & $0.54^{+0.01}_{-0.01}$ & $1.33^{+0.04}_{-0.04}$ & $0.79^{+0.02}_{-0.02}$ & $1.26^{+0.03}_{-0.03}$ & $0.66^{+0.01}_{-0.01}$ \\
                \addlinespace[1ex]
                \rhostar{} (\rhosun{}) & $0.21^{+0.04}_{-0.03}$ & $0.42^{+0.11}_{-0.07}$ & $0.77^{+0.14}_{-0.09}$ & $0.51^{+0.03}_{-0.04}$ & $1.29^{+0.10}_{-0.10}$ & $0.58^{+0.04}_{-0.03}$ & $2.50^{+0.04}_{-0.05}$ \\
                \midrule
                \multicolumn{8}{c}{PARSEC} \\
                \midrule
                \mstar{} (\msun{}) & $0.67^{+0.02}_{-0.02}$ & $0.51^{+0.03}_{-0.03}$ & $0.22^{+0.01}_{-0.01}$ & $1.18^{+0.03}_{-0.03}$ & $0.68^{+0.02}_{-0.01}$ & $1.15^{+0.01}_{-0.02}$ & \phantom{} \\
                \addlinespace[1ex]
                \rstar{} (\rsun{}) & $1.46^{+0.01}_{-0.01}$ & $1.02^{+0.04}_{-0.04}$ & $0.65^{+0.03}_{-0.03}$ & $1.31^{+0.04}_{-0.04}$ & $0.83^{+0.02}_{-0.02}$ & $1.24^{+0.02}_{-0.03}$ & \phantom{} \\
                \addlinespace[1ex]
                \rhostar{} (\rhosun{}) & $0.214^{+0.001}_{-0.001}$ & $0.46^{+0.10}_{-0.06}$ & $0.79^{+0.12}_{-0.12}$ & $0.53^{+0.03}_{-0.03}$ & $1.17^{+0.08}_{-0.09}$ & $0.61^{+0.03}_{-0.02}$ & \phantom{} \\
                \midrule
                \multicolumn{8}{c}{Empirical (Literature)\tnote{b}} \\
                \midrule
                \mstar{} (\msun{}) & $0.70 \pm 0.04$ & $0.56^{+0.09}_{-0.09}$ & $0.170 \pm 0.015$ & $1.22 \pm 0.05$ & $0.50 \pm 0.03$ & $1.10 \pm 0.05$ & $0.74 \pm 0.02$ \\
                \addlinespace[1ex]
                \rstar{} (\rsun{}) & $1.45 \pm 0.10$ & $1.05^{+0.07}_{-0.07}$ & $0.56 \pm 0.03$ & $1.38 \pm 0.06$ & $0.75 \pm 0.03$ & $1.305 \pm 0.070$ & $0.66 \pm 0.02$ \\
                \addlinespace[1ex]
                \rhostar{} (\rhosun{}) & $0.22 \pm 0.03$ & $0.51^{+0.04}_{-0.07}$ & $0.94 \pm 0.18$ & $0.46 \pm 0.06$ & $1.18 \pm 0.16$ & $0.46 \pm 0.08$ & $2.50^{+0.13}_{-0.12}$ \\
                Ref.\tnote{c} & 1 & 2 & 3 & 4 & 5 & 6 & 7, 8 \\
                \bottomrule
            \end{tabular}
            
            \begin{tablenotes}
                \item[a] RA and DEC values from \gaia{} DR3.
                \item[b] \citet{barber_giant_2024} and \citet{mann_tess_2022} used an ensemble of techniques to derive \mstar{} and \rstar{} for IRAS 04125+2902 and TOI-1227, respectively, which included \rstarcode{}.
                \item[c] Literature references: 
                [1] \citet{barber_giant_2024};
                [2] \citet{mann_zodiacal_2016};
                [3] \citet{mann_tess_2022};
                [4] \citet{rizzuto_tess_2020};
                [5] \citet{plavchan_planet_2020};
                [6] \citet{david_warm_2019};
                [7] \citet{brandt_age_2015};
                and [8] \citet{mann_zodiacal_2018}.
            \end{tablenotes}
            
        \end{threeparttable}
    }
\end{table*}

\subsection{Stellar Rotation Period} \label{sec:prot}

Young stars ($\lesssim$10\,Myr) tend to have \prot{} values from 0.1 to 30\,days. The long-period end of this distribution is challenging for \ktwo{} and \tess{} due to the narrow observing windows (27--80\,days). However, the peak in the period distribution is $\simeq$2\,days, and only a small percentage of stars have \prot{} values longer than 10\,days \citep{rebull_rotation_2018}. 
Therefore, we used a search grid from  0.2--20\,days, which captures the vast majority of rotators. We adopted the \prot{} value that corresponds to the highest peak in the LS power spectrum, with an eye inspection to check for multiple periods and aliases.

For each star we generated \tess{} lightcurves by using a causal pixel model (CPM) as implemented in the \texttt{unpopular} package \citep{hattori_unpopular_2022} and detailed in \citet{barber_transit_2022}. \texttt{unpopular} creates lightcurves using the pixels outside the target aperture to model the pixel response of the pixels within the aperture. We subtracted the systematic model from the raw aperture lightcurve, which results in the CPM lightcurve. 

Boyle et al. (in prep.) estimated uncertainties on \prot{} values measured in this way. Their method was to compare periods estimated from different \tess\ sectors as well as between \ktwo\ and \tess{} data of the same set of stars. In principle, this method includes effects like spot evolution and differential rotation, provided the gap between sectors and between \ktwo\ and \tess\ is long compared to the evolutionary timescale. They found that \prot{} can be determined to better than 5\% for stars with rotation periods below 10\,days, and $\simeq$2\% for typical rotators in the disk-bearing sample ($P_{\mathrm{rot}} \simeq 2$\,days). Thus, we expect \prot{} will have little impact on the total error budget of the final result.

\subsection{Stellar Inclination} \label{sec:cosi}

Deriving \istar{} from the \vsini{}, \rstar{}, and \prot{} values we measured/inferred requires more attention beyond a direct application of Equation~\ref{eqn:sini}. It is possible, for instance, to measure $v\sin i_{\star} > v_{\mathrm{eq}}$ (which is nonphysical) simply due to random measurement uncertainties \citep[see, e.g.,][]{morton_obliquities_2014}. 

\citet{masuda_inference_2020} presented a framework for measuring \cosi{} (rather than \istar{}) that accounts for the statistical co-dependence of \veq{} and \vsini{}. Briefly, the authors state that it is incorrect to sample \veq{} and \vsini{} independently. Instead, it is more appropriate to sample \veq{} and \cosi{} independently and derive \vsini{} from the combination of the two. Following \citet{morton_obliquities_2014} and \citet{masuda_inference_2020}, we estimate \cosi{} within an \mcmc{} framework using \texttt{emcee}.\footnote{Our Python implementation of \citet{masuda_inference_2020} is available at \url{https://github.com/mjfields/cosi}.}

Uncertainties on \cosi{} can be large, corresponding to inclination uncertainties of 10--30$^\circ$, with significant variation with both \veq{} and \istar. In this sense it is challenging to determine \istar{} of any given target beyond broad categories. However, we can still combine many such measurements to study the overall alignment frequency. 

\subsection{Disk Inclination}\label{sec:idisk}

\idisk{} derived from \alma{} observations are widely available in the literature. To determine typical uncertainties ($\sigma_{i_{\mathrm{d}}}$) from such measurements, we drew from multiple prior surveys which used \alma{}. Specifically, we drew \idisk{} measurements from \citet{ansdell_alma_2016,ansdell_are_2020}, \citet{barenfeld_measurement_2017}, \citet{tazzari_physical_2017}, \citet{tripathi_millimeter_2017}, \citet{huang_disk_2018}, and \citet{long_compact_2019}. Collectively, this provided 138 measurements from 107 disks spanning the protoplanetary disk lifetime ($\sim$1--10\,Myr). The systems were surveyed from four regions: the $\rho$\,Ophiuchus star-forming region \citep[$\sim$1\,Myr;][]{andrews_submillimeter_2007}, the Lupus star-forming complex \citep[$\sim$3\,Myr;][]{alcala_x-shooter_2017}, the Taurus Molecular Cloud \citep[$\lesssim$6\,Myr;][]{krolikowski_gaia_2021}, and the Upper Scorpius OB association \citep[$\sim$10\,Myr;][]{david_age_2019}. 

\begin{figure*}
    \centering
    \includegraphics[width=\textwidth]{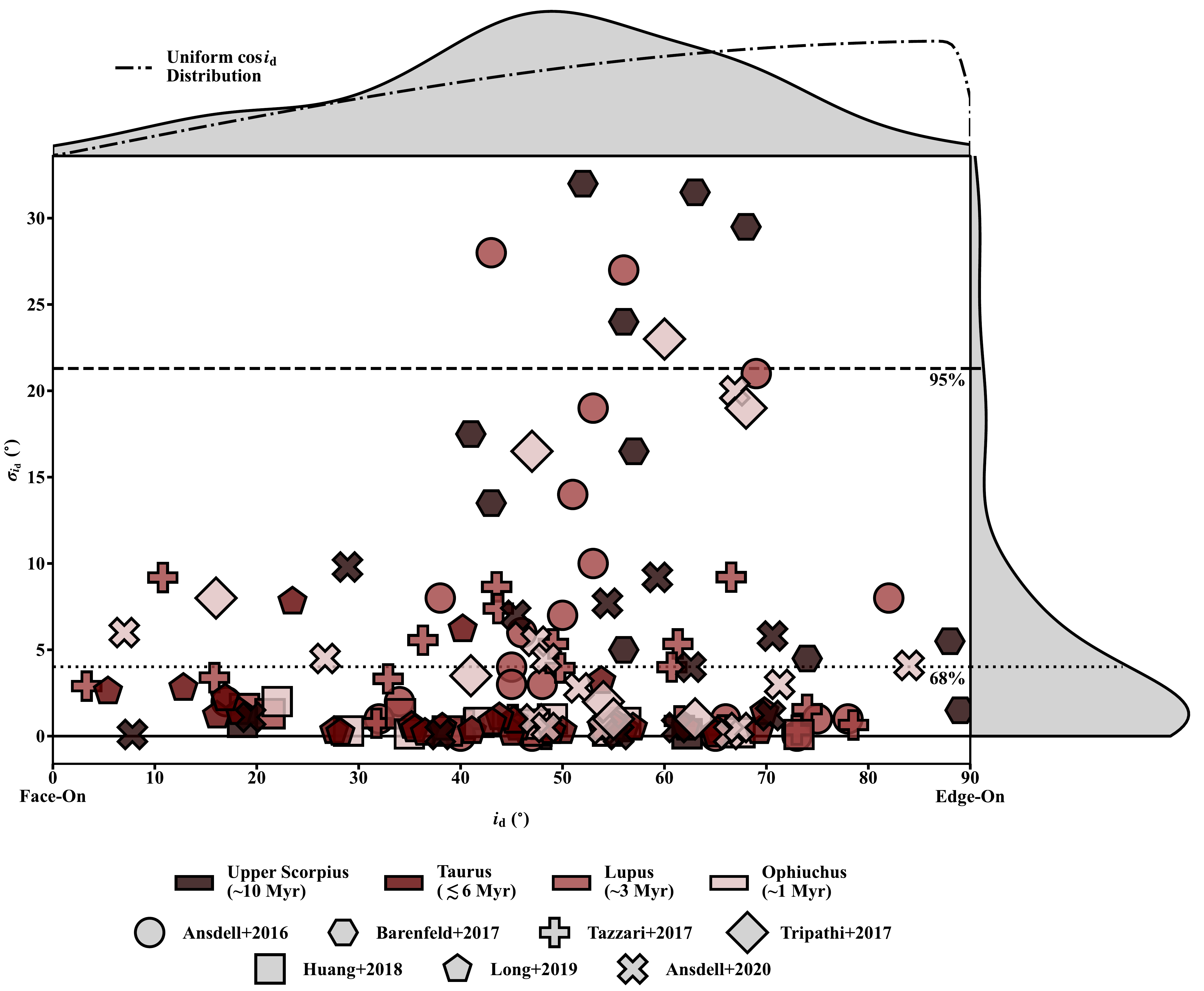}
    \caption{Uncertainties on \idisk{} ($\sigma_{i_{\mathrm{d}}}$) vs. \idisk{} from seven previous surveys that used \alma{} data (138 total measurements): \citet[][circles]{ansdell_alma_2016}, \citet[][hexagons]{barenfeld_measurement_2017}, \citet[][plusses]{tazzari_physical_2017}, \citet[][diamonds]{tripathi_millimeter_2017}, \citet[][squares]{huang_disk_2018}, \citet[][pentagons]{long_compact_2019}, and \citet[][exes]{ansdell_are_2020}. The regions are colored from dark to light maroon according to descending age (Upper Scorpius, Taurus, Lupus, and $\rho$\,Ophiuchus); those lacking association memberships were excluded. On the right is a histogram of $\sigma_{i_{\mathrm{d}}}$ values (gray). Thresholds marking 68\% and 95\% of the data are shown as dotted and dashed lines, respectively. On the top is a histogram of \idisk{} values (gray) compared to the theoretical distribution of uniform $\cos i_{\mathrm{d}}$ values (dashed-dotted line). The deficit at edge-on inclinations is observational bias. The `bump' at $i_{\mathrm{d}} \simeq 45^{\circ}$ is consistent with random noise. Most measurements (68\%) have uncertainties $\lesssim$4$^{\circ}$, meaning that literature \idisk{} values will likely have a small impact on the total error of final alignment results.}
    \label{fig:disk_uncertainties}
\end{figure*}

We show the results of our analysis in Figure~\ref{fig:disk_uncertainties}. For 68\% and 95\% of measurements, uncertainties were $\lesssim$4$^{\circ}$ and $\lesssim$21$^{\circ}$, respectively. There was no significant trend in the uncertainties with disk orientation or age, although the overall detection rate drops with increasing age. There is an almost uniform spread of measurements between face-on and edge-on disks, as expected for (nearly) complete surveys. The exception is a slight deficit of systems with $80^{\circ} < i_{\mathrm{d}} < 90^{\circ}$; this is likely due to the fact that edge-on disks will occult their host and therefore be less likely to be included in the \alma{} observing list.

\begin{figure}
    \centering
    \includegraphics[width=0.5\textwidth]{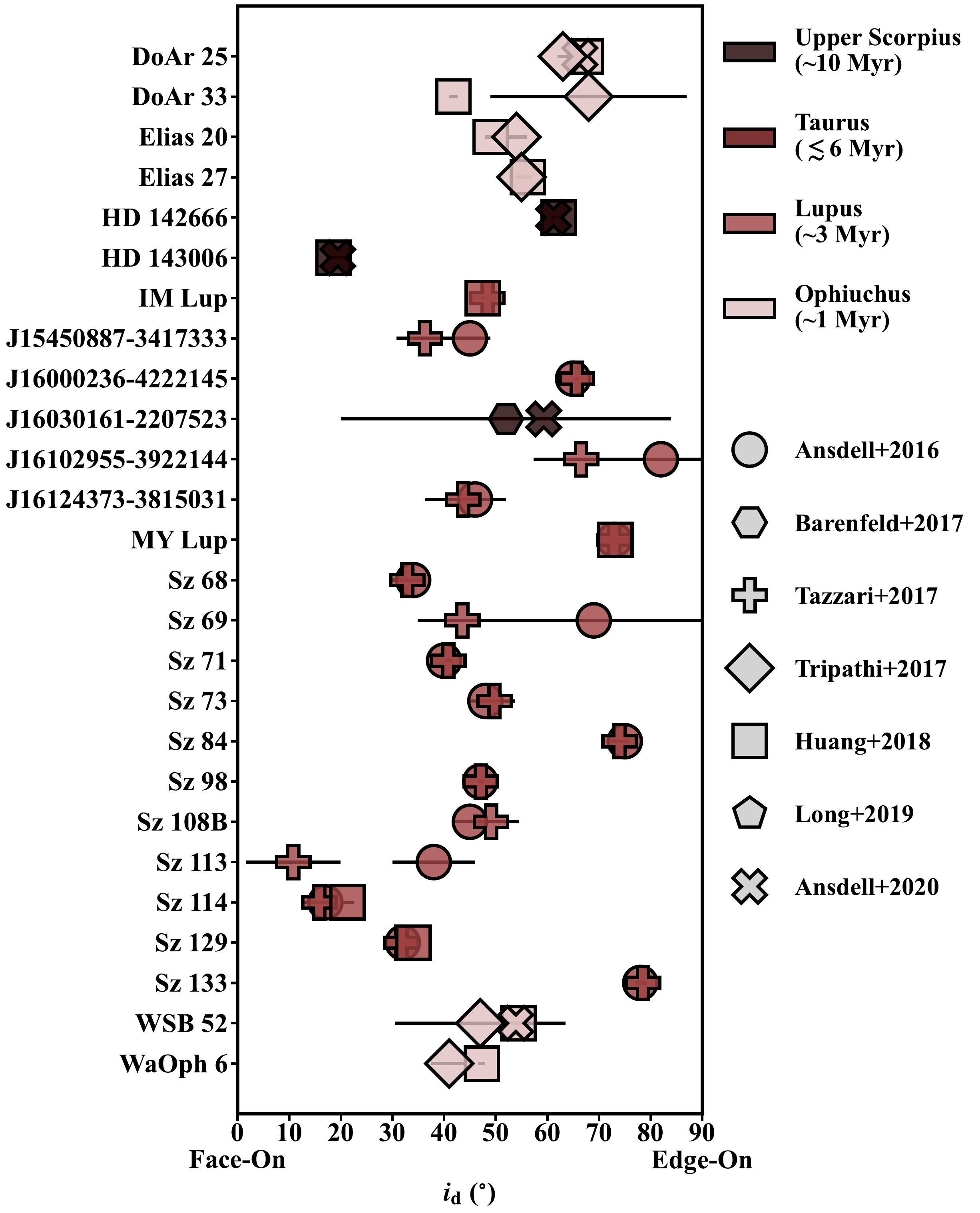}
    \caption{A comparison of \idisk{} for 26 systems that have two or more independent measurements across seven previous surveys that used \alma{} data. The coloring and marker shapes follow the same style as in Figure~\ref{fig:disk_uncertainties}. 24 of the 26 disks have \idisk{} measurements that agree within 1$\sigma$ (and the outlier measurements agree within 2$\sigma$).}
    \label{fig:disk_duplicates}
\end{figure}

As a check on the literature \idisk{} measurements, we compared 26 systems that had independent \idisk{} measurements for the same system in at least two or more surveys. These are not necessarily independent measurements; many are using the same underlying \alma{} data. Rather, this is a test of sensitivity to the fitting method. We show the results of our literature comparison in Figure~\ref{fig:disk_duplicates}. Of the 26 overlapping systems, 24 had measurements that agreed across multiple surveys within 1$\sigma$, and the other systems' measurements agreed within 2$\sigma$.  

The distribution seen in Figure~\ref{fig:disk_uncertainties} shows a deficit of edge-on disks and a surplus of intermediate-angle ($i_{\mathrm{d}}\simeq45^\circ$) disks compared to the expected uniform $\cos{i_{\mathrm{d}}}$ distribution. The former is due to observational bias. Disks flare outward by 10--25$^\circ$, so those with $i_{\mathrm{d}}\gtrsim70^\circ$ are more likely to be occulted/extincted by the disk. The lack of a visible optical component means these were more likely to be skipped in the sub-mm or \textit{Spitzer} survey that initially identified the infrared excesses. This bias against edge-on disks can be included in the model (Section~\ref{sec:hbm}). 

The excess at moderate inclinations appears to be a combination of disk measurements with high uncertainties and random noise. Given Poisson errors, the observed and expected distributions are marginally consistent with each other out to $i_{\mathrm{d}}=75^\circ$, particularly if we only consider disks with $\sigma_{i_{\mathrm{d}}}<10^\circ$.

Overall, this indicates \idisk{} is a small component of the total error budget (particularly compared to \istar{}), and that using literature determinations from \alma{} data and assuming the reported Gaussian uncertainties will not introduce a bias in the overall disk-star alignment measurement. Although the bias against edge-on systems likely needs to be included in the model.

\subsection{Fitting for Global Alignment}\label{sec:hbm}

As noted above, uncertainties on \istar{} can be large in any individual system.  However, we can still use these to draw conclusions about the overall alignment distribution within a population, provided we can control for systematic effects. To this end, we combine \istar{} and \idisk{} for all systems within a \hbm{}, where each disk-star offset ($\alpha^{\prime}_n$) is a free parameter evolving under the observational constraints (e.g., \vsini{}, \rstar{}, \idisk{}) simultaneously with global parameters that describe the population-level disk-star alignment.

The method is flexible to the assumed model of disk-star alignment. For simplicity, we assume the overall (global) alignment can be modeled as a Gaussian described by two parameters ($\mu$ and $\sigma$). Similarly, we also assume the alignments of the individual systems can be modeled as Gaussians (see Section~\ref{sec:synthetic_sample} for more detail). Combined with the individual systems the assumed Gaussian global alignment yields $N+2$ fit parameters. Physically, this corresponds to an offset in the alignment ($\mu$) and a spread around that offset ($\sigma$). It is just as simple to describe it with a Fisher distribution following \citet{morton_obliquities_2014}, a
uniform distribution (e.g., random alignment), or a combination of the two with a mixture amplitude (e.g., a mix of randomly aligned and aligned systems). 

We define the log-probability ($\log\mathcal{P}$) in two parts that are summed together: the sums of the log-likelihoods ($\log\mathcal{L}$) of observing the measured ensemble of alignments given $\alpha^{\prime}_n$ and of observing $\alpha^{\prime}_n$ given $\mu$ and $\sigma$:\footnote{In reality, we use the \textit{negative} log-probability to marginalize over the fit parameters, i.e., we maximize $-\log\mathcal{P} = -\sum\log\mathcal{L}{.}$}
\begin{equation}
    \label{eqn:hbm}
    \log\mathcal{P} = \sum_{n=1}^{N}\log\mathcal{L}(i_{\star, n}, i_{d, n}|\alpha^{\prime}_n) + \sum_{n=1}^{N}\log\mathcal{L}(\alpha^{\prime}_n|\mu, \sigma),
\end{equation}
where $i_{\star, n}$ and $i_{d, n}$ represent the measured inclinations of each system and $N$ is the total number of systems.

In the first part of Equation~\ref{eqn:hbm}, we compare the estimated \istar{} (converted from \cosi{}; Section~\ref{sec:cosi}) and \idisk{} (Section~\ref{sec:idisk}) with $\alpha^{\prime}_n$ and scale them by the combined uncertainties on \istar{} and \idisk{}, i.e.,
\begin{align*}
    \sum_{n=1}^{N}\log\mathcal{L}(i_{\star, n}, i_{d, n}|\alpha^{\prime}_n) &\propto \sum_{n=1}^{N}\frac{[\alpha^{\prime}_n - (i_{\star, n} - i_{d, n})]^2}{\sigma_{i_{\star, n}}^2 + \sigma_{i_{d, n}}^2}\\
    &+ \sum_{n=1}^{N}\log\left[2\pi(\sigma_{i_{\star, n}}^2 + \sigma_{i_{d, n}}^2)\right] \mathrm{.}
\end{align*}
In the second part, we compare $\mu$ to $\alpha^{\prime}_n$, scaled by $\sigma$, i.e.,
\begin{equation*}
    \sum_{n=1}^{N}\log\mathcal{L}(\alpha^{\prime}_n|\mu, \sigma) \propto \sum_{n=1}^{N}\left[\frac{(\mu - \alpha^{\prime}_n)^2}{\sigma^2} + \log\left(2\pi\sigma^2\right)\right] \mathrm{.}
\end{equation*}
In this way, $\mu$ and $\sigma$ are constrained by $\alpha^{\prime}_n$, which themselves are modulated by the previously-derived \istar{} and \idisk{}.

\subsubsection{Additional Effects}\label{sec:additional_effects}

Our treatment above does not consider the full three-dimensional alignment, only the line-of-sight-projected alignment. This issue is discussed in prior studies on both disk-star alignment \citep{davies_stardisc_2019} and more extensively on planet-star alignment \citep[e.g.,][]{dong_hierarchical_2023}. Since there is a missing angle, a misaligned system may appear aligned because most/all of the misalignment is hidden in the unseen angle. Similarly, systems mostly aligned in three dimensions might look preferentially misaligned if all of the misalignment is between the observed \istar{} and \idisk{}.

While this limits our ability to measure overall alignment in any given system, we could infer that a two-dimensional treatment would only affect global $\sigma$ in an ensemble analysis. Assuming there is no preferred direction to the misalignment or the observer angle, then the amount that any misalignment is hidden in the unseen angle should be effectively random. Thus, when \istar{} and \idisk{} are aligned, any misalignment must come from the difference in sky-projected stellar and disk angles ($\lambda_{\star}$ and $\lambda_d$, respectively). Indeed, modeling $\lambda_{\star}$ and $\lambda_d$ (or their difference) does not meaningfully change our \hbm{} results except that the true misalignment uncertainty is larger. The impact is also smaller than many of the other effects discussed in this paper (e.g., inclination uncertainties).

This breaks down if there is a preferred direction of misalignment, such as if disk warping were driven by the overall angular momentum of the parent star-forming cloud. In that case, the degree of misalignment in the observable direction is correlated between systems. The solution here is to observe over many star-forming regions and compare, since any preferred direction would be different between the populations.

\begin{figure}
    \centering
    \includegraphics[width=0.5\textwidth]{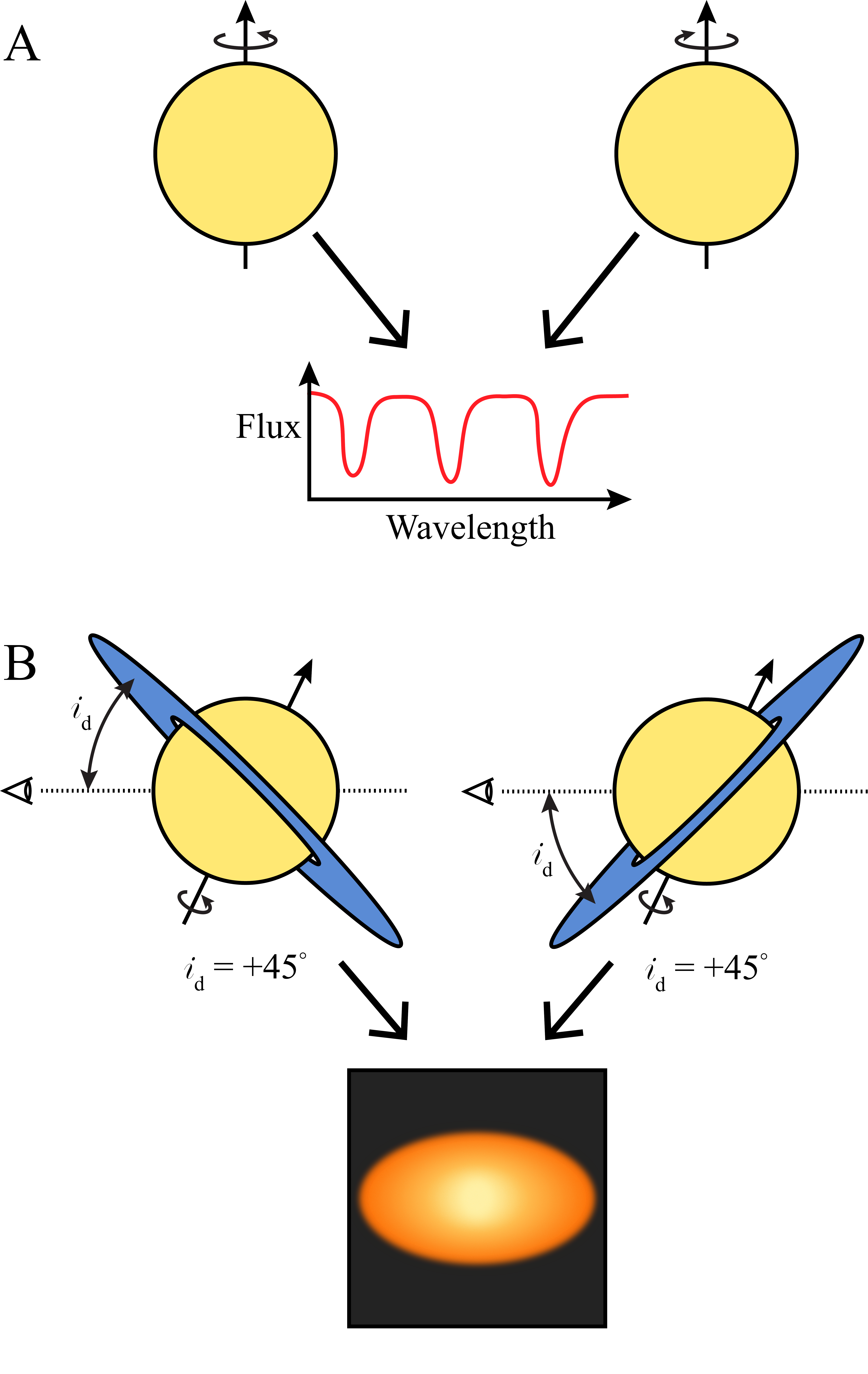}
    \caption{Diagram illustrating two potential issues when measuring stellar and disk parameters. A: two identical stars (yellow) rotating in opposite directions with the same velocity will produce identical spectral profiles (red curve below), leading to matching \vsini{} values. B: two stars with disks (blue) oriented with opposite (mirrored) angles with respect to the line-of-site (dotted line) will have equal \idisk{} values. This is shown by the black panel below, as it is unclear whether the `top' of the disk (orange) is facing toward or away from the observer. Since both stars have identical axes of rotation, the left system is nearly aligned and the right system has a large misalignment; both systems would be measured as consistent with alignment.}
    \label{fig:alias_diagram}
\end{figure}

As we show in Figure~\ref{fig:alias_diagram}, measurements of both the disk and star projected onto the sky may lead us to thinking misaligned systems are aligned. For the star, \vsini{} only measures the tilt of the star toward or away from the observer, not the direction. For the disk; there is no way to tell which part of an inclined disk is in front of versus behind a star, so identical protoplanetary disks with line-of-sight-flipped orientations will have the same measured \idisk{} values.

For an ensemble analysis, the orientation mirroring has a minor impact on our ability to measure the alignment distribution, other than requiring a larger sample. A more subtle but related effect is the barrier at inclinations of $0^\circ$ and $90^\circ$ (both star and disk). This could make systems look more aligned than they are. This effect is included in our \hbm{} example, and the major impact is to systematically overestimate $\mu$ and underestimate $\sigma$ (see Figure~\ref{fig:hbm_vs_N}). The size of this bias is small compared to other effects discussed in this paper, but larger than the uncertainties for sample sizes more than $\simeq$30. The solution for this problem is either to model a simulated population and apply a correction, or include the effect as a parameter of the simulation.

\subsection{Multi-Star Systems}\label{sec:binaries}

Multi-star systems can complicate estimating \rstar{} (unresolved systems will appear brighter), \vsini{} (unresolved but separated lines will increase apparent broadening), and \prot{} (there may be two rotation signatures in the lightcurve). Modeling these effects can be difficult because the impact depends on the separation and contrast of the companion. Fortunately, these generally only impact binaries where the components are unresolved and have low mass ratios where there is still significant flux from the companion. These represent a small fraction of star systems. Double-line spectroscopic binaries (SB2s) and high-order multiples, for example, make up only $\sim$3\% of main-sequence stars \citep{kounkel_double-lined_2021}. Equal-mass binaries, while over-represented in the binary population \citep{el-badry_discovery_2019}, are also the easiest to identify. 

Separately, the disk-star alignment distribution is likely different for close binaries as it is for single stars and wide binaries. Disks in close and intermediate binaries are expected to be shorter-lived, and their orientations should be heavily influenced by the companion through disk truncation \citep[e.g.,][]{artymowicz_dynamics_1994, williams_protoplanetary_2011, andrews_truncated_2010, jang-condell_likelihood_2015}, increased accretion \citep[e.g.,][]{artymowicz_dynamics_1994, jensen_periodic_2007}, and increased photoevaporation \citep[e.g.,][]{alexander_dispersal_2012, rosotti_evolution_2018}. Disks orbiting both binary components (circumbinary disks) are also believed to be rare and may only live for a few Myr \citep[e.g.,][]{akeson_resolved_2019, czekala_degree_2019, offner_origin_2023}. 

Fortunately, if the disk is not present it would automatically not be included in a disk-star alignment sample; the input is necessarily the list of targets with a resolved disk. Thus, most of the binaries that would impact our stellar parameters will be removed because the disk has dissipated below detectable levels prior to any observations. 

The best solution is to remove all close-in binaries from the sample, although complete removal is challenging. \gaia{} \ruwe{} is an indicator of unresolved companions \citep[e.g.,][]{stassun_parallax_2021}. \ruwe{} is most sensitive to a specific range of binary orbits \citep{wood_characterizing_2021}, but this includes most near-equal binaries that are problematic for our work. Young stars have higher \ruwe{} than their older counterparts \citep{fitton_disk_2022}, so we adopt a more generous cutoff of $\mathrm{RUWE}<10$. Separately, we remove any system that shows two sets of lines in the spectrum or two peaks in the CCFs from the \vsini{} analysis (Section~\ref{sec:vsini}). We remove any target where \rstarcode{} yields an age inconsistent with the group age; this is typically a sign of an elevated color-magnitude diagram (CMD) position due to an unresolved binary. For stars with \alma{} data, we can also remove any target with a clearly resolved tight companion. 

To explore the effectiveness of these cuts, we adopt the \ruwe{} cut above and \alma{} sensitivity and beam size typically used for disk morphology \citep{ansdell_alma_2017}. We generated a population of binaries using the \texttt{MOLUSC} code \citep{wood_characterizing_2021}, which generates a realistic sample of stellar companions and determines which survive a set of input data (including \ruwe). The problematic binaries that may hamper measurements but not deplete the disk or be detected in the above data are generally those with orbits $<$0.1\,au, which are most likely to be detected as spectroscopic binaries. These also represent a small fraction of the overall binary population ($<$1\%). We conclude these cuts are sufficient to make a clean sample.

\section{Application to Pre-Main-Sequence Stars Without Disks} \label{sec:betapic}

As a test, we calculated stellar parameters (following Sections~\ref{sec:vsini}--\ref{sec:prot}) for a sample of members mostly in \bpmg{}, as well as \tuchor{} and \carext{}, and compared them to literature estimates. Members of \bpmg{}, \tuchor{}, and \carext{} are ideal to test our methods for stellar parameters because they are \pms{} with ages $\sim$11--26\,Myr \citep[][]{couture_addressing_2023}, $\sim$40\,Myr \citep{kraus_slar_2014}, and $\sim$34--44\,Myr \citep{luhman_census_2024, wood_lithium_2023}, respectively. These stars are nearby, and many have data from \igrins. Additionally, stars in this association should not have protoplanetary disks, eliminating the bias against high \idisk{} values. The result is that \cosi{} should be uniformly distributed. 

We chose our sample based on available \igrins{} data. We downloaded archival \igrins{} \textit{K}-band spectra for 25 stars with spectral classes ranging from early G- to late M-type. We provide general information for the sample in Table~\ref{tab:betapic-info}. We removed 5 of 25 stars that were SB2s and/or had very high \ruwe{} values ($>$20).

\begin{table*}
    \footnotesize
    \centering
    \caption{General information for the \pms{} star test sample. Columns include target name, 2MASS ID, RA, DEC, parallax, association, multiplicity characterization, spectral type, and literature reference(s).}
    \label{tab:betapic-info}
    \begin{threeparttable}
        
        \begin{tabular}{lcccrcclcp{2cm}}
            \toprule
            Name & 2MASS & RA (J2000)\tnote{a} & DEC (J2000) & $\pi$ & Assoc. & Mult.\tnote{b,c} & SpT & Ref.\tnote{d} & Note \\
                 &       & (h:m:s) & (d:m:s) & (mas) &     &     &     &     &     \\
            \midrule
            TYC 585313181 & J01071194$-$1935359 & 01:07:12.02 & $-$19:35:36.47 & $17.64 \pm 0.17$ & \bpmg{} & S? & M1V & 1 & \phantom{} \\
            TIC 1031299 & J02155892$-$0929121 & 02:15:58.85 & $-$09:29:11.42 & $22.62 \pm 0.25$ & Tuc-Hor & Bw & M2.5V & 2, 3 & \phantom{} \\
            UCAC4 513-003622 & J02175601+1225266 & 02:17:56.08 & +12:25:25.67 & $15.90 \pm 0.03$ & \bpmg{} & S & M3.5V & 1 & \phantom{} \\
            BD+05 378 & J02412589+0559181 & 02:41:25.97 & +05:59:17.49 & $22.58 \pm 0.02$ & \bpmg{} & S & K6Ve & 1 & \phantom{} \\
            HIP 12787 & J02442137+1057411 & 02:44:21.45 & +10:57:40.19 & $20.78 \pm 0.13$ & \bpmg{} & T & M0Ve & 4 & \phantom{} \\
            TIC 10932072 & J02501167$-$0151295 & 02:50:11.76 & $-$01:51:30.38 & $19.66 \pm 0.11$ & Tuc-Hor & S & M7V & 2, 5 & \phantom{} \\
            TIC 26126812 & J03350208+2342356 & 03:35:02.15 & +23:42:34.41 & $19.72 \pm 0.09$ & \bpmg{} & Bc? & M8.5V & 1 & \phantom{} \\
            TIC 55441420 & J03550477$-$1032415 & 03:55:04.85 & $-$10:32:42.14 & $19.14 \pm 0.17$ & Tuc-Hor & S & M8.5V & 2, 6 & \phantom{} \\
            GJ 3305 & J04373746$-$0229282 & 04:37:37.51 & $-$02:29:29.71 & $36.01 \pm 0.48$ & \bpmg{} & Bc & M1V & 1 & $RUWE=22.9$ \\
            TIC 299007548 & J04433761+0002051 & 04:43:37.67 & +00:02:03.39 & $47.62 \pm 0.14$ & \bpmg{} & S & M9V & 7 & \phantom{} \\
            V1005 Ori & J04593483+0147007 & 04:59:34.88 & +01:46:59.15 & $40.99 \pm 0.01$ & \bpmg{} & SB1 & M0.5Ve & 1 & \phantom{} \\
            HD 49855 & J06434625$-$7158356 & 06:43:46.27 & $-$71:58:34.42 & $18.05 \pm 0.01$ & Car-Ext & S & G6V & 2, 8 & \phantom{} \\
            TWA 22 & J10172689$-$5354265 & 10:17:26.58 & $-$53:54:26.54 & $50.52 \pm 0.20$ & \bpmg{} & Bc & M5V & 1 & \phantom{} \\
            HIP 76629 & J15385757$-$5742273 & 15:38:57.45 & $-$57:42:28.83 & $25.83 \pm 0.20$ & \bpmg{} & SB1 & K0V & 1 & \phantom{} \\
            HD 319139 & J18141047$-$3247344 & 18:14:10.49 & $-$32:47:35.36 & $13.99 \pm 0.02$ & \bpmg{} & SB2 & K5Ve & 1 & \phantom{} \\
            GSC 07396-00759 & J18142207$-$3246100 & 18:14:22.08 & $-$32:46:10.98 & $13.92 \pm 0.02$ & \bpmg{} & SB? & M1.5V & 1 & Likely SB2; companion to HD 319139 \\
            TYC 907724891 & J18453704$-$6451460 & 18:45:37.10 & $-$64:51:48.36 & $35.16 \pm 0.18$ & \bpmg{} & SB2 & K8Ve & 9 & \phantom{} \\
            UCAC3 116-474938 & J19560294$-$3207186 & 19:56:02.98 & $-$32:07:19.83 & $19.54 \pm 0.73$ & \bpmg{} & Tc & M4V & 1 & $RUWE=29.7$ \\
            HD 196982B & & 20:41:51.44 & $-$32:26:13.33 & $101.97 \pm 0.08$ & \bpmg{} & B & M4.5Ve & 1, 8 & \phantom{} \\
            AU Mic & J20450949$-$3120266 & 20:45:09.88 & $-$31:20:33.00 & $102.94 \pm 0.02$ & \bpmg{} & S & M1Ve & 1 & Debris disk \\
            HD 358623 & J20560274$-$1710538 & 20:56:02.90 & $-$17:10:56.48 & $21.70 \pm 0.02$ & \bpmg{} & Bw & K6Ve & 1 & \phantom{} \\
            GSC 00543-00620 & J21374019+0137137 & 21:37:40.28 & +01:37:12.69 & $27.85 \pm 0.12$ & \bpmg{} & Bc & M5V & 1 & \phantom{} \\
            
            HIP 107345 & J21443012$-$6058389 & 21:44:30.21 & $-$60:58:40.39 & $21.55 \pm 0.01$ & Tuc-Hor & S & M0Ve & 2, 8 & \phantom{} \\
            WW PsA & J22445794$-$3315015 & 22:44:58.19 & $-$33:15:03.72 & $47.92 \pm 0.03$ & \bpmg{} & Bw & M4IVe & 1 & \phantom{} \\
            TX PsA & J22450004$-$3315258 & 22:45:00.29 & $-$33:15:28.03 & $48.00 \pm 0.03$ & \bpmg{} & Bw & M5Ve & 1 & \phantom{} \\
            \bottomrule
            \end{tabular}
    
            \begin{tablenotes}
                \item[a] RA and DEC values from \gaia{} DR3.
                \item[b]{Multiplicity classifications from \citet{messina__2017}:
                [S] single;
                [B] binary;
                [T] triple;
                [SB1] single-line spectroscopic binary;
                [SB2] double-line spectroscopic binary;
                [c] close orbit ($<$60\,au);
                [w] wide orbit ($>$60\,au);
                and
                [?] uncertain.}
                \item[c]{We apply the multiplicity nomenclature uniformly to all stars even if a star was not in the \citet{messina__2017} sample.}
                \item[d]{Literature references: 
                [1] \citet{messina__2017};
                [2] \citet{luhman_census_2024};
                [3] \citet{bowler_rotation_2023};
                [4] \citet{sperauskas_spectroscopic_2019};
                [5] \citet{gagne_banyan_2015};
                [6] \citet{shkolnik_all-sky_2017};
                [7] \citet{deshpande_intermediate_2012};
                [8] \citet{torres_search_2006};
                and
                [9] \citet{zuniga-fernandez_search_2021}.
                }
            \end{tablenotes}
            
    \end{threeparttable}
\end{table*}

We calculated \vsini{} and \rstar{} values for the 20 remaining stars following Section~\ref{sec:vsini} and Section~\ref{sec:rstar}, respectively. We generated lightcurves using \tess{} data downloaded from \mast{} for all but one star in our sample (HD 358623), which did not have \tess{} data, and calculated \prot{} values and errors (following Section~\ref{sec:prot}). We provide our estimated stellar parameters alongside literature values in Table~\ref{tab:betapic-values}.

\begin{table*}
    \footnotesize
    \centering
    \rotatebox{90}{
        \begin{threeparttable}
            \caption{Quantities estimated in this work and from the literature for the final \pms{} star test sample.}
            \label{tab:betapic-values}
            
            \begin{tabular}{lrccrccccrcc}
                \toprule
                Name & \vsini{} & Lit. \vsini{} & \vsini{} & \rstar{} & Lit. \rstar{} & \rstar{} & \prot{} & TESS & LS\tnote{c} & Lit. \prot{} & \prot{} \\
                     & (\kms{}) & (\kms{}) & Ref.\tnote{a} & (\rsun{}) & (\rsun{}) & Ref. & (d) & Sec.\tnote{b} & & (d) & Ref. \\
                \midrule
                TYC 585313181 & $9.08 \pm 0.40$ & $11.5 \pm 1.4$ & 1 & $1.09 \pm 0.04$ & $1.25 \pm 0.26$ & 3 & $5.24 \pm 0.15$ & 3 & 0.73 & $7.26 \pm 0.07$ & 4 \\
                \phantom{} & & $6.8 \pm 0.6$ & 2 & & & & & & & & \phantom{} \\
                TIC 1031299 & $14.47 \pm 0.31$ & $12.5 \pm 0.7$ & 5 & $0.86 \pm 0.02$ & $0.39 \pm 0.06$ & 5 & $1.44 \pm 0.01$ & 4 & 0.82 & $1.44 \pm 0.03$ & 5 \\
                \phantom{} & & $15.7 \pm 1.3$ & 6 & & & & & & & & \phantom{} \\
                UCAC4 513-003622 & $16.25 \pm 0.44$ & $22.6 \pm 3.0$ & 7, 8 & $0.68 \pm 0.03$ & $1.16 \pm 0.12$ & 3 & $2.15 \pm 0.03$ & 70 & 0.72 & $1.995 \pm 0.005$ & 3 \\
                BD+05 378 & $7.99 \pm 0.14$ & 9.0 & 3 & $1.12 \pm 0.04$ & $0.92 \pm 0.25$ & 3 & $4.84 \pm 0.13$ & 42 & 0.85 & $4.83 \pm 0.03$ & 3 \\
                HIP 12787 & $24.91 \pm 0.34$ & & & $1.13 \pm 0.03$ & $1.08 \pm 0.11$ & 9 & $1.67 \pm 0.02$ & 43 & 0.92 & & \phantom{} \\
                TIC 10932072 & $6.94 \pm 0.19$ & & & $0.27 \pm 0.01$ & & & $1.80 \pm 0.02$ & 4 & 0.02 & & \phantom{} \\
                TIC 26126812 & $41.64 \pm 0.25$ & 30.0 & 10 & $0.30 \pm 0.02$ & $0.12 \pm 0.04$ & 3 & $0.2185 \pm 0.0003$ & 44 & 0.14 & $0.472 \pm 0.005$ & 3 \\
                TIC 55441420 & $22.99 \pm 0.38$ & & & $0.27 \pm 0.01$ & & & $0.485 \pm 0.002$ & 5 & 0.04 & & \\
                TIC 299007548 & $11.68 \pm 0.19$ & $13.1 \pm 2.0$ & 11 & $0.30 \pm 0.02$ & $0.121 \pm 0.004$ & 9 & $0.529 \pm 0.002$ & 32 & 0.06 & & \phantom{} \\
                V1005 Ori & $9.61 \pm 0.18$ & 14.0 & 12, 13 & $0.93 \pm 0.02$ & $0.868 \pm 0.106$ & 9 & $4.37 \pm 0.11$ & 32 & 0.78 & $4.43 \pm 0.03$ & 3 \\
                \phantom{} & & 8.7 & 14 & & & & & & & & \phantom{} \\
                HD 49855 & $11.06 \pm 0.36$ & $12.930 \pm 0.009$ & 15 & $1.12 \pm 0.05$ & $0.86 \pm 0.05$ & 17 & $3.88 \pm 0.09$ & 37 & 0.96 & $3.87 \pm 0.08$ & 18 \\
                \phantom{} & & 17.38 & 16 & & & & & & & & \phantom{} \\
                \phantom{} & & $11.6 \pm 1.1$ & 16 & & & & & & & & \phantom{} \\
                \phantom{} & & $12.4 \pm 0.3$ & 12 & & & & & & & & \phantom{} \\
                TWA 22 & $8.44 \pm 0.23$ & 8.7 & 19, 20 & $0.41 \pm 0.03$ & $0.411 \pm 0.012$ & 9 & $0.731 \pm 0.003$ & 10 & 0.91 & $0.83 \pm 0.01$ & 3 \\
                HIP 76629 & $16.75 \pm 0.44$ & 11.0 & 21 & $1.29 \pm 0.03$ & $1.455 \pm 0.099$ & 9 & $4.31 \pm 0.10$ & 39 & 0.93 & $4.27 \pm 0.10$ & 23 \\
                \phantom{} & & 16.6 & 12 & & & & & & & & \phantom{} \\
                \phantom{} & & 17.0 & 20 & & & & & & & & \phantom{} \\
                \phantom{} & & 18.2 & 22 & & & & & & & & \phantom{} \\
                HD 196982B & $14.28 \pm 0.28$ & $15.8 \pm 1.2$ & 12 & $0.55 \pm 0.05$ & $0.59 \pm 0.09$ & 3 & $1.20 \pm 0.01$ & 27 & 0.58 & $0.781 \pm 0.002$ & 3 \\
                \phantom{} & & 17.0 & 20 & & & & & & & & \phantom{} \\
                AU Mic & $9.95 \pm 0.24$ & 9.3 & 12 & $0.83 \pm 0.02$ & $0.82 \pm 0.08$ & 3 & $4.84 \pm 0.13$ & 1 & 0.52 & $4.86 \pm 0.02$ & 3 \\
                HD 358623 & $13.36 \pm 0.36$ & 15.6 & 12 & $1.09 \pm 0.04$ & $1.11 \pm 0.1$ & 3 & & & & $3.41 \pm 0.05$ & 3, 23 \\
                \phantom{} & & 14.6 & 19 & & & & & & & & \phantom{} \\
                \phantom{} & & 12.0 & 21 & & & & & & & & \phantom{} \\
                \phantom{} & & $20.0 \pm 2.0$ & 24 & & & & & & & & \phantom{} \\
                GSC 00543-00620 & $42.33 \pm 0.89$ & 55.0 & 25 & $0.62 \pm 0.04$ & $0.283 \pm 0.052$ & 9 & $0.3719 \pm 0.0008$ & 55 & 0.86 & $0.202 \pm 0.001$ & 3 \\
                \phantom{} & & $45.0 \pm 5.0$ & 26 & & & & & & & & \phantom{} \\
                \phantom{} & & $66.0 \pm 9.0$ & 1 & & & & & & & & \phantom{} \\
                HIP 107345 & $8.17 \pm 0.44$ & $7.9 \pm 1.5$ & 15 & $0.85 \pm 0.02$ & $0.70 \pm 0.02$ & 17 & $4.49 \pm 0.12$ & 28 & 0.76 & $4.563 \pm 0.001$ & 27 \\
                \phantom{} & & $5.9 \pm 0.5$ & 2 & & & & & & & & \phantom{} \\
                \phantom{} & & $8.2 \pm 0.1$ & 12 & & & & & & & & \phantom{} \\
                WW PsA & $12.52 \pm 0.48$ & 12.1 & 12 & $0.62 \pm 0.02$ & $0.82 \pm 0.08$ & 3 & $2.35 \pm 0.03$ & 28 & 0.86 & $2.37 \pm 0.01$ & 3 \\
                \phantom{} & & $14.00 \pm 1.73$ & 19 & & & & & & & & \phantom{} \\
                TX PsA & $21.70 \pm 0.34$ & 16.8 & 12 & $0.46 \pm 0.02$ & $0.59 \pm 0.09$ & 3 & $1.08 \pm 0.01$ & 28 & 0.89 & $1.086 \pm 0.005$ & 3 \\
                \phantom{} & & $24.30 \pm 4.93$ & 19 & & & & & & & & \phantom{} \\
                \bottomrule
            \end{tabular}
            
            \begin{tablenotes}
                \item[a] Literature references: 
                {[1]} \citet{malo_banyan_2014}; 
                {[2]} \citet{kraus_slar_2014}; 
                {[3]} \citet{messina__2017}; 
                {[4]} \citet{messina_race-oc_2011}; 
                {[5]} \citet{bowler_rotation_2023};
                {[6]} \citet{fouque_spirou_2018};
                {[7]} \citet{binks_lithium_2014}; 
                {[8]} \citet{binks_spectroscopic_2016}; 
                {[9]} \citet{stassun_revised_2019}; 
                {[10]} \citet{reid_high-resolution_2002}; 
                {[11]} \citet{deshpande_intermediate_2012};
                {[12]} \citet{torres_search_2006}; 
                {[13]} \citet{favata_lithium_1995};
                {[14]} \citet{vogt_rotational_1983};
                {[15]} \citet{zuniga-fernandez_search_2021};
                {[16]} \citet{desidera_vltnaco_2015};
                {[17]} \citet{fernandes_using_2023};
                {[18]} \citet{colman_methods_2024};
                {[19]} \citet{jayawardhana_accretion_2006};
                {[20]} \citet{scholz_rotation_2007};
                {[21]} \citet{de_la_reza_rotation_2004};
                {[22]} \citet{weise_rotational_2010};
                {[23]} \citet{messina_race-oc_2010};
                {[24]} \citet{lepine_nearby_2009};
                {[25]} \citet{mochnacki_spectroscopic_2002}; 
                {[26]} \citet{schlieder_cool_2012};
                and {[27]} \citet{howard_evryflare_2020}.
                \item[b] TESS sector whose lightcurve gave the highest Lomb-Scargle peak.
                \item[c] The Lomb-Scargle peak power.
            \end{tablenotes}
            
        \end{threeparttable}
    }
\end{table*}

\begin{figure}
    \centering
    \includegraphics[width=0.5\textwidth]{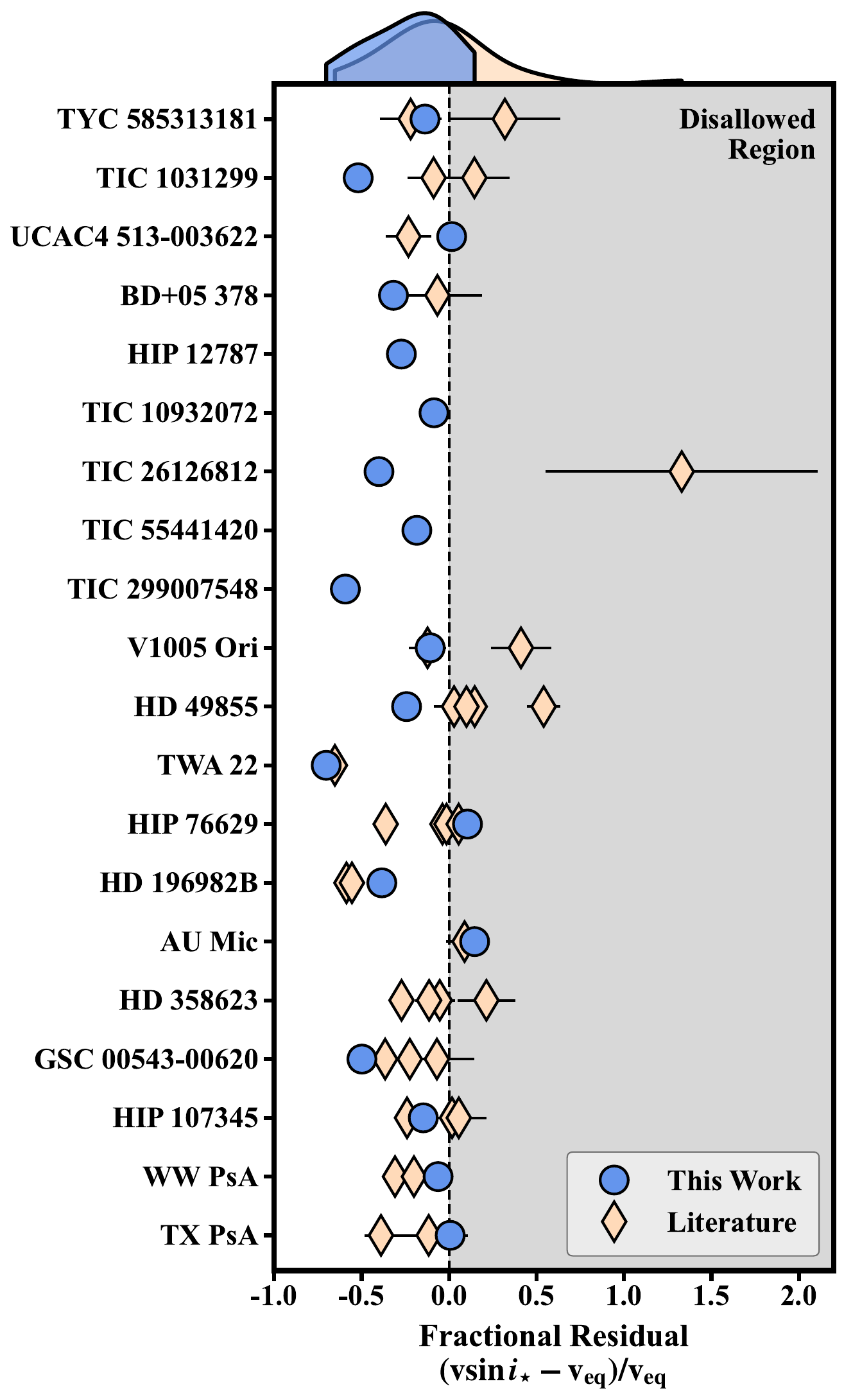}
    \caption{Fractional residuals comparing \vsini{} and \veq{} for a sample of systems in \bpmg{}, \tuchor{}, and \carext{}. Fractional residuals we estimated are shown as blue circles and those calculated with literature values are shown as orange diamonds. On the top of the main plot are the distributions of fractional residuals from this work (blue) and the literature (orange). Targets without a result from either this work or the literature had insufficient data for the full comparison. Our estimated values are consistent with expectations for a realistic, randomly-generated synthetic sample. In contrast, the literature results are more extreme; fewer points are at low values (near -1) and more are above zero than expected for a random distribution.}
    \label{fig:betapic_vsini}
\end{figure}

We compare our results to those from the literature in Figure~\ref{fig:betapic_vsini}. Specifically, we show the calculated equatorial velocities (${v_{\mathrm{eq}} = 2\pi R_{\star} / P_\mathrm{rot}}$), using \rstar{} and \prot{} from the literature and from our own estimates, and compared them to literature \vsini{} and our estimated \vsini{}, respectively, via fractional residual $(v \sin i_{\star} - v_{\mathrm{eq}})/v_{\mathrm{eq}}$.

Since there are more edge-on inclinations than face-on ones, we expect there to be more stars near zero (equality) than at $\simeq$$-$1 and some stars may be over one due to random errors. However, the literature results are extreme even considering this.

As a test, we generated a fake sample of systems assuming random \cosi{} and properties (including measurement uncertainties) matching the sample in Figure~\ref{fig:betapic_vsini}. For our measurements, the K-S test gives a p-value of 90\%, suggesting values are consistent with the expected distribution. For the literature, the result depends on how we select the sample. Using just the best measurements yield a p-value of just 0.3\% (inconsistent). Combining all measurements gives 9\% although this double counts the same stars (not done in the synthetic sample; see Section~\ref{sec:synthetic_sample}). Averaging multiple measurements for the same star (many of which do not agree with each other) dropped this to 0.2\%.

This effect has been noted in prior studies \citep[e.g.,][]{newton_rotation_2016}. Literature \vsini{} values tend to be overestimated, likely due to incomplete consideration of effects discussed in Section~\ref{sec:vsini}. This would lead to an overestimate in the number of edge-on stars. Our measurements do not seem impacted by this, which offers an observational confirmation of the methods outlined here.

\section{Application to a Synthetic Sample}\label{sec:synthetic_sample}

As a test of the issues discussed in Sections~\ref{sec:hbm} and~\ref{sec:additional_effects}, we constructed a synthetic sample of systems with realistic stellar and disk parameters and applied the methods of Sections~\ref{sec:cosi} and~\ref{sec:hbm}. First, we created a parent alignment distribution, which we assumed to be a normal distribution centered at a true mean ($\mu_\mathrm{true}$) and scaled by a true intrinsic scatter ($\sigma_\mathrm{true}$), both of which we manually selected. These are defined such that a sample with perfect disk-star alignment would have $\mu_\mathrm{true}=\sigma_\mathrm{true}=0$.

Next, we drew a sample of $N$ systems from the parent population. We defined a \cosi{} distribution as a uniform distribution of $N$ points in the interval [-1, 1], and we calculated \idisk{} values from the difference between the generated \istar{} values and the Gaussian alignment distribution. Since \istar{} and \idisk{} are measured from 0--90$^{\circ}$ (0--1 in cosine), as we discussed in Section~\ref{sec:additional_effects}, we forced \istar{} and \idisk{} to 0--90$^{\circ}$ by taking the absolute value of the cosine of the angle.

We defined \rstar{} and \prot{} distributions as log-uniform distributions of $N$ points in the intervals [0.1, 1.5]\,\rsun{} and [0.2, 12]\,days, respectively. While not a perfect match to an observed star-forming population, the results were not sensitive to changes in the \rstar{} and \prot{} distributions. We then calculated \veq{} values from the assigned \rstar{} and \prot{} values. By chance, some stars had unphysically large \veq{} (above breakup speeds), so these were adjusted manually. Last, we calculated \vsini{} from a combination of \cosi{}, \rstar{}, and \prot{}. 

At this point all assigned parameters have no uncertainties (i.e., these are true values instead of measured ones). So, we assigned each measured parameter an uncertainty based on the empirical tests in prior sections. Following Section~\ref{sec:vsini}, we used 1\,\kms{} uncertainties for all \vsini{} values. For \rstar{} we used 5\% uncertainties. For \prot{} we calculated uncertainties following Boyle et al. (in prep.). We gave all \idisk{} values an uncertainty of 4$^{\circ}$, following our analysis in Section~\ref{sec:idisk}. 

\subsection{Test: Stellar Inclination} \label{sec:test_cosi}

One concern from our approach is that our likelihood assumes the \cosi{} distributions can be approximated as Gaussians. We can test this with the synthetic sample. If we do not remove edge-on disks, the resulting stellar inclinations should be uniformly distributed in \cosi{}, while a non-uniform distribution would suggest problems (especially near the extreme inclinations) with the Gaussian assumption. 

We drew 25 synthetic systems from the above population, from which we derived \cosi{} values. We chose 25 systems for this test arbitrarily, and we anticipated similar results for a larger number of systems, which we discuss further in Section~\ref{sec:test_number_systems}. 

\begin{figure}
    \centering
    \includegraphics[width=0.5\textwidth]{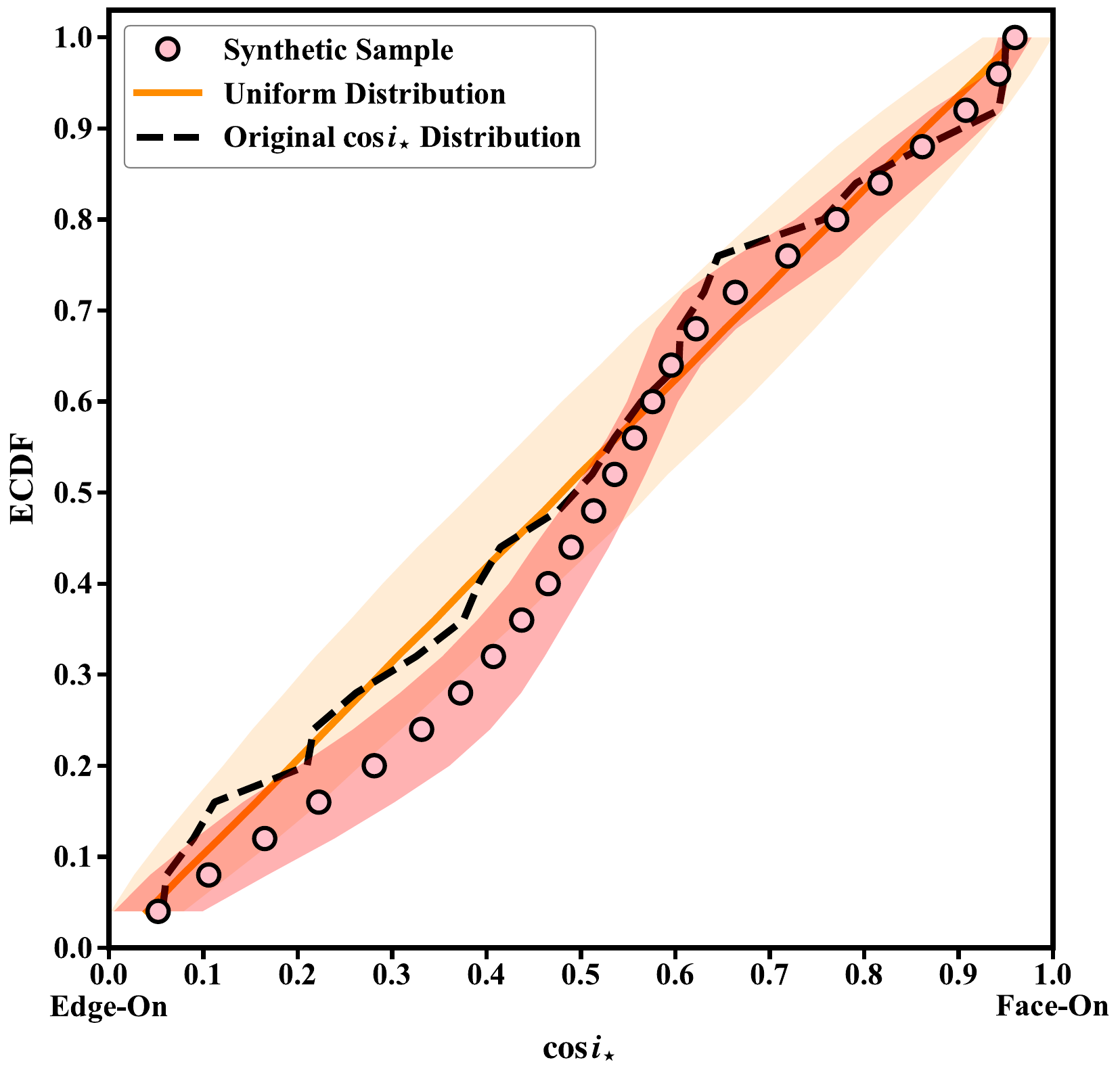}
    \caption{ECDF of \cosi{} values compared to that of a uniform distribution for 25 synthetic systems. The pink samples and shaded region are mean values and standard deviations, respectively, of 1000 random draws from the estimated \cosi{} posteriors for each star in the synthetic sample. The orange solid line and shaded region are mean values and standard deviations, respectively, of 1000 random draws from a perfect uniform distribution. Shown for comparison is the ECDF of the original, unperturbed \cosi{} distribution (black dashed line), which itself was sampled from a uniform distribution. The estimated \cosi{} values are consistent with a uniform distribution, which is what we expect.}
    \label{fig:cosi_ecdf}
\end{figure}

We compared our estimated values to a uniform distribution using empirical cumulative distribution functions (ECDFs). We took 1000 random draws from each \cosi{} posterior and calculated the ECDF at each draw (i.e., we had 1000 ECDFs, each derived from 25 random \cosi{} values). Similarly, we took 1000 random draws of 25 data points from a uniform distribution and calculated the ECDFs at each draw. We calculated the mean and standard deviation of each ensemble of ECDFs, which we show in Figure~\ref{fig:cosi_ecdf}.

For comparison, we also calculated the ECDF of the originally-defined \cosi{} distribution (before any perturbation), which was essentially a single random draw of 25 data points from a uniform distribution. Figure~\ref{fig:cosi_ecdf} shows that the ECDFs calculated from the \cosi{} posteriors and the uniform distribution are in agreement with one another, which is what we expect. 

\subsection{Test: Alignment Distribution} \label{sec:test_alignment}

As a test on our \hbm{} (Section~\ref{sec:hbm}), we used the same setup as in the previous test and inferred $\mu$ and $\sigma$ for the parent alignment distribution, and the alignment values ($\alpha_n^{\prime}$) for the individual systems. In this case, we did not account for the fact that edge-on disks are disfavored observationally. 

\begin{figure*}
    \centering
    \includegraphics[width=0.45\textwidth]{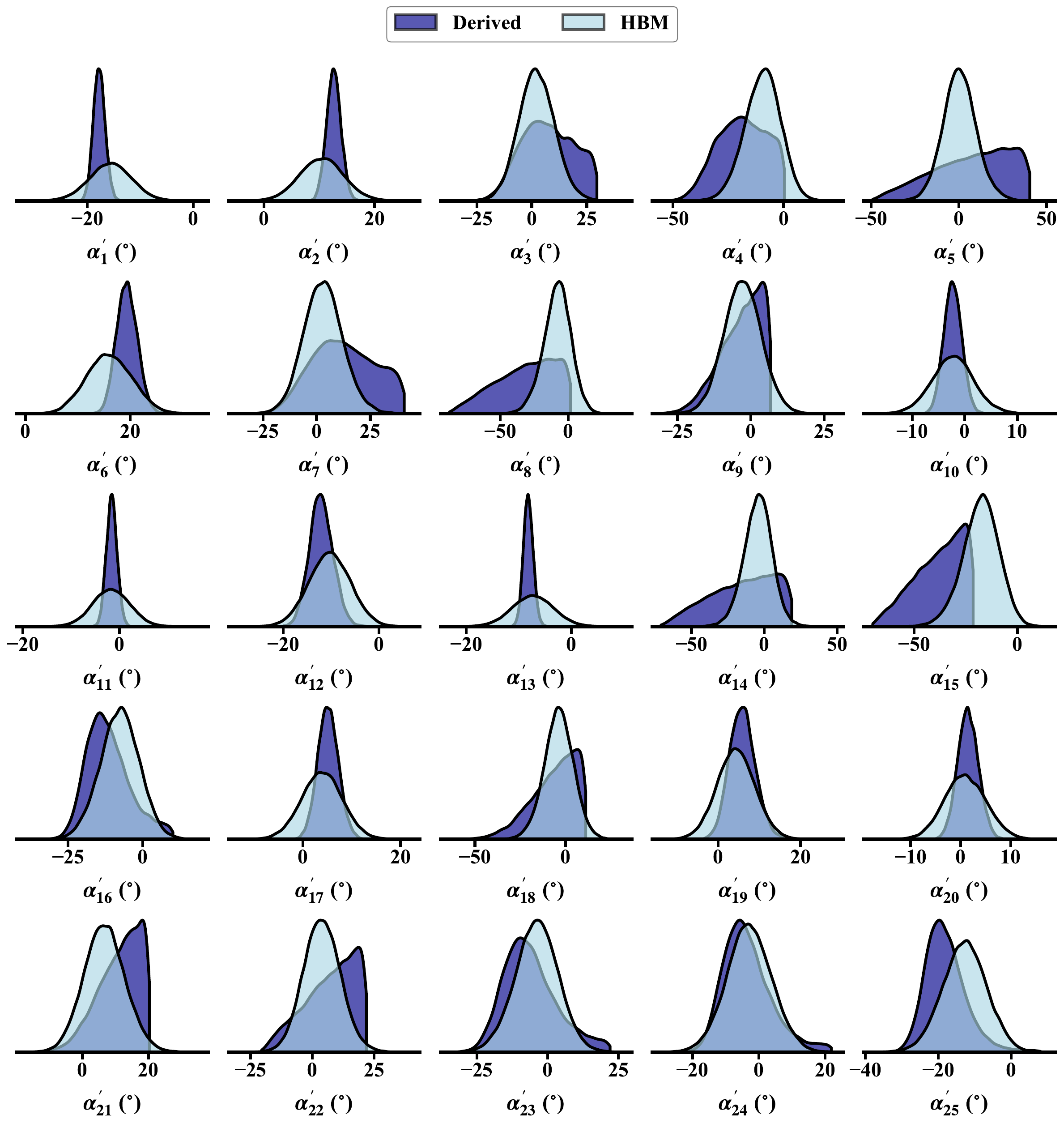}
    \includegraphics[width=0.45\textwidth]{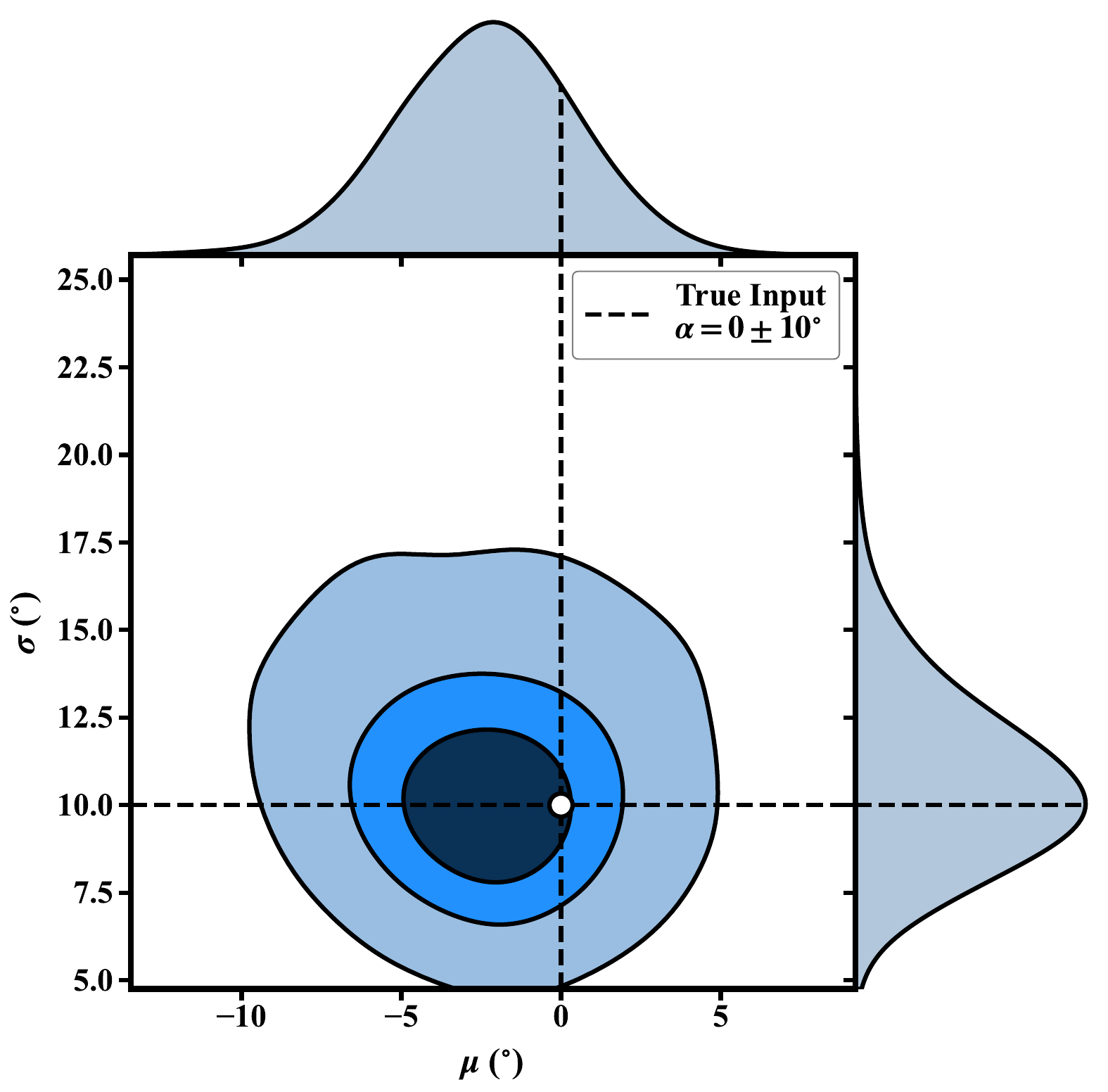}
    \caption{Example \hbm{} result for a synthetic sample of 25 systems. Left panels: individual system disk-star differences ($\alpha^{\prime}_n$). Synthetic alignments we derived from the combination of \istar{} and \idisk{} (following Sections~\ref{sec:cosi} and~\ref{sec:idisk}, respectively) are shown in dark blue. The light blue distributions are posterior probabilities from the \hbm{} fit. Note how the model (incorrectly) assumes the distributions are Gaussian. Right panel: the global mean ($\mu$) and standard deviation ($\sigma$) of the parent alignment distribution.  The dashed lines are the injected values. The estimated global values are consistent with the `true' input values to within 1$\sigma$.}
    \label{fig:hbm_joint}
\end{figure*}

In Figure~\ref{fig:hbm_joint}, we show our result for the inferred $\mu$ and $\sigma$ as a 2-dimensional contour, accompanied by 1-dimensional probability densities for each $\alpha_n^{\prime}$. The estimated $\mu$ and $\sigma$ agreed with the initial alignment distribution to within 1$\sigma$.

We show the \hbm{} fit to the individual systems in Figure~\ref{fig:hbm_joint}, where they were all in agreement with the alignments derived through the combination of \istar{} (following Section~\ref{sec:cosi}) and \idisk{} (following Section~\ref{sec:idisk}). Note that the method assumes Gaussian distributions, which is not accurate for some systems. As we discuss below, this does not prevent recovering the global $\mu$ and $\sigma$, but likely leads to underestimated uncertainties on these parameters.

\subsection{Dependence on the Number of Systems}\label{sec:test_number_systems}

To explore what kind of sample size is required to retrieve the disk-star alignment distribution, we varied the number of systems from 5--150 and re-ran our \hbm{} to estimate $\mu$ and $\sigma$. For this, we used the same setup as in previous tests and kept the input $\mu$ and $\sigma$ constant (except we increased $\sigma$ from previous tests), although the result shows only weak dependence on the exact $\mu$ and $\sigma$ used. The resulting underestimation is more easily seen at larger $N$ where the uncertainties on $\mu$ and $\sigma$ are smaller.

\begin{figure*}
    \centering
    \includegraphics[width=\textwidth]{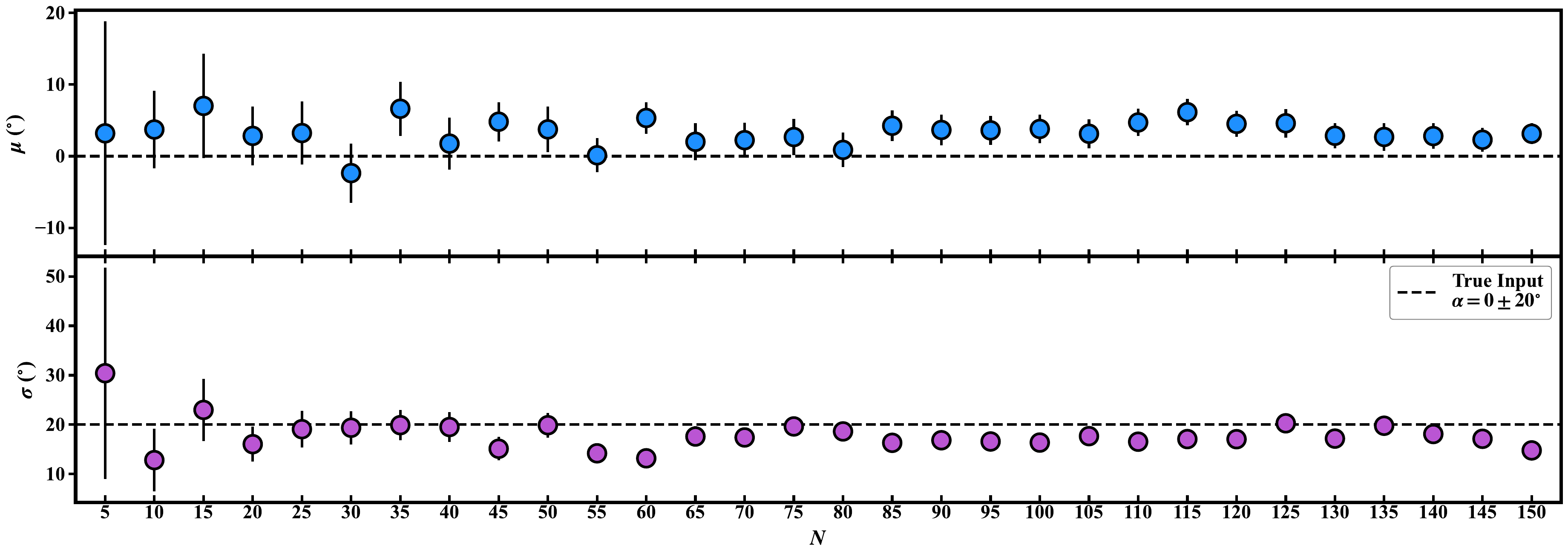}
    \caption{The effect of the number of systems on the \hbm{} results for a synthetic sample whose properties were generated with realistic uncertainties. Here, we enforced ${0<i_{\mathrm{d}}<90^{\circ}}$ following Section~\ref{sec:additional_effects}. The panels show $\mu$ (top) and $\sigma$ (bottom) versus $N$ for 5--150 systems. The dotted lines mark the input $\mu$ and $\sigma$ used to generate the parent alignment distribution of $0^{\circ}$ and $20^{\circ}$, respectively. The underestimation is most clearly seen in $\sigma$ at large $N$ where uncertainties are smaller. Both $\mu$ and $\sigma$ vary around their expected values by more than their uncertainties, especially at larger $N$. This is likely because we assume all the individual $\alpha_n^{\prime}$ estimates are Gaussian.}
    \label{fig:hbm_vs_N}
\end{figure*}

We show these results in Figure~\ref{fig:hbm_vs_N}. At lower $N$, the variation around the input values is expected given the uncertainties. However, past $N\simeq50$, the recovered $\mu$ and $\sigma$ vary around the input values by more than the expected uncertainties. 
This is driven by the assumption that the individual $\alpha^{\prime}_n$ estimates are Gaussian, while many are asymmetric (Figure~\ref{fig:hbm_joint}). 

The bias caused by enforcing inclinations to 0--90$^{\circ}$ (see Section~\ref{sec:additional_effects}) is most apparent in $\sigma$, where points are statistically $\simeq3^\circ$ below the input values. On a real dataset, this can be corrected either by including it in the model, or generating a synthetic sample like this one and applying a correction. In either case, the bias depends on the underlying model. A tight distribution of aligned systems, for example, shows an almost negligible bias. This suggests exploring a few different possible distributions in the final fit as a test of sensitivity to such assumptions.

\section{Summary and Conclusions}\label{sec:conclusion}

In this paper, we considered the important factors that contribute to measuring the distribution of disk-star alignment angles for a large number of stars hosting protoplanetary disks. We explored what uncertainties are realistic for the input parameters, including \rstar, \vsini\, \prot, and \idisk{} using existing observations of infant stars. We applied our methodology to both real and synthetic datasets to evaluate the impact of assumptions in our methods. 

We summarize the main results as follows:
\begin{enumerate}

    \item \textit{\vsini{}:} Following \citet{kesseli_magnetic_2018} we were able to estimate \vsini{} with uncertainties as good as $\simeq$1\,\kms{} for G- through M-type \pms{} stars, including disk-bearing stars.
    \item \textit{\rstar{} and \rstarcode{}:} With \rstarcode{}, we are able to estimate \rstar{} for \pms{} stars to $\simeq$5\%. Given the complexities of estimating the fundamental properties of young stars, this is surprisingly precise. However, this method reproduces empirical densities from transits and eclipsing binaries, including for at least one disk-bearing star (IRAS 04125+2902). 
    \item \textit{Archival \idisk{} Measurements:} \alma{} \idisk{} measurements already exist in the literature \citep[e.g.,][]{huang_disk_2018} spanning the protoplanetary disk lifetime (e.g., $\rho$\,Ophiuchus, Lupus, Taurus, Upper Scorpius). Of these, 68\% had uncertainties $\lesssim$4$^{\circ}$. Overlapping measurements between surveys agree within reported uncertainties. While there is an observational bias against edge-on disks, the \idisk{} values follow the expected distribution for $i_{\mathrm{d}}<75^\circ$. We conclude that most previous \idisk{} measurements are precise and consistent between studies, and uncertainties/systematics are small compared to measurements related to \istar{}. 
    \item \textit{\pms{} Star Literature Comparison:} We calculated stellar parameters for 20 young ($\sim$11--44\,Myr) \pms{} stars (without disks) within \bpmg{}, \tuchor{}, and \carext{}. We found that literature \vsini{} estimates are systematically overestimated, yielding a non-random stellar inclination distribution. Our estimates reproduce the expected distribution. 
    \item \textit{Global Alignment Distribution:} Based on a synthetic sample, we find that an \hbm{} analysis is sufficient to reproduce the input values within uncertainties, despite the necessary assumptions made along the way. Modest samples ($N\simeq20$) are sufficient to identify any significant population of misaligned systems. At $N \gtrsim 70$, systematics dominate over random samples, requiring more sophisticated modeling (e.g., consideration of non-Gaussian uncertainties).  

\end{enumerate}

We aim to use this method to estimate the disk-star alignment distributions for nearby young populations with protoplanetary disks. With increasingly available data from \alma{}, \tess{}, \ktwo{}, and \gaia{}, we need only \igrins{} or similar NIR spectra of the sample. Many such samples already exist in the literature \citep[e.g.][]{lopez-valdivia_igrins_2021}, suggesting this study may be possible with largely existing data. 

Right now results on the alignment between disks and stars are limited primarily by the quality of the stellar parameters, the small sample of targets with high-quality data (high-resolution spectra with high signal-to-noise ratios and precise disk inclinations, e.g., from \alma{}), and the methods used to turn these into population statistics. Our work has focused on the larger sources of bias and uncertainty, hence we can make significant improvements over prior studies \citep[e.g.,][]{davies_stardisc_2019, hurt_evidence_2023} while still making a number of simplifying assumptions in our \hbm{} (e.g., ignoring the missing angle and assuming Gaussian uncertainties on \istar). As more data becomes available and astronomers develop better methods to measure fundamental stellar parameters, it will be more important to properly account for the non-Gaussian posteriors on $\alpha$ and use a full three-dimensional misalignment (i.e., marginalize over the missing angle).

\section*{Acknowledgements}

The authors thank the two referees for their comments; their feedback helped improve the paper. The authors thank Bernie and Bandit for their useful contributions. This work used The Immersion Grating Infrared Spectrometer (IGRINS). M.J.F. acknowledges support from the NC Space Grant Graduate Research Fellowship. This research was completed with funding from the NASA Exoplanet Research Program (XRP) award \#80NSSC21K0393.

\section*{Data Availability}

The \igrins{} spectra are available in the Raw \& Reduced IGRINS Spectral Archive at \url{https://igrinscontact.github.io/}. 

Photometry from \twomass{}, \gaia{}, \sdss{}, \apass{}, \tycho{}, and \hipparcos{} are available in online databases, e.g., in the VizieR archive at \url{https://vizier.cds.unistra.fr/viz-bin/VizieR-2}.

\tess{} photometry is availble in \mast{} at \url{https://archive.stsci.edu/}.

\idisk{} measurements were taken from literature references (cited above), but the underlying data is availble in the \alma{} archive at \url{https://almascience.nrao.edu/aq/} or \url{https://almascience.eso.org/aq/}.

All new data derived in this work are available in the tables herein.

\clearpage

\bibliographystyle{rasti}
\bibliography{DSA1.bib}

\begin{thebibliography}{160}
\expandafter\ifx\csname natexlab\endcsname\relax\def\natexlab#1{#1}\fi

\bibitem[Ahumada et~al.(2020)Ahumada, Prieto, Almeida, Anders, Anderson, Andrews, Anguiano, Arcodia, Armengaud, Aubert, Avila, Avila-Reese, Badenes, Balland, Barger, Barrera-Ballesteros, Basu, Bautista, Beaton, Beers, Benavides, Bender, Bernardi, Bershady, Beutler, Bidin, Bird, Bizyaev, Blanc, Blanton, Boquien, Borissova, Bovy, Brandt, Brinkmann, Brownstein, Bundy, Bureau, Burgasser, Burtin, Cano-Díaz, Capasso, Cappellari, Carrera, Chabanier, Chaplin, Chapman, Cherinka, Chiappini, Choi, Chojnowski, Chung, Clerc, Coffey, Comerford, Comparat, Costa, Cousinou, Covey, Crane, Cunha, Ilha, Dai, Damsted, Darling, Davidson, Davies, Dawson, De, Macorra, Lee, Queiroz, Machado, Torre, Dell’Agli, Bourboux, Diamond-Stanic, Dillon, Donor, Drory, Duckworth, Dwelly, Ebelke, Eftekharzadeh, Eigenbrot, Elsworth, Eracleous, Erfanianfar, Escoffier, Fan, Farr, Fernández-Trincado, Feuillet, Finoguenov, Fofie, Fraser-McKelvie, Frinchaboy, Fromenteau, Fu, Galbany, Garcia, García-Hernández, Oehmichen, Ge, Maia, Geisler, Gelfand,
  Goddy, Gonzalez-Perez, Grabowski, Green, Grier, Guo, Guy, Harding, Hasselquist, Hawken, Hayes, Hearty, Hekker, Hogg, Holtzman, Horta, Hou, Hsieh, Huber, Hunt, Chitham, Imig, Jaber, Angel, Johnson, Jones, Jönsson, Jullo, Kim, Kinemuchi, IV, Kite, Klaene, Kneib, Kollmeier, Kong, Kounkel, Krishnarao, Lacerna, Lan, Lane, Law, Goff, Leung, Lewis, Li, Lian, Lin, Long, Longa-Peña, Lundgren, Lyke, Mackereth, MacLeod, Majewski, Manchado, Maraston, Martini, Masseron, Masters, Mathur, McDermid, Merloni, Merrifield, Mészáros, Miglio, Minniti, Minsley, Miyaji, Mohammad, Mosser, Mueller, Muna, Muñoz-Gutiérrez, Myers, Nadathur, Nair, Nandra, Nascimento, Nevin, Newman, Nidever, Nitschelm, Noterdaeme, O’Connell, Olmstead, Oravetz, Oravetz, Osorio, Pace, Padilla, Palanque-Delabrouille, Palicio, Pan, Pan, Parker, Paviot, Peirani, Ramŕez, Penny, Percival, Perez-Fournon, Pérez-Ràfols, Petitjean, Pieri, Pinsonneault, Poovelil, Povick, Prakash, Price-Whelan, Raddick, Raichoor, Ray, Rembold, Rezaie, Riffel, Riffel, Rix,
  Robin, Roman-Lopes, Román-Zúñiga, Rose, Ross, Rossi, Rowlands, Rubin, Salvato, Sánchez, Sánchez-Menguiano, Sánchez-Gallego, Sayres, Schaefer, Schiavon, Schimoia, Schlafly, Schlegel, Schneider, Schultheis, Schwope, Seo, Serenelli, Shafieloo, Shamsi, Shao, Shen, Shetrone, Shirley, Aguirre, Simon, Skrutskie, Slosar, Smethurst, Sobeck, Sodi, Souto, Stark, Stassun, Steinmetz, Stello, Stermer, Storchi-Bergmann, Streblyanska, Stringfellow, Stutz, Suárez, Sun, Taghizadeh-Popp, Talbot, Tayar, Thakar, Theriault, Thomas, Thomas, Tinker, Tojeiro, Toledo, Tremonti, Troup, Tuttle, Unda-Sanzana, Valentini, Vargas-González, Vargas-Magaña, Vázquez-Mata, Vivek, Wake, Wang, Weaver, Weijmans, Wild, Wilson, Wilson, Wolthuis, Wood-Vasey, Yan, Yang, Yèche, Zamora, Zarrouk, Zasowski, Zhang, Zhao, Zhao, Zheng, Zheng, Zhu, \& Zou]{ahumada_16th_2020}
Ahumada, R., Prieto, C.~A., Almeida, A., Anders, F., Anderson, S.~F., Andrews, B.~H., Anguiano, B., Arcodia, R., Armengaud, E., Aubert, M., Avila, S., Avila-Reese, V., Badenes, C., Balland, C., Barger, K., Barrera-Ballesteros, J.~K., Basu, S., Bautista, J., Beaton, R.~L., Beers, T.~C., Benavides, B. I.~T., Bender, C.~F., Bernardi, M., Bershady, M., Beutler, F., Bidin, C.~M., Bird, J., Bizyaev, D., Blanc, G.~A., Blanton, M.~R., Boquien, M., Borissova, J., Bovy, J., Brandt, W.~N., Brinkmann, J., Brownstein, J.~R., Bundy, K., Bureau, M., Burgasser, A., Burtin, E., Cano-Díaz, M., Capasso, R., Cappellari, M., Carrera, R., Chabanier, S., Chaplin, W., Chapman, M., Cherinka, B., Chiappini, C., Choi, P.~D., Chojnowski, S.~D., Chung, H., Clerc, N., Coffey, D., Comerford, J.~M., Comparat, J., Costa, L.~d., Cousinou, M.-C., Covey, K., Crane, J.~D., Cunha, K., Ilha, G. d.~S., Dai, Y.~S., Damsted, S.~B., Darling, J., Davidson, J.~W., Davies, R., Dawson, K., De, N., Macorra, A. d.~l., Lee, N.~D., Queiroz, A. B. d.~A.,
  Machado, A.~D., Torre, S. d.~l., Dell’Agli, F., Bourboux, H. d. M.~d., Diamond-Stanic, A.~M., Dillon, S., Donor, J., Drory, N., Duckworth, C., Dwelly, T., Ebelke, G., Eftekharzadeh, S., Eigenbrot, A.~D., Elsworth, Y.~P., Eracleous, M., Erfanianfar, G., Escoffier, S., Fan, X., Farr, E., Fernández-Trincado, J.~G., Feuillet, D., Finoguenov, A., Fofie, P., Fraser-McKelvie, A., Frinchaboy, P.~M., Fromenteau, S., Fu, H., Galbany, L., Garcia, R.~A., García-Hernández, D.~A., Oehmichen, L. A.~G., Ge, J., Maia, M. A.~G., Geisler, D., Gelfand, J., Goddy, J., Gonzalez-Perez, V., Grabowski, K., Green, P., Grier, C.~J., Guo, H., Guy, J., Harding, P., Hasselquist, S., Hawken, A.~J., Hayes, C.~R., Hearty, F., Hekker, S., Hogg, D.~W., Holtzman, J.~A., Horta, D., Hou, J., Hsieh, B.-C., Huber, D., Hunt, J. A.~S., Chitham, J.~I., Imig, J., Jaber, M., Angel, C. E.~J., Johnson, J.~A., Jones, A.~M., Jönsson, H., Jullo, E., Kim, Y., Kinemuchi, K., IV, C. C.~K., Kite, G.~W., Klaene, M., Kneib, J.-P., Kollmeier, J.~A., Kong,
  H., Kounkel, M., Krishnarao, D., Lacerna, I., Lan, T.-W., Lane, R.~R., Law, D.~R., Goff, J.-M.~L., Leung, H.~W., Lewis, H., Li, C., Lian, J., Lin, L., Long, D., Longa-Peña, P., Lundgren, B., Lyke, B.~W., Mackereth, J.~T., MacLeod, C.~L., Majewski, S.~R., Manchado, A., Maraston, C., Martini, P., Masseron, T., Masters, K.~L., Mathur, S., McDermid, R.~M., Merloni, A., Merrifield, M., Mészáros, S., Miglio, A., Minniti, D., Minsley, R., Miyaji, T., Mohammad, F.~G., Mosser, B., Mueller, E.-M., Muna, D., Muñoz-Gutiérrez, A., Myers, A.~D., Nadathur, S., Nair, P., Nandra, K., Nascimento, J. C.~d., Nevin, R.~J., Newman, J.~A., Nidever, D.~L., Nitschelm, C., Noterdaeme, P., O’Connell, J.~E., Olmstead, M.~D., Oravetz, D., Oravetz, A., Osorio, Y., Pace, Z.~J., Padilla, N., Palanque-Delabrouille, N., Palicio, P.~A., Pan, H.-A., Pan, K., Parker, J., Paviot, R., Peirani, S., Ramŕez, K.~P., Penny, S., Percival, W.~J., Perez-Fournon, I., Pérez-Ràfols, I., Petitjean, P., Pieri, M.~M., Pinsonneault, M., Poovelil,
  V.~J., Povick, J.~T., Prakash, A., Price-Whelan, A.~M., Raddick, M.~J., Raichoor, A., Ray, A., Rembold, S.~B., Rezaie, M., Riffel, R.~A., Riffel, R., Rix, H.-W., Robin, A.~C., Roman-Lopes, A., Román-Zúñiga, C., Rose, B., Ross, A.~J., Rossi, G., Rowlands, K., Rubin, K. H.~R., Salvato, M., Sánchez, A.~G., Sánchez-Menguiano, L., Sánchez-Gallego, J.~R., Sayres, C., Schaefer, A., Schiavon, R.~P., Schimoia, J.~S., Schlafly, E., Schlegel, D., Schneider, D.~P., Schultheis, M., Schwope, A., Seo, H.-J., Serenelli, A., Shafieloo, A., Shamsi, S.~J., Shao, Z., Shen, S., Shetrone, M., Shirley, R., Aguirre, V.~S., Simon, J.~D., Skrutskie, M.~F., Slosar, A., Smethurst, R., Sobeck, J., Sodi, B.~C., Souto, D., Stark, D.~V., Stassun, K.~G., Steinmetz, M., Stello, D., Stermer, J., Storchi-Bergmann, T., Streblyanska, A., Stringfellow, G.~S., Stutz, A., Suárez, G., Sun, J., Taghizadeh-Popp, M., Talbot, M.~S., Tayar, J., Thakar, A.~R., Theriault, R., Thomas, D., Thomas, Z.~C., Tinker, J., Tojeiro, R., Toledo, H.~H.,
  Tremonti, C.~A., Troup, N.~W., Tuttle, S., Unda-Sanzana, E., Valentini, M., Vargas-González, J., Vargas-Magaña, M., Vázquez-Mata, J.~A., Vivek, M., Wake, D., Wang, Y., Weaver, B.~A., Weijmans, A.-M., Wild, V., Wilson, J.~C., Wilson, R.~F., Wolthuis, N., Wood-Vasey, W.~M., Yan, R., Yang, M., Yèche, C., Zamora, O., Zarrouk, P., Zasowski, G., Zhang, K., Zhao, C., Zhao, G., Zheng, Z., Zheng, Z., Zhu, G., \& Zou, H., 2020.
\newblock The 16th {Data} {Release} of the {Sloan} {Digital} {Sky} {Surveys}: {First} {Release} from the {APOGEE}-2 {Southern} {Survey} and {Full} {Release} of {eBOSS} {Spectra}, {\it The Astrophysical Journal Supplement Series\/}, {\bf 249}(1), 3, Publisher: The American Astronomical Society.

\bibitem[Akeson et~al.(2019)Akeson, Jensen, Carpenter, Ricci, Laos, Nogueira, \& Suen-Lewis]{akeson_resolved_2019}
Akeson, R.~L., Jensen, E. L.~N., Carpenter, J., Ricci, L., Laos, S., Nogueira, N.~F., \& Suen-Lewis, E.~M., 2019.
\newblock Resolved {Young} {Binary} {Systems} and {Their} {Disks}, {\it The Astrophysical Journal\/}, {\bf 872}(2), 158, Publisher: The American Astronomical Society.

\bibitem[Albrecht et~al.(2022)Albrecht, Dawson, \& Winn]{albrecht_stellar_2022}
Albrecht, S.~H., Dawson, R.~I., \& Winn, J.~N., 2022.
\newblock Stellar obliquities in exoplanetary systems, {\it arXiv:2203.05460 [astro-ph]\/}, arXiv: 2203.05460.

\bibitem[Alcalá et~al.(2017)Alcalá, Manara, Natta, Frasca, Testi, Nisini, Stelzer, Williams, Antoniucci, Biazzo, Covino, Esposito, Getman, \& Rigliaco]{alcala_x-shooter_2017}
Alcalá, J.~M., Manara, C.~F., Natta, A., Frasca, A., Testi, L., Nisini, B., Stelzer, B., Williams, J.~P., Antoniucci, S., Biazzo, K., Covino, E., Esposito, M., Getman, F., \& Rigliaco, E., 2017.
\newblock X-shooter spectroscopy of young stellar objects in {Lupus} - {Accretion} properties of class {II} and transitional objects, {\it Astronomy \& Astrophysics\/}, {\bf 600}, A20, Publisher: EDP Sciences.

\bibitem[Alexander(2012)]{alexander_dispersal_2012}
Alexander, R., 2012.
\newblock {THE} {DISPERSAL} {OF} {PROTOPLANETARY} {DISKS} {AROUND} {BINARY} {STARS}, {\it The Astrophysical Journal Letters\/}, {\bf 757}(2), L29, Publisher: The American Astronomical Society.

\bibitem[Allard et~al.(2013)Allard, Homeier, Freytag, Schaffenberger, \& Rajpurohit]{allard_progress_2013}
Allard, F., Homeier, D., Freytag, B., Schaffenberger, W., \& Rajpurohit, A.~S., 2013.
\newblock Progress in modeling very low mass stars, brown dwarfs, and planetary mass objects., {\it Memorie della Societa Astronomica Italiana Supplementi\/}, {\bf 24}, 128, ADS Bibcode: 2013MSAIS..24..128A.

\bibitem[Andrews \& Williams(2007)]{andrews_submillimeter_2007}
Andrews, S.~M. \& Williams, J.~P., 2007.
\newblock A {Submillimeter} {View} of {Circumstellar} {Dust} {Disks} in $\rho$ {Ophiuchi}, {\it The Astrophysical Journal\/}, {\bf 671}(2), 1800, Publisher: IOP Publishing.

\bibitem[Andrews et~al.(2010)Andrews, Czekala, Wilner, Espaillat, Dullemond, \& Hughes]{andrews_truncated_2010}
Andrews, S.~M., Czekala, I., Wilner, D.~J., Espaillat, C., Dullemond, C.~P., \& Hughes, A.~M., 2010.
\newblock {TRUNCATED} {DISKS} {IN} {TW} {Hya} {ASSOCIATION} {MULTIPLE} {STAR} {SYSTEMS}, {\it The Astrophysical Journal\/}, {\bf 710}(1), 462, Publisher: The American Astronomical Society.

\bibitem[Ansdell et~al.(2016)Ansdell, Williams, van~der Marel, Carpenter, Guidi, Hogerheijde, Mathews, Manara, Miotello, Natta, Oliveira, Tazzari, Testi, van Dishoeck, \& van Terwisga]{ansdell_alma_2016}
Ansdell, M., Williams, J.~P., van~der Marel, N., Carpenter, J.~M., Guidi, G., Hogerheijde, M., Mathews, G.~S., Manara, C.~F., Miotello, A., Natta, A., Oliveira, I., Tazzari, M., Testi, L., van Dishoeck, E.~F., \& van Terwisga, S.~E., 2016.
\newblock {ALMA} {Survey} of {Lupus} {Protoplanetary} {Disks}. {I}. {Dust} and {Gas} {Masses}, {\it The Astrophysical Journal\/}, {\bf 828}, 46, ADS Bibcode: 2016ApJ...828...46A.

\bibitem[Ansdell et~al.(2017)Ansdell, Williams, Manara, Miotello, Facchini, Marel, Testi, \& Dishoeck]{ansdell_alma_2017}
Ansdell, M., Williams, J.~P., Manara, C.~F., Miotello, A., Facchini, S., Marel, N. v.~d., Testi, L., \& Dishoeck, E. F.~v., 2017.
\newblock An {ALMA} {Survey} of {Protoplanetary} {Disks} in the $\sigma$ {Orionis} {Cluster}, {\it The Astronomical Journal\/}, {\bf 153}(5), 240, Publisher: The American Astronomical Society.

\bibitem[Ansdell et~al.(2020)Ansdell, Gaidos, Hedges, Tazzari, Kraus, Wyatt, Kennedy, Williams, Mann, Angelo, Dûchene, Mamajek, Carpenter, Esplin, \& Rizzuto]{ansdell_are_2020}
Ansdell, M., Gaidos, E., Hedges, C., Tazzari, M., Kraus, A.~L., Wyatt, M.~C., Kennedy, G.~M., Williams, J.~P., Mann, A.~W., Angelo, I., Dûchene, G., Mamajek, E.~E., Carpenter, J., Esplin, T.~L., \& Rizzuto, A.~C., 2020.
\newblock Are inner disc misalignments common? {ALMA} reveals an isotropic outer disc inclination distribution for young dipper stars, {\it Monthly Notices of the Royal Astronomical Society\/}, {\bf 492}(1), 572--588.

\bibitem[Artymowicz \& Lubow(1994)]{artymowicz_dynamics_1994}
Artymowicz, P. \& Lubow, S.~H., 1994.
\newblock Dynamics of binary-disk interaction. 1: {Resonances} and disk gap sizes, {\it The Astrophysical Journal\/}, {\bf 421}, 651.

\bibitem[Avenhaus et~al.(2014)Avenhaus, Quanz, Schmid, Meyer, Garufi, Wolf, \& Dominik]{avenhaus_structures_2014}
Avenhaus, H., Quanz, S.~P., Schmid, H.~M., Meyer, M.~R., Garufi, A., Wolf, S., \& Dominik, C., 2014.
\newblock Structures in the {Protoplanetary} {Disk} of {HD142527} {Seen} in {Polarized} {Scattered} {Light}, {\it The Astrophysical Journal\/}, {\bf 781}, 87, Publisher: IOP ADS Bibcode: 2014ApJ...781...87A.

\bibitem[Baraffe et~al.(2015)Baraffe, Homeier, Allard, \& Chabrier]{baraffe_new_2015}
Baraffe, I., Homeier, D., Allard, F., \& Chabrier, G., 2015.
\newblock New evolutionary models for pre-main sequence and main sequence low-mass stars down to the hydrogen-burning limit, {\it Astronomy and Astrophysics\/}, {\bf 577}, A42, ADS Bibcode: 2015A\&A...577A..42B.

\bibitem[Barber et~al.(2022)Barber, Mann, Bush, Tofflemire, Kraus, Krolikowski, Vanderburg, Fields, Newton, Owens, \& Thao]{barber_transit_2022}
Barber, M.~G., Mann, A.~W., Bush, J.~L., Tofflemire, B.~M., Kraus, A.~L., Krolikowski, D.~M., Vanderburg, A., Fields, M.~J., Newton, E.~R., Owens, D.~A., \& Thao, P.~C., 2022.
\newblock Transit {Hunt} for {Young} and {Maturing} {Exoplanets} ({THYME}). {VIII}. {A} {Pleiades}-age {Association} {Harboring} {Two} {Transiting} {Planetary} {Systems} from {Kepler}, {\it The Astronomical Journal\/}, {\bf 164}(3), 88, Publisher: The American Astronomical Society.

\bibitem[Barber et~al.(2024{\natexlab{a}})Barber, Mann, Vanderburg, Krolikowski, Kraus, Ansdell, Pearce, Mace, Andrews, Boyle, Collins, De~Furio, Dragomir, Espaillat, Feinstein, Fields, Jaffe, Lopez~Murillo, Murgas, Newton, Palle, Sawczynec, Schwarz, Thao, Tofflemire, Watkins, Jenkins, Latham, Ricker, Seager, Vanderspek, Winn, Charbonneau, Essack, Rodriguez, Shporer, Twicken, \& Villaseñor]{barber_giant_2024}
Barber, M.~G., Mann, A.~W., Vanderburg, A., Krolikowski, D., Kraus, A., Ansdell, M., Pearce, L., Mace, G.~N., Andrews, S.~M., Boyle, A.~W., Collins, K.~A., De~Furio, M., Dragomir, D., Espaillat, C., Feinstein, A.~D., Fields, M., Jaffe, D., Lopez~Murillo, A.~I., Murgas, F., Newton, E.~R., Palle, E., Sawczynec, E., Schwarz, R.~P., Thao, P.~C., Tofflemire, B.~M., Watkins, C.~N., Jenkins, J.~M., Latham, D.~W., Ricker, G., Seager, S., Vanderspek, R., Winn, J.~N., Charbonneau, D., Essack, Z., Rodriguez, D.~R., Shporer, A., Twicken, J.~D., \& Villaseñor, J.~N., 2024{\natexlab{a}}.
\newblock A giant planet transiting a 3-{Myr} protostar with a misaligned disk, {\it Nature\/}, {\bf 635}, 574--577, ADS Bibcode: 2024Natur.635..574B.

\bibitem[Barber et~al.(2024{\natexlab{b}})Barber, Thao, Mann, Vanderburg, Mori, Livingston, Fukui, Narita, Kraus, Tofflemire, Newton, Winn, Jenkins, Seager, Collins, \& Twicken]{barber_tess_2024}
Barber, M.~G., Thao, P.~C., Mann, A.~W., Vanderburg, A., Mori, M., Livingston, J.~H., Fukui, A., Narita, N., Kraus, A.~L., Tofflemire, B.~M., Newton, E.~R., Winn, J.~N., Jenkins, J.~M., Seager, S., Collins, K.~A., \& Twicken, J.~D., 2024{\natexlab{b}}.
\newblock {TESS} {Investigation}—{Demographics} of {Young} {Exoplanets} ({TI}-{DYE}). {II}. {A} {Second} {Giant} {Planet} in the 17 {Myr} {System} {HIP} 67522, {\it The Astrophysical Journal\/}, {\bf 973}, L30, Publisher: IOP ADS Bibcode: 2024ApJ...973L..30B.

\bibitem[Barenfeld et~al.(2017)Barenfeld, Carpenter, Sargent, Isella, \& Ricci]{barenfeld_measurement_2017}
Barenfeld, S.~A., Carpenter, J.~M., Sargent, A.~I., Isella, A., \& Ricci, L., 2017.
\newblock Measurement of {Circumstellar} {Disk} {Sizes} in the {Upper} {Scorpius} {OB} {Association} with {ALMA}, {\it The Astrophysical Journal\/}, {\bf 851}, 85, Publisher: IOP ADS Bibcode: 2017ApJ...851...85B.

\bibitem[Bate(2018)]{bate_diversity_2018}
Bate, M.~R., 2018.
\newblock On the diversity and statistical properties of protostellar discs, {\it Monthly Notices of the Royal Astronomical Society\/}, {\bf 475}(4), 5618--5658.

\bibitem[Bate et~al.(2010)Bate, Lodato, \& Pringle]{bate_chaotic_2010}
Bate, M.~R., Lodato, G., \& Pringle, J.~E., 2010.
\newblock Chaotic star formation and the alignment of stellar rotation with disc and planetary orbital axes, {\it Monthly Notices of the Royal Astronomical Society\/}, {\bf 401}(3), 1505--1513.

\bibitem[Batygin(2012)]{batygin_primordial_2012}
Batygin, K., 2012.
\newblock A primordial origin for misalignments between stellar spin axes and planetary orbits, {\it Nature\/}, {\bf 491}, 418--420, ADS Bibcode: 2012Natur.491..418B.

\bibitem[Beck \& Giles(2005)]{2005ApJ...621L.153B}
Beck, J.~G. \& Giles, P., 2005.
\newblock Helioseismic {Determination} of the {Solar} {Rotation} {Axis}, {\it The Astrophysical Journal\/}, {\bf 621}, L153--L156, ADS Bibcode: 2005ApJ...621L.153B.

\bibitem[Binks \& Jeffries(2014)]{binks_lithium_2014}
Binks, A.~S. \& Jeffries, R.~D., 2014.
\newblock A lithium depletion boundary age of 21 {Myr} for the {Beta} {Pictoris} moving group, {\it Monthly Notices of the Royal Astronomical Society\/}, {\bf 438}, L11--L15, Publisher: OUP ADS Bibcode: 2014MNRAS.438L..11B.

\bibitem[Binks \& Jeffries(2016)]{binks_spectroscopic_2016}
Binks, A.~S. \& Jeffries, R.~D., 2016.
\newblock Spectroscopic confirmation of {M}-dwarf candidate members of the {Beta} {Pictoris} and {AB} {Doradus} {Moving} {Groups}, {\it Monthly Notices of the Royal Astronomical Society\/}, {\bf 455}, 3345--3358, Publisher: OUP ADS Bibcode: 2016MNRAS.455.3345B.

\bibitem[Blackwell \& Shallis(1977)]{blackwell_stellar_1977}
Blackwell, D.~E. \& Shallis, M.~J., 1977.
\newblock Stellar angular diameters from infrared photometry. {Application} to {Arcturus} and other stars; with effective temperatures, {\it Monthly Notices of the Royal Astronomical Society\/}, {\bf 180}(2), 177--191.

\bibitem[Bowler et~al.(2023)Bowler, Tran, Zhang, Morgan, Ashok, Blunt, Bryan, Evans, Franson, Huber, Nagpal, Wu, \& Zhou]{bowler_rotation_2023}
Bowler, B.~P., Tran, Q.~H., Zhang, Z., Morgan, M., Ashok, K.~B., Blunt, S., Bryan, M.~L., Evans, A.~E., Franson, K., Huber, D., Nagpal, V., Wu, Y.-L., \& Zhou, Y., 2023.
\newblock Rotation {Periods}, {Inclinations}, and {Obliquities} of {Cool} {Stars} {Hosting} {Directly} {Imaged} {Substellar} {Companions}: {Spin}-{Orbit} {Misalignments} {Are} {Common}, {\it The Astronomical Journal\/}, {\bf 165}, 164, Publisher: IOP ADS Bibcode: 2023AJ....165..164B.

\bibitem[Brandt \& Huang(2015)]{brandt_age_2015}
Brandt, T.~D. \& Huang, C.~X., 2015.
\newblock {THE} {AGE} {AND} {AGE} {SPREAD} {OF} {THE} {PRAESEPE} {AND} {HYADES} {CLUSTERS}: {A} {CONSISTENT}, $\sim$800 {Myr} {PICTURE} {FROM} {ROTATING} {STELLAR} {MODELS}, {\it The Astrophysical Journal\/}, {\bf 807}(1), 24, Publisher: The American Astronomical Society.

\bibitem[Bressan et~al.(2012)Bressan, Marigo, Girardi, Salasnich, Dal~Cero, Rubele, \& Nanni]{bressan_parsec_2012}
Bressan, A., Marigo, P., Girardi, L., Salasnich, B., Dal~Cero, C., Rubele, S., \& Nanni, A., 2012.
\newblock {PARSEC}: stellar tracks and isochrones with the {PAdova} and {TRieste} {Stellar} {Evolution} {Code}, {\it Monthly Notices of the Royal Astronomical Society\/}, {\bf 427}(1), 127--145.

\bibitem[Campbell \& Garrison(1985)]{campbell_inclination_1985}
Campbell, B. \& Garrison, R.~F., 1985.
\newblock On the inclination of extra-solar planetary orbits, {\it Publications of the Astronomical Society of the Pacific\/}, {\bf 97}, 180.

\bibitem[Cardelli et~al.(1989)Cardelli, Clayton, \& Mathis]{cardelli_relationship_1989}
Cardelli, J.~A., Clayton, G.~C., \& Mathis, J.~S., 1989.
\newblock The {Relationship} between {Infrared}, {Optical}, and {Ultraviolet} {Extinction}, {\it The Astrophysical Journal\/}, {\bf 345}, 245.

\bibitem[Casagrande et~al.(2010)Casagrande, Ramírez, Meléndez, Bessell, \& Asplund]{casagrande_absolutely_2010}
Casagrande, L., Ramírez, I., Meléndez, J., Bessell, M., \& Asplund, M., 2010.
\newblock An absolutely calibrated {Teff} scale from the infrared flux method. {Dwarfs} and subgiants, {\it Astronomy and Astrophysics\/}, {\bf 512}, A54, ADS Bibcode: 2010A\&A...512A..54C.

\bibitem[Casassus et~al.(2018)Casassus, Avenhaus, Pérez, Navarro, Cárcamo, Marino, Cieza, Quanz, Alarcón, Zurlo, Osses, Rannou, Román, \& Barraza]{casassus_inner_2018}
Casassus, S., Avenhaus, H., Pérez, S., Navarro, V., Cárcamo, M., Marino, S., Cieza, L., Quanz, S.~P., Alarcón, F., Zurlo, A., Osses, A., Rannou, F.~R., Román, P.~E., \& Barraza, M., 2018.
\newblock An inner warp in the {DoAr} 44 {T} {Tauri} transition disc, {\it Monthly Notices of the Royal Astronomical Society\/}, {\bf 477}(4), 5104--5114.

\bibitem[Chaboyer et~al.(2001)Chaboyer, Fenton, Nelan, Patnaude, \& Simon]{chaboyer_heavy-element_2001}
Chaboyer, B., Fenton, W.~H., Nelan, J.~E., Patnaude, D.~J., \& Simon, F.~E., 2001.
\newblock Heavy-{Element} {Diffusion} in {Metal}-poor {Stars}, {\it The Astrophysical Journal\/}, {\bf 562}(1), 521, Publisher: IOP Publishing.

\bibitem[Chambers \& {Pan-STARRS Team}(2018)]{chambers_pan-starrs1_2018}
Chambers, K. \& {Pan-STARRS Team}, 2018.
\newblock The {Pan}-{STARRS1} {Surveys}, {\bf 231}, 102.01, Conference Name: American Astronomical Society Meeting Abstracts \#231 ADS Bibcode: 2018AAS...23110201C.

\bibitem[Choi et~al.(2016)Choi, Dotter, Conroy, Cantiello, Paxton, \& Johnson]{choi_mesa_2016}
Choi, J., Dotter, A., Conroy, C., Cantiello, M., Paxton, B., \& Johnson, B.~D., 2016.
\newblock Mesa {Isochrones} and {Stellar} {Tracks} ({MIST}). {I}. {Solar}-scaled {Models}, {\it The Astrophysical Journal\/}, {\bf 823}, 102, Publisher: IOP ADS Bibcode: 2016ApJ...823..102C.

\bibitem[Colman et~al.(2024)Colman, Angus, David, Curtis, Hattori, \& Lu]{colman_methods_2024}
Colman, I.~L., Angus, R., David, T., Curtis, J., Hattori, S., \& Lu, Y.~L., 2024.
\newblock Methods for the {Detection} of {Stellar} {Rotation} {Periods} in {Individual} {TESS} {Sectors} and {Results} from the {Prime} {Mission}, {\it The Astronomical Journal\/}, {\bf 167}, 189, Publisher: IOP ADS Bibcode: 2024AJ....167..189C.

\bibitem[Couture et~al.(2023)Couture, Gagné, \& Doyon]{couture_addressing_2023}
Couture, D., Gagné, J., \& Doyon, R., 2023.
\newblock Addressing {Systematics} in the {Traceback} {Age} of the $\beta$ {Pictoris} {Moving} {Group}, {\it The Astrophysical Journal\/}, {\bf 946}(1), 6, Publisher: The American Astronomical Society.

\bibitem[Czekala et~al.(2019)Czekala, Chiang, Andrews, Jensen, Torres, Wilner, Stassun, \& Macintosh]{czekala_degree_2019}
Czekala, I., Chiang, E., Andrews, S.~M., Jensen, E. L.~N., Torres, G., Wilner, D.~J., Stassun, K.~G., \& Macintosh, B., 2019.
\newblock The {Degree} of {Alignment} between {Circumbinary} {Disks} and {Their} {Binary} {Hosts}, {\it The Astrophysical Journal\/}, {\bf 883}(1), 22, Publisher: The American Astronomical Society.

\bibitem[David et~al.(2019{\natexlab{a}})David, Cody, Hedges, Mamajek, Hillenbrand, Ciardi, Beichman, Petigura, Fulton, Isaacson, Howard, Gagné, Saunders, Rebull, Stauffer, Vasisht, \& Hinkley]{david_warm_2019}
David, T.~J., Cody, A.~M., Hedges, C.~L., Mamajek, E.~E., Hillenbrand, L.~A., Ciardi, D.~R., Beichman, C.~A., Petigura, E.~A., Fulton, B.~J., Isaacson, H.~T., Howard, A.~W., Gagné, J., Saunders, N.~K., Rebull, L.~M., Stauffer, J.~R., Vasisht, G., \& Hinkley, S., 2019{\natexlab{a}}.
\newblock A {Warm} {Jupiter}-sized {Planet} {Transiting} the {Pre}-main-sequence {Star} {V1298} {Tau}, {\it The Astronomical Journal\/}, {\bf 158}(2), 79, Publisher: American Astronomical Society.

\bibitem[David et~al.(2019{\natexlab{b}})David, Hillenbrand, Gillen, Cody, Howell, Isaacson, \& Livingston]{david_age_2019}
David, T.~J., Hillenbrand, L.~A., Gillen, E., Cody, A.~M., Howell, S.~B., Isaacson, H.~T., \& Livingston, J.~H., 2019{\natexlab{b}}.
\newblock Age {Determination} in {Upper} {Scorpius} with {Eclipsing} {Binaries}, {\it The Astrophysical Journal\/}, {\bf 872}(2), 161, Publisher: American Astronomical Society.

\bibitem[Davies(2019)]{davies_stardisc_2019}
Davies, C.~L., 2019.
\newblock Star–disc (mis-)alignment in {Rho} {Oph} and {Upper} {Sco}: insights from spatially resolved disc systems with {K2} rotation periods, {\it Monthly Notices of the Royal Astronomical Society\/}, {\bf 484}(2), 1926--1935.

\bibitem[de~la Reza \& Pinzón(2004)]{de_la_reza_rotation_2004}
de~la Reza, R. \& Pinzón, G., 2004.
\newblock On the {Rotation} of {Post}-{T} {Tauri} {Stars} in {Associations}, {\it The Astronomical Journal\/}, {\bf 128}, 1812--1824, Publisher: IOP ADS Bibcode: 2004AJ....128.1812D.

\bibitem[Deshpande et~al.(2012)Deshpande, Martín, Montgomery, Osorio, Rodler, Burgo, Bao, Lyubchik, Tata, Bouy, \& Pavlenko]{deshpande_intermediate_2012}
Deshpande, R., Martín, E.~L., Montgomery, M.~M., Osorio, M. R.~Z., Rodler, F., Burgo, C.~d., Bao, N.~P., Lyubchik, Y., Tata, R., Bouy, H., \& Pavlenko, Y., 2012.
\newblock {INTERMEDIATE} {RESOLUTION} {NEAR}-{INFRARED} {SPECTROSCOPY} {OF} 36 {LATE} {M} {DWARFS}, {\it The Astronomical Journal\/}, {\bf 144}(4), 99, Publisher: The American Astronomical Society.

\bibitem[Desidera et~al.(2015)Desidera, Covino, Messina, Carson, Hagelberg, Schlieder, Biazzo, Alcalá, Chauvin, Vigan, Beuzit, Bonavita, Bonnefoy, Delorme, D'Orazi, Esposito, Feldt, Girardi, Gratton, Henning, Lagrange, Lanzafame, Launhardt, Marmier, Melo, Meyer, Mouillet, Moutou, Segransan, Udry, \& Zaidi]{desidera_vltnaco_2015}
Desidera, S., Covino, E., Messina, S., Carson, J., Hagelberg, J., Schlieder, J.~E., Biazzo, K., Alcalá, J.~M., Chauvin, G., Vigan, A., Beuzit, J.~L., Bonavita, M., Bonnefoy, M., Delorme, P., D'Orazi, V., Esposito, M., Feldt, M., Girardi, L., Gratton, R., Henning, T., Lagrange, A.~M., Lanzafame, A.~C., Launhardt, R., Marmier, M., Melo, C., Meyer, M., Mouillet, D., Moutou, C., Segransan, D., Udry, S., \& Zaidi, C.~M., 2015.
\newblock The {VLT}/{NaCo} large program to probe the occurrence of exoplanets and brown dwarfs in wide orbits. {I}. {Sample} definition and characterization, {\it Astronomy and Astrophysics\/}, {\bf 573}, A126, Publisher: EDP ADS Bibcode: 2015A\&A...573A.126D.

\bibitem[Dong \& Foreman-Mackey(2023)]{dong_hierarchical_2023}
Dong, J. \& Foreman-Mackey, D., 2023.
\newblock A {Hierarchical} {Bayesian} {Framework} for {Inferring} the {Stellar} {Obliquity} {Distribution}, {\it The Astronomical Journal\/}, {\bf 166}, 112, Publisher: IOP ADS Bibcode: 2023AJ....166..112D.

\bibitem[Dotter et~al.(2008)Dotter, Chaboyer, Jevremović, Kostov, Baron, \& Ferguson]{dotter_dartmouth_2008}
Dotter, A., Chaboyer, B., Jevremović, D., Kostov, V., Baron, E., \& Ferguson, J.~W., 2008.
\newblock The {Dartmouth} {Stellar} {Evolution} {Database}, {\it The Astrophysical Journal Supplement Series\/}, {\bf 178}(1), 89--101.

\bibitem[El-Badry et~al.(2019)El-Badry, Rix, Tian, Duchêne, \& Moe]{el-badry_discovery_2019}
El-Badry, K., Rix, H.-W., Tian, H., Duchêne, G., \& Moe, M., 2019.
\newblock Discovery of an equal-mass `twin' binary population reaching 1000 + au separations, {\it Monthly Notices of the Royal Astronomical Society\/}, {\bf 489}, 5822--5857, Publisher: OUP ADS Bibcode: 2019MNRAS.489.5822E.

\bibitem[{ESA}(1997)]{esa_hipparcos_1997}
{ESA}, 1997.
\newblock The {HIPPARCOS} and {TYCHO} catalogues. {Astrometric} and photometric star catalogues derived from the {ESA} {HIPPARCOS} {Space} {Astrometry} {Mission}, {\it ESA Special Publication\/}, {\bf 1200}, ADS Bibcode: 1997ESASP1200.....E.

\bibitem[Fabrycky \& Winn(2009)]{fabrycky_exoplanetary_2009}
Fabrycky, D.~C. \& Winn, J.~N., 2009.
\newblock Exoplanetary {Spin}-{Orbit} {Alignment}: {Results} from the {Ensemble} of {Rossiter}-{McLaughlin} {Observations}, {\it The Astrophysical Journal\/}, {\bf 696}, 1230--1240, Publisher: IOP ADS Bibcode: 2009ApJ...696.1230F.

\bibitem[Favata et~al.(1995)Favata, Barbera, Micela, \& Sciortino]{favata_lithium_1995}
Favata, F., Barbera, M., Micela, G., \& Sciortino, S., 1995.
\newblock Lithium, {X}-ray activity and rotation in an {X}-ray selected sample of solar-type stars., {\it Astronomy and Astrophysics\/}, {\bf 295}, 147--160, ADS Bibcode: 1995A\&A...295..147F.

\bibitem[Feiden(2016)]{feiden_magnetic_2016}
Feiden, G.~A., 2016.
\newblock Magnetic inhibition of convection and the fundamental properties of low-mass stars: {III}. {A} consistent 10 {Myr} age for the {Upper} {Scorpius} {OB} association, {\it Astronomy \& Astrophysics\/}, {\bf 593}, A99.

\bibitem[Feiden \& Chaboyer(2012)]{feiden_self-consistent_2012}
Feiden, G.~A. \& Chaboyer, B., 2012.
\newblock {SELF}-{CONSISTENT} {MAGNETIC} {STELLAR} {EVOLUTION} {MODELS} {OF} {THE} {DETACHED}, {SOLAR}-{TYPE} {ECLIPSING} {BINARY} {EF} {AQUARII}, {\it The Astrophysical Journal\/}, {\bf 761}(1), 30, Publisher: The American Astronomical Society.

\bibitem[Fernandes et~al.(2023)Fernandes, Hardegree-Ullman, Pascucci, Bergsten, Mulders, Cunha, Mamajek, Pearson, Feiden, \& Curtis]{fernandes_using_2023}
Fernandes, R.~B., Hardegree-Ullman, K.~K., Pascucci, I., Bergsten, G.~J., Mulders, G.~D., Cunha, K., Mamajek, E.~E., Pearson, K.~A., Feiden, G.~A., \& Curtis, J.~L., 2023.
\newblock Using {Photometrically} {Derived} {Properties} of {Young} {Stars} to {Refine} {TESS}'s {Transiting} {Young} {Planet} {Survey} {Completeness}, {\it The Astronomical Journal\/}, {\bf 166}, 175, Publisher: IOP ADS Bibcode: 2023AJ....166..175F.

\bibitem[Fielding et~al.(2015)Fielding, McKee, Socrates, Cunningham, \& Klein]{fielding_turbulent_2015}
Fielding, D.~B., McKee, C.~F., Socrates, A., Cunningham, A.~J., \& Klein, R.~I., 2015.
\newblock The turbulent origin of spin–orbit misalignment in planetary systems, {\it Monthly Notices of the Royal Astronomical Society\/}, {\bf 450}(3), 3306--3318.

\bibitem[Fitton et~al.(2022)Fitton, Tofflemire, \& Kraus]{fitton_disk_2022}
Fitton, S., Tofflemire, B.~M., \& Kraus, A.~L., 2022.
\newblock Disk {Material} {Inflates} {Gaia} {RUWE} {Values} in {Single} {Stars}, {\it Research Notes of the AAS\/}, {\bf 6}(1), 18, Publisher: The American Astronomical Society.

\bibitem[Ford \& Rasio(2008)]{ford_origins_2008}
Ford, E.~B. \& Rasio, F.~A., 2008.
\newblock Origins of {Eccentric} {Extrasolar} {Planets}: {Testing} the {Planet}-{Planet} {Scattering} {Model}, {\it The Astrophysical Journal\/}, {\bf 686}, 621--636, Publisher: IOP ADS Bibcode: 2008ApJ...686..621F.

\bibitem[Foreman-Mackey et~al.(2013)Foreman-Mackey, Hogg, Lang, \& Goodman]{foreman-mackey_emcee_2013}
Foreman-Mackey, D., Hogg, D.~W., Lang, D., \& Goodman, J., 2013.
\newblock emcee: {The} {MCMC} {Hammer}, {\it Publications of the Astronomical Society of the Pacific\/}, {\bf 125}(925), 306, Publisher: IOP Publishing.

\bibitem[Fouqué et~al.(2018)Fouqué, Moutou, Malo, Martioli, Lim, Rajpurohit, Artigau, Delfosse, Donati, Forveille, Morin, Allard, Delage, Doyon, Hébrard, \& Neves]{fouque_spirou_2018}
Fouqué, P., Moutou, C., Malo, L., Martioli, E., Lim, O., Rajpurohit, A., Artigau, E., Delfosse, X., Donati, J.-F., Forveille, T., Morin, J., Allard, F., Delage, R., Doyon, R., Hébrard, E., \& Neves, V., 2018.
\newblock {SPIRou} {Input} {Catalogue}: global properties of 440 {M} dwarfs observed with {ESPaDOnS} at {CFHT}, {\it Monthly Notices of the Royal Astronomical Society\/}, {\bf 475}, 1960--1986, Publisher: OUP ADS Bibcode: 2018MNRAS.475.1960F.

\bibitem[Gagné et~al.(2015)Gagné, Faherty, Cruz, Lafreniére, Doyon, Malo, Burgasser, Naud, Artigau, Bouchard, Gizis, \& Albert]{gagne_banyan_2015}
Gagné, J., Faherty, J.~K., Cruz, K.~L., Lafreniére, D., Doyon, R., Malo, L., Burgasser, A.~J., Naud, M.-E., Artigau, E., Bouchard, S., Gizis, J.~E., \& Albert, L., 2015.
\newblock {BANYAN}. {VII}. {A} {New} {Population} of {Young} {Substellar} {Candidate} {Members} of {Nearby} {Moving} {Groups} from the {BASS} {Survey}, {\it The Astrophysical Journal Supplement Series\/}, {\bf 219}, 33, Publisher: IOP ADS Bibcode: 2015ApJS..219...33G.

\bibitem[Gillen et~al.(2017)Gillen, Hillenbrand, David, Aigrain, Rebull, Stauffer, Cody, \& Queloz]{gillen_new_2017}
Gillen, E., Hillenbrand, L.~A., David, T.~J., Aigrain, S., Rebull, L., Stauffer, J., Cody, A.~M., \& Queloz, D., 2017.
\newblock New {Low}-mass {Eclipsing} {Binary} {Systems} in {Praesepe} {Discovered} by {K2}, {\it The Astrophysical Journal\/}, {\bf 849}(1), 11, Publisher: The American Astronomical Society.

\bibitem[Greaves et~al.(2014)Greaves, Kennedy, Thureau, Eiroa, Marshall, Maldonado, Matthews, Olofsson, Barlow, Moro-Martín, Sibthorpe, Absil, Ardila, Booth, Broekhoven-Fiene, Brown, Cameron, Burgo, Francesco, Eislöffel, Duchêne, Ertel, Holland, Horner, Kalas, Kavelaars, Lestrade, Vican, Wilner, Wolf, \& Wyatt]{greaves_alignment_2014}
Greaves, J.~S., Kennedy, G.~M., Thureau, N., Eiroa, C., Marshall, J.~P., Maldonado, J., Matthews, B.~C., Olofsson, G., Barlow, M.~J., Moro-Martín, A., Sibthorpe, B., Absil, O., Ardila, D.~R., Booth, M., Broekhoven-Fiene, H., Brown, D. J.~A., Cameron, A.~C., Burgo, C.~d., Francesco, J.~D., Eislöffel, J., Duchêne, G., Ertel, S., Holland, W.~S., Horner, J., Kalas, P., Kavelaars, J.~J., Lestrade, J.-F., Vican, L., Wilner, D.~J., Wolf, S., \& Wyatt, M.~C., 2014.
\newblock Alignment in star–debris disc systems seen by {Herschel}, {\it Monthly Notices of the Royal Astronomical Society: Letters\/}, {\bf 438}(1), L31--L35.

\bibitem[Gully-Santiago et~al.(2017)Gully-Santiago, Herczeg, Czekala, Somers, Grankin, Covey, Donati, Alencar, Hussain, Shappee, Mace, Lee, Holoien, Jose, \& Liu]{gully-santiago_placing_2017}
Gully-Santiago, M.~A., Herczeg, G.~J., Czekala, I., Somers, G., Grankin, K., Covey, K.~R., Donati, J.~F., Alencar, S. H.~P., Hussain, G. A.~J., Shappee, B.~J., Mace, G.~N., Lee, J.-J., Holoien, T. W.~S., Jose, J., \& Liu, C.-F., 2017.
\newblock Placing the {Spotted} {T} {Tauri} {Star} {LkCa} 4 on an {HR} {Diagram}, {\it The Astrophysical Journal\/}, {\bf 836}, 200, Publisher: IOP ADS Bibcode: 2017ApJ...836..200G.

\bibitem[Harris et~al.(2020)Harris, Millman, van~der Walt, Gommers, Virtanen, Cournapeau, Wieser, Taylor, Berg, Smith, Kern, Picus, Hoyer, van Kerkwijk, Brett, Haldane, del Río, Wiebe, Peterson, Gérard-Marchant, Sheppard, Reddy, Weckesser, Abbasi, Gohlke, \& Oliphant]{harris_array_2020}
Harris, C.~R., Millman, K.~J., van~der Walt, S.~J., Gommers, R., Virtanen, P., Cournapeau, D., Wieser, E., Taylor, J., Berg, S., Smith, N.~J., Kern, R., Picus, M., Hoyer, S., van Kerkwijk, M.~H., Brett, M., Haldane, A., del Río, J.~F., Wiebe, M., Peterson, P., Gérard-Marchant, P., Sheppard, K., Reddy, T., Weckesser, W., Abbasi, H., Gohlke, C., \& Oliphant, T.~E., 2020.
\newblock Array programming with {NumPy}, {\it Nature\/}, {\bf 585}(7825), 357--362.

\bibitem[Hattori et~al.(2022)Hattori, Foreman-Mackey, Hogg, Montet, Angus, Pritchard, Curtis, \& Schölkopf]{hattori_unpopular_2022}
Hattori, S., Foreman-Mackey, D., Hogg, D.~W., Montet, B.~T., Angus, R., Pritchard, T.~A., Curtis, J.~L., \& Schölkopf, B., 2022.
\newblock The unpopular {Package}: {A} {Data}-driven {Approach} to {Detrending} {TESS} {Full}-frame {Image} {Light} {Curves}, {\it The Astronomical Journal\/}, {\bf 163}(6), 284.

\bibitem[Henden et~al.(2009)Henden, Welch, Terrell, \& Levine]{henden_aavso_2009}
Henden, A.~A., Welch, D.~L., Terrell, D., \& Levine, S.~E., 2009.
\newblock The {AAVSO} {Photometric} {All}-{Sky} {Survey} ({APASS}), {\bf 214}, 407.02, Conference Name: American Astronomical Society Meeting Abstracts \#214 ADS Bibcode: 2009AAS...21440702H.

\bibitem[Henden et~al.(2015)Henden, Levine, Terrell, \& Welch]{henden_apass_2015}
Henden, A.~A., Levine, S., Terrell, D., \& Welch, D.~L., 2015.
\newblock {APASS} - {The} {Latest} {Data} {Release}, {\bf 225}, 336.16, Conference Name: American Astronomical Society Meeting Abstracts \#225 ADS Bibcode: 2015AAS...22533616H.

\bibitem[Herczeg \& Hillenbrand(2008)]{herczeg_uv_2008}
Herczeg, G.~J. \& Hillenbrand, L.~A., 2008.
\newblock {UV} {Excess} {Measures} of {Accretion} onto {Young} {Very} {Low} {Mass} {Stars} and {Brown} {Dwarfs}, {\it The Astrophysical Journal\/}, {\bf 681}, 594--625, Publisher: IOP ADS Bibcode: 2008ApJ...681..594H.

\bibitem[Howard et~al.(2020)Howard, Corbett, Law, Ratzloff, Galliher, Glazier, Fors, del Ser, \& Haislip]{howard_evryflare_2020}
Howard, W.~S., Corbett, H., Law, N.~M., Ratzloff, J.~K., Galliher, N., Glazier, A., Fors, O., del Ser, D., \& Haislip, J., 2020.
\newblock {EvryFlare}. {II}. {Rotation} {Periods} of the {Cool} {Flare} {Stars} in {TESS} across {Half} the {Southern} {Sky}, {\it The Astrophysical Journal\/}, {\bf 895}, 140, Publisher: IOP ADS Bibcode: 2020ApJ...895..140H.

\bibitem[Huang et~al.(2018)Huang, Andrews, Dullemond, Isella, Pérez, Guzmán, Öberg, Zhu, Zhang, Bai, Benisty, Birnstiel, Carpenter, Hughes, Ricci, Weaver, \& Wilner]{huang_disk_2018}
Huang, J., Andrews, S.~M., Dullemond, C.~P., Isella, A., Pérez, L.~M., Guzmán, V.~V., Öberg, K.~I., Zhu, Z., Zhang, S., Bai, X.-N., Benisty, M., Birnstiel, T., Carpenter, J.~M., Hughes, A.~M., Ricci, L., Weaver, E., \& Wilner, D.~J., 2018.
\newblock The {Disk} {Substructures} at {High} {Angular} {Resolution} {Project} ({DSHARP}). {II}. {Characteristics} of {Annular} {Substructures}, {\it The Astrophysical Journal\/}, {\bf 869}(2), L42, Publisher: American Astronomical Society.

\bibitem[Hurt \& MacGregor(2023)]{hurt_evidence_2023}
Hurt, S.~A. \& MacGregor, M.~A., 2023.
\newblock Evidence for {Misalignment} between {Debris} {Disks} and {Their} {Host} {Stars}, {\it The Astrophysical Journal\/}, {\bf 954}(1), 10, Publisher: The American Astronomical Society.

\bibitem[Husser et~al.(2013)Husser, Berg, Dreizler, Homeier, Reiners, Barman, \& Hauschildt]{husser_new_2013}
Husser, T.-O., Berg, S. W.-v., Dreizler, S., Homeier, D., Reiners, A., Barman, T., \& Hauschildt, P.~H., 2013.
\newblock A new extensive library of {PHOENIX} stellar atmospheres and synthetic spectra, {\it Astronomy \& Astrophysics\/}, {\bf 553}, A6, Publisher: EDP Sciences.

\bibitem[Høg et~al.(2000)Høg, Fabricius, Makarov, Urban, Corbin, Wycoff, Bastian, Schwekendiek, \& Wicenec]{hog_tycho-2_2000}
Høg, E., Fabricius, C., Makarov, V.~V., Urban, S., Corbin, T., Wycoff, G., Bastian, U., Schwekendiek, P., \& Wicenec, A., 2000.
\newblock The {Tycho}-2 catalogue of the 2.5 million brightest stars, {\it Astronomy and Astrophysics\/}, {\bf 355}, L27--L30, ADS Bibcode: 2000A\&A...355L..27H.

\bibitem[Jang-Condell(2015)]{jang-condell_likelihood_2015}
Jang-Condell, H., 2015.
\newblock {ON} {THE} {LIKELIHOOD} {OF} {PLANET} {FORMATION} {IN} {CLOSE} {BINARIES}, {\it The Astrophysical Journal\/}, {\bf 799}(2), 147, Publisher: The American Astronomical Society.

\bibitem[Jayawardhana et~al.(2006)Jayawardhana, Coffey, Scholz, Brandeker, \& van Kerkwijk]{jayawardhana_accretion_2006}
Jayawardhana, R., Coffey, J., Scholz, A., Brandeker, A., \& van Kerkwijk, M.~H., 2006.
\newblock Accretion {Disks} around {Young} {Stars}: {Lifetimes}, {Disk} {Locking}, and {Variability}, {\it The Astrophysical Journal\/}, {\bf 648}, 1206--1218, Publisher: IOP ADS Bibcode: 2006ApJ...648.1206J.

\bibitem[Jensen et~al.(2007)Jensen, Dhital, Stassun, Patience, Herbst, Walter, Simon, \& Basri]{jensen_periodic_2007}
Jensen, E. L.~N., Dhital, S., Stassun, K.~G., Patience, J., Herbst, W., Walter, F.~M., Simon, M., \& Basri, G., 2007.
\newblock Periodic {Accretion} from a {Circumbinary} {Disk} in the {Young} {Binary} {UZ} {Tau} {E}, {\it The Astronomical Journal\/}, {\bf 134}(1), 241, Publisher: IOP Publishing.

\bibitem[Kado-Fong et~al.(2016)Kado-Fong, Williams, Mann, Berger, Burgett, Chambers, Huber, Kaiser, Kudritzki, Magnier, Rest, Wainscoat, \& Waters]{kado-fong_m_2016}
Kado-Fong, E., Williams, P. K.~G., Mann, A.~W., Berger, E., Burgett, W.~S., Chambers, K.~C., Huber, M.~E., Kaiser, N., Kudritzki, R.~P., Magnier, E.~A., Rest, A., Wainscoat, R.~J., \& Waters, C., 2016.
\newblock M {Dwarf} {Activity} in the {Pan}-{STARRS1} {Medium}-{Deep} {Survey}: {First} {Catalog} and {Rotation} {Periods}, {\it The Astrophysical Journal\/}, {\bf 833}, 281, Publisher: IOP ADS Bibcode: 2016ApJ...833..281K.

\bibitem[Kennedy et~al.(2019)Kennedy, Matrà, Facchini, Milli, Panić, Price, Wilner, Wyatt, \& Yelverton]{kennedy_circumbinary_2019}
Kennedy, G.~M., Matrà, L., Facchini, S., Milli, J., Panić, O., Price, D., Wilner, D.~J., Wyatt, M.~C., \& Yelverton, B.~M., 2019.
\newblock A circumbinary protoplanetary disk in a polar configuration, {\it Nature Astronomy\/}, {\bf 3}(3), 230--235, Publisher: Nature Publishing Group.

\bibitem[Kesseli et~al.(2018)Kesseli, Muirhead, Mann, \& Mace]{kesseli_magnetic_2018}
Kesseli, A.~Y., Muirhead, P.~S., Mann, A.~W., \& Mace, G., 2018.
\newblock Magnetic {Inflation} and {Stellar} {Mass}. {II}. {On} the {Radii} of {Single}, {Rapidly} {Rotating}, {Fully} {Convective} {M}-{Dwarf} {Stars}, {\it The Astronomical Journal\/}, {\bf 155}(6), 225, Publisher: The American Astronomical Society.

\bibitem[Koposov(2023)]{koposov_segasaiminimint_2023}
Koposov, S., 2023.
\newblock segasai/minimint: {Minimint} v0.4.1.

\bibitem[Kounkel et~al.(2021)Kounkel, Covey, Stassun, Price-Whelan, Holtzman, Chojnowski, Longa-Peña, Román-Zúñiga, Hernandez, Serna, Badenes, Lee, Majewski, Stringfellow, Kratter, Moe, Frinchaboy, Beaton, Fernández-Trincado, Mahadevan, Minniti, Beers, Schneider, Barba, Brownstein, García-Hernández, Pan, \& Bizyaev]{kounkel_double-lined_2021}
Kounkel, M., Covey, K.~R., Stassun, K.~G., Price-Whelan, A.~M., Holtzman, J., Chojnowski, D., Longa-Peña, P., Román-Zúñiga, C.~G., Hernandez, J., Serna, J., Badenes, C., Lee, N.~D., Majewski, S., Stringfellow, G.~S., Kratter, K.~M., Moe, M., Frinchaboy, P.~M., Beaton, R.~L., Fernández-Trincado, J.~G., Mahadevan, S., Minniti, D., Beers, T.~C., Schneider, D.~P., Barba, R., Brownstein, J.~R., García-Hernández, D.~A., Pan, K., \& Bizyaev, D., 2021.
\newblock Double-lined {Spectroscopic} {Binaries} in the {APOGEE} {DR16} and {DR17} {Data}, {\it The Astronomical Journal\/}, {\bf 162}(5), 184, Publisher: The American Astronomical Society.

\bibitem[Kraus et~al.(2014)Kraus, Shkolnik, Allers, \& Liu]{kraus_slar_2014}
Kraus, A.~L., Shkolnik, E.~L., Allers, K.~N., \& Liu, M.~C., 2014.
\newblock A {STELLAR} {CENSUS} {OF} {THE} {TUCANA}–{HOROLOGIUM} {MOVING} {GROUP}, {\it The Astronomical Journal\/}, {\bf 147}(6), 146, Publisher: The American Astronomical Society.

\bibitem[Kraus et~al.(2015)Kraus, Cody, Covey, Rizzuto, Mann, \& Ireland]{kraus_massradius_2015}
Kraus, A.~L., Cody, A.~M., Covey, K.~R., Rizzuto, A.~C., Mann, A.~W., \& Ireland, M.~J., 2015.
\newblock {THE} {MASS}–{RADIUS} {RELATION} {OF} {YOUNG} {STARS}. {I}. {USCO} 5, {AN} {M4}.5 {ECLIPSING} {BINARY} {IN} {UPPER} {SCORPIUS} {OBSERVED} {BY} {K2}, {\it The Astrophysical Journal\/}, {\bf 807}(1), 3, Publisher: American Astronomical Society.

\bibitem[Kraus(2020)]{kraus_gw_2020}
Kraus, S., 2020.
\newblock {GW} {Orionis}: {A} pre-main-sequence triple with a warped disk and a torn-apart ring as benchmark for disk hydrodynamics, {\it arXiv:2012.06578 [astro-ph]\/}, arXiv: 2012.06578.

\bibitem[Krolikowski et~al.(2021)Krolikowski, Kraus, \& Rizzuto]{krolikowski_gaia_2021}
Krolikowski, D.~M., Kraus, A.~L., \& Rizzuto, A.~C., 2021.
\newblock Gaia {EDR3} {Reveals} the {Substructure} and {Complicated} {Star} {Formation} {History} of the {Greater} {Taurus}-{Auriga} {Star}-forming {Complex}, {\it The Astronomical Journal\/}, {\bf 162}(3), 110, Publisher: The American Astronomical Society.

\bibitem[Kuffmeier et~al.(2021)Kuffmeier, Dullemond, Reissl, \& Goicovic]{kuffmeier_misaligned_2021}
Kuffmeier, M., Dullemond, C.~P., Reissl, S., \& Goicovic, F.~G., 2021.
\newblock Misaligned disks induced by infall, {\it Astronomy \& Astrophysics\/}, {\bf 656}, A161, Publisher: EDP Sciences.

\bibitem[Lam et~al.(2015)Lam, Pitrou, \& Seibert]{lam_numba_2015}
Lam, S.~K., Pitrou, A., \& Seibert, S., 2015.
\newblock Numba: a {LLVM}-based {Python} {JIT} compiler, in {\em Proceedings of the {Second} {Workshop} on the {LLVM} {Compiler} {Infrastructure} in {HPC}\/}, {LLVM} '15, pp. 1--6, Association for Computing Machinery, New York, NY, USA.

\bibitem[Lavail et~al.(2019)Lavail, Kochukhov, \& Hussain]{lavail_characterising_2019}
Lavail, A., Kochukhov, O., \& Hussain, G. A.~J., 2019.
\newblock Characterising the surface magnetic fields of {T} {Tauri} stars with high-resolution near-infrared spectroscopy, {\it Astronomy and Astrophysics\/}, {\bf 630}, A99, ADS Bibcode: 2019A\&A...630A..99L.

\bibitem[Loaiza-Tacuri et~al.(2023)Loaiza-Tacuri, Cunha, Smith, Martinez, Ghezzi, Schuler, Teske, \& Howell]{loaiza-tacuri_spectroscopic_2023}
Loaiza-Tacuri, V., Cunha, K., Smith, V.~V., Martinez, C.~F., Ghezzi, L., Schuler, S.~C., Teske, J., \& Howell, S.~B., 2023.
\newblock A {Spectroscopic} {Analysis} of a {Sample} of {K2} {Planet}-host {Stars}: {Stellar} {Parameters}, {Metallicities} and {Planetary} {Radii}, {\it The Astrophysical Journal\/}, {\bf 946}(2), 61, Publisher: The American Astronomical Society.

\bibitem[Long et~al.(2019)Long, Herczeg, Harsono, Pinilla, Tazzari, Manara, Pascucci, Cabrit, Nisini, Johnstone, Edwards, Salyk, Menard, Lodato, Boehler, Mace, Liu, Mulders, Hendler, Ragusa, Fischer, Banzatti, Rigliaco, Plas, Dipierro, Gully-Santiago, \& Lopez-Valdivia]{long_compact_2019}
Long, F., Herczeg, G.~J., Harsono, D., Pinilla, P., Tazzari, M., Manara, C.~F., Pascucci, I., Cabrit, S., Nisini, B., Johnstone, D., Edwards, S., Salyk, C., Menard, F., Lodato, G., Boehler, Y., Mace, G.~N., Liu, Y., Mulders, G.~D., Hendler, N., Ragusa, E., Fischer, W.~J., Banzatti, A., Rigliaco, E., Plas, G. v.~d., Dipierro, G., Gully-Santiago, M., \& Lopez-Valdivia, R., 2019.
\newblock Compact {Disks} in a {High}-resolution {ALMA} {Survey} of {Dust} {Structures} in the {Taurus} {Molecular} {Cloud}, {\it The Astrophysical Journal\/}, {\bf 882}(1), 49, Publisher: American Astronomical Society.

\bibitem[Luhman(2024)]{luhman_census_2024}
Luhman, K.~L., 2024.
\newblock A {Census} of the $\beta$ {Pic} {Moving} {Group} and {Other} {Nearby} {Associations} with {Gaia}, {\it The Astronomical Journal\/}, {\bf 168}, 159, Publisher: IOP ADS Bibcode: 2024AJ....168..159L.

\bibitem[Lépine \& Simon(2009)]{lepine_nearby_2009}
Lépine, S. \& Simon, M., 2009.
\newblock Nearby {Young} {Stars} {Selected} by {Proper} {Motion}. {I}. {Four} {New} {Members} of the $\beta$ {Pictoris} {Moving} {Group} {From} {The} {Tycho}-2 {Catalog}, {\it The Astronomical Journal\/}, {\bf 137}, 3632--3645, Publisher: IOP ADS Bibcode: 2009AJ....137.3632L.

\bibitem[López-Valdivia et~al.(2021)López-Valdivia, Sokal, Mace, Kidder, Hussaini, Nofi, Prato, Johns-Krull, Oh, Lee, Park, Oh, Kraus, Kaplan, Llama, Mann, Kim, Gully-Santiago, Lee, Pak, Hwang, \& Jaffe]{lopez-valdivia_igrins_2021}
López-Valdivia, R., Sokal, K.~R., Mace, G.~N., Kidder, B.~T., Hussaini, M., Nofi, L., Prato, L., Johns-Krull, C.~M., Oh, H., Lee, J.-J., Park, C., Oh, J.~S., Kraus, A., Kaplan, K.~F., Llama, J., Mann, A.~W., Kim, H., Gully-Santiago, M.~A., Lee, H.-I., Pak, S., Hwang, N., \& Jaffe, D.~T., 2021.
\newblock The {IGRINS} {YSO} {Survey}. {I}. {Stellar} {Parameters} of {Pre}-main-sequence {Stars} in {Taurus}-{Auriga}, {\it The Astrophysical Journal\/}, {\bf 921}(1), 53, Publisher: The American Astronomical Society.

\bibitem[Malo et~al.(2014)Malo, Artigau, Doyon, Lafrenière, Albert, \& Gagné]{malo_banyan_2014}
Malo, L., Artigau, E., Doyon, R., Lafrenière, D., Albert, L., \& Gagné, J., 2014.
\newblock {BANYAN}. {III}. {RADIAL} {VELOCITY}, {ROTATION}, {AND} {X}-{RAY} {EMISSION} {OF} {LOW}-{MASS} {STAR} {CANDIDATES} {IN} {NEARBY} {YOUNG} {KINEMATIC} {GROUPS}, {\it The Astrophysical Journal\/}, {\bf 788}(1), 81, Publisher: The American Astronomical Society.

\bibitem[Mamajek(2009)]{mamajek_initial_2009}
Mamajek, E.~E., 2009.
\newblock Initial {Conditions} of {Planet} {Formation}: {Lifetimes} of {Primordial} {Disks}, {\bf 1158}, 3--10, Conference Name: Exoplanets and Disks: Their Formation and Diversity Place: eprint: arXiv:0906.5011 Publisher: AIP ADS Bibcode: 2009AIPC.1158....3M.

\bibitem[Mann et~al.(2013)Mann, Gaidos, \& Ansdell]{mann_spectro-thermometry_2013}
Mann, A.~W., Gaidos, E., \& Ansdell, M., 2013.
\newblock Spectro-thermometry of {M} {Dwarfs} and {Their} {Candidate} {Planets}: {Too} {Hot}, {Too} {Cool}, or {Just} {Right}?, {\it The Astrophysical Journal\/}, {\bf 779}, 188, Publisher: IOP ADS Bibcode: 2013ApJ...779..188M.

\bibitem[Mann et~al.(2015)Mann, Feiden, Gaidos, Boyajian, \& Braun]{mann_how_2015}
Mann, A.~W., Feiden, G.~A., Gaidos, E., Boyajian, T., \& Braun, K.~v., 2015.
\newblock {HOW} {TO} {CONSTRAIN} {YOUR} {M} {DWARF}: {MEASURING} {EFFECTIVE} {TEMPERATURE}, {BOLOMETRIC} {LUMINOSITY}, {MASS}, {AND} {RADIUS}, {\it The Astrophysical Journal\/}, {\bf 804}(1), 64, Publisher: The American Astronomical Society.

\bibitem[Mann et~al.(2016)Mann, Newton, Rizzuto, Irwin, Feiden, Gaidos, Mace, Kraus, James, Ansdell, Charbonneau, Covey, Ireland, Jaffe, Johnson, Kidder, \& Vanderburg]{mann_zodiacal_2016}
Mann, A.~W., Newton, E.~R., Rizzuto, A.~C., Irwin, J., Feiden, G.~A., Gaidos, E., Mace, G.~N., Kraus, A.~L., James, D.~J., Ansdell, M., Charbonneau, D., Covey, K.~R., Ireland, M.~J., Jaffe, D.~T., Johnson, M.~C., Kidder, B., \& Vanderburg, A., 2016.
\newblock {ZODIACAL} {EXOPLANETS} {IN} {TIME} ({ZEIT}). {III}. {A} {SHORT}-{PERIOD} {PLANET} {ORBITING} {A} {PRE}-{MAIN}-{SEQUENCE} {STAR} {IN} {THE} {UPPER} {SCORPIUS} {OB} {ASSOCIATION}, {\it The Astronomical Journal\/}, {\bf 152}(3), 61, Publisher: The American Astronomical Society.

\bibitem[Mann et~al.(2018)Mann, Vanderburg, Rizzuto, Kraus, Berlind, Bieryla, Calkins, Esquerdo, Latham, Mace, Morris, Quinn, Sokal, \& Stefanik]{mann_zodiacal_2018}
Mann, A.~W., Vanderburg, A., Rizzuto, A.~C., Kraus, A.~L., Berlind, P., Bieryla, A., Calkins, M.~L., Esquerdo, G.~A., Latham, D.~W., Mace, G.~N., Morris, N.~R., Quinn, S.~N., Sokal, K.~R., \& Stefanik, R.~P., 2018.
\newblock Zodiacal {Exoplanets} in {Time} ({ZEIT}). {VI}. {A} {Three}-planet {System} in the {Hyades} {Cluster} {Including} an {Earth}-sized {Planet}, {\it The Astronomical Journal\/}, {\bf 155}, 4, ADS Bibcode: 2018AJ....155....4M.

\bibitem[Mann et~al.(2022)Mann, Wood, Schmidt, Barber, Owen, Tofflemire, Newton, Mamajek, Bush, Mace, Kraus, Thao, Vanderburg, Llama, Johns-Krull, Prato, Stahl, Tang, Fields, Collins, Collins, Gan, Jensen, Kamler, Schwarz, Furlan, Gnilka, Howell, Lester, Owens, Suarez, Mekarnia, Guillot, Abe, Triaud, Johnson, Milburn, Rizzuto, Quinn, Kerr, Ricker, Vanderspek, Latham, Seager, Winn, Jenkins, Guerrero, Shporer, Schlieder, McLean, \& Wohler]{mann_tess_2022}
Mann, A.~W., Wood, M.~L., Schmidt, S.~P., Barber, M.~G., Owen, J.~E., Tofflemire, B.~M., Newton, E.~R., Mamajek, E.~E., Bush, J.~L., Mace, G.~N., Kraus, A.~L., Thao, P.~C., Vanderburg, A., Llama, J., Johns-Krull, C.~M., Prato, L., Stahl, A.~G., Tang, S.-Y., Fields, M.~J., Collins, K.~A., Collins, K.~I., Gan, T., Jensen, E. L.~N., Kamler, J., Schwarz, R.~P., Furlan, E., Gnilka, C.~L., Howell, S.~B., Lester, K.~V., Owens, D.~A., Suarez, O., Mekarnia, D., Guillot, T., Abe, L., Triaud, A. H. M.~J., Johnson, M.~C., Milburn, R.~P., Rizzuto, A.~C., Quinn, S.~N., Kerr, R., Ricker, G.~R., Vanderspek, R., Latham, D.~W., Seager, S., Winn, J.~N., Jenkins, J.~M., Guerrero, N.~M., Shporer, A., Schlieder, J.~E., McLean, B., \& Wohler, B., 2022.
\newblock {TESS} {Hunt} for {Young} and {Maturing} {Exoplanets} ({THYME}). {VI}. {An} 11 {Myr} {Giant} {Planet} {Transiting} a {Very}-low-mass {Star} in {Lower} {Centaurus} {Crux}, {\it The Astronomical Journal\/}, {\bf 163}(4), 156, Publisher: American Astronomical Society.

\bibitem[Masuda \& Winn(2020)]{masuda_inference_2020}
Masuda, K. \& Winn, J.~N., 2020.
\newblock On the {Inference} of a {Star}’s {Inclination} {Angle} from its {Rotation} {Velocity} and {Projected} {Rotation} {Velocity}, {\it The Astronomical Journal\/}, {\bf 159}(3), 81, Publisher: The American Astronomical Society.

\bibitem[Mayo et~al.(2018)Mayo, Vanderburg, Latham, Bieryla, Morton, Buchhave, Dressing, Beichman, Berlind, Calkins, Ciardi, Crossfield, Esquerdo, Everett, Gonzales, Hirsch, Horch, Howard, Howell, Livingston, Patel, Petigura, Schlieder, Scott, Schumer, Sinukoff, Teske, \& Winters]{mayo_275_2018}
Mayo, A.~W., Vanderburg, A., Latham, D.~W., Bieryla, A., Morton, T.~D., Buchhave, L.~A., Dressing, C.~D., Beichman, C., Berlind, P., Calkins, M.~L., Ciardi, D.~R., Crossfield, I. J.~M., Esquerdo, G.~A., Everett, M.~E., Gonzales, E.~J., Hirsch, L.~A., Horch, E.~P., Howard, A.~W., Howell, S.~B., Livingston, J., Patel, R., Petigura, E.~A., Schlieder, J.~E., Scott, N.~J., Schumer, C.~F., Sinukoff, E., Teske, J., \& Winters, J.~G., 2018.
\newblock 275 {Candidates} and 149 {Validated} {Planets} {Orbiting} {Bright} {Stars} in {K2} {Campaigns} 0–10, {\it The Astronomical Journal\/}, {\bf 155}(3), 136, Publisher: The American Astronomical Society.

\bibitem[Messina et~al.(2010)Messina, Desidera, Turatto, Lanzafame, \& Guinan]{messina_race-oc_2010}
Messina, S., Desidera, S., Turatto, M., Lanzafame, A.~C., \& Guinan, E.~F., 2010.
\newblock {RACE}-{OC} project: {Rotation} and variability of young stellar associations within 100 pc, {\it Astronomy and Astrophysics\/}, {\bf 520}, A15, ADS Bibcode: 2010A\&A...520A..15M.

\bibitem[Messina et~al.(2011)Messina, Desidera, Lanzafame, Turatto, \& Guinan]{messina_race-oc_2011}
Messina, S., Desidera, S., Lanzafame, A.~C., Turatto, M., \& Guinan, E.~F., 2011.
\newblock {RACE}-{OC} project: rotation and variability in the $\epsilon$ {Chamaeleontis}, {Octans}, and {Argus} stellar associations, {\it Astronomy and Astrophysics\/}, {\bf 532}, A10, ADS Bibcode: 2011A\&A...532A..10M.

\bibitem[Messina et~al.(2017)Messina, Millward, Buccino, Zhang, Medhi, Jofré, Petrucci, Pi, Hambsch, Kehusmaa, Harlingten, Artemenko, Curtis, Hentunen, Malo, Mauas, Monard, Serrano, Naves, Santallo, Savuskin, \& Tan]{messina__2017}
Messina, S., Millward, M., Buccino, A., Zhang, L., Medhi, B.~J., Jofré, E., Petrucci, R., Pi, Q., Hambsch, F.-J., Kehusmaa, P., Harlingten, C., Artemenko, S., Curtis, I., Hentunen, V.-P., Malo, L., Mauas, P., Monard, B., Serrano, M.~M., Naves, R., Santallo, R., Savuskin, A., \& Tan, T.~G., 2017.
\newblock The $\beta$ {Pictoris} association: {Catalog} of photometric rotational periods of low-mass members and candidate members, {\it Astronomy \& Astrophysics\/}, {\bf 600}, A83, Publisher: EDP Sciences.

\bibitem[Mochnacki et~al.(2002)Mochnacki, Gladders, Thomson, Lu, Ehlers, Guler, Hussain, Kameda, King, Mitchell, Rowe, Schindler, \& Scott]{mochnacki_spectroscopic_2002}
Mochnacki, S.~W., Gladders, M.~D., Thomson, J.~R., Lu, W., Ehlers, P., Guler, M., Hussain, A., Kameda, Q., King, K., Mitchell, P., Rowe, J., Schindler, P., \& Scott, H., 2002.
\newblock A {Spectroscopic} {Survey} of a {Sample} of {Active} {M} {Dwarfs}, {\it The Astronomical Journal\/}, {\bf 124}, 2868--2882, Publisher: IOP ADS Bibcode: 2002AJ....124.2868M.

\bibitem[Morbidelli et~al.(2012)Morbidelli, Lunine, O'Brien, Raymond, \& Walsh]{morbidelli_building_2012}
Morbidelli, A., Lunine, J., O'Brien, D., Raymond, S., \& Walsh, K., 2012.
\newblock Building {Terrestrial} {Planets}, {\it Annual Review of Earth and Planetary Sciences\/}, {\bf 40}(1), 251--275, \_eprint: https://doi.org/10.1146/annurev-earth-042711-105319.

\bibitem[Mori et~al.(2024)Mori, Ikuta, Fukui, Narita, de~Leon, Livingston, Ikoma, Kawai, Kawauchi, Murgas, Palle, Parviainen, Fernández~Rodríguez, Terada, Watanabe, \& Tamura]{mori_characterization_2024}
Mori, M., Ikuta, K., Fukui, A., Narita, N., de~Leon, J.~P., Livingston, J.~H., Ikoma, M., Kawai, Y., Kawauchi, K., Murgas, F., Palle, E., Parviainen, H., Fernández~Rodríguez, G., Terada, Y., Watanabe, N., \& Tamura, M., 2024.
\newblock Characterization of starspots on a young {M}-dwarf {K2}-25: multiband observations of stellar photometric variability and planetary transits, {\it Monthly Notices of the Royal Astronomical Society\/}, {\bf 530}, 167--189, Publisher: OUP ADS Bibcode: 2024MNRAS.530..167M.

\bibitem[Morton(2015)]{morton_isochrones_2015}
Morton, T.~D., 2015.
\newblock isochrones: {Stellar} model grid package, {\it Astrophysics Source Code Library\/}, p. ascl:1503.010.

\bibitem[Morton \& Winn(2014)]{morton_obliquities_2014}
Morton, T.~D. \& Winn, J.~N., 2014.
\newblock {OBLIQUITIES} {OF} {KEPLER} {STARS}: {COMPARISON} {OF} {SINGLE}- {AND} {MULTIPLE}-{TRANSIT} {SYSTEMS}, {\it The Astrophysical Journal\/}, {\bf 796}(1), 47, Publisher: The American Astronomical Society.

\bibitem[Muirhead et~al.(2012)Muirhead, Hamren, Schlawin, Rojas-Ayala, Covey, \& Lloyd]{muirhead_characterizing_2012}
Muirhead, P.~S., Hamren, K., Schlawin, E., Rojas-Ayala, B., Covey, K.~R., \& Lloyd, J.~P., 2012.
\newblock {CHARACTERIZING} {THE} {COOL} {KEPLER} {OBJECTS} {OF} {INTERESTS}. {NEW} {EFFECTIVE} {TEMPERATURES}, {METALLICITIES}, {MASSES}, {AND} {RADII} {OF} {LOW}-{MASS} {KEPLER} {PLANET}-{CANDIDATE} {HOST} {STARS}, {\it The Astrophysical Journal Letters\/}, {\bf 750}(2), L37, Publisher: The American Astronomical Society.

\bibitem[Muirhead et~al.(2013)Muirhead, Vanderburg, Shporer, Becker, Swift, Lloyd, Fuller, Zhao, Hinkley, Pineda, Bottom, Howard, Braun, Boyajian, Law, Baranec, Riddle, Ramaprakash, Tendulkar, Bui, Burse, Chordia, Das, Dekany, Punnadi, \& Johnson]{muirhead_characterizing_2013}
Muirhead, P.~S., Vanderburg, A., Shporer, A., Becker, J., Swift, J.~J., Lloyd, J.~P., Fuller, J., Zhao, M., Hinkley, S., Pineda, J.~S., Bottom, M., Howard, A.~W., Braun, K.~v., Boyajian, T.~S., Law, N., Baranec, C., Riddle, R., Ramaprakash, A.~N., Tendulkar, S.~P., Bui, K., Burse, M., Chordia, P., Das, H., Dekany, R., Punnadi, S., \& Johnson, J.~A., 2013.
\newblock {CHARACTERIZING} {THE} {COOL} {KOIs}. {V}. {KOI}-256: {A} {MUTUALLY} {ECLIPSING} {POST}-{COMMON} {ENVELOPE} {BINARY}, {\it The Astrophysical Journal\/}, {\bf 767}(2), 111, Publisher: American Astronomical Society.

\bibitem[Naoz et~al.(2011)Naoz, Farr, Lithwick, Rasio, \& Teyssandier]{naoz_hot_2011}
Naoz, S., Farr, W.~M., Lithwick, Y., Rasio, F.~A., \& Teyssandier, J., 2011.
\newblock Hot {Jupiters} from secular planet-planet interactions, {\it Nature\/}, {\bf 473}, 187--189, ADS Bibcode: 2011Natur.473..187N.

\bibitem[Newton et~al.(2016)Newton, Irwin, Charbonneau, Berta-Thompson, Dittmann, \& West]{newton_rotation_2016}
Newton, E.~R., Irwin, J., Charbonneau, D., Berta-Thompson, Z.~K., Dittmann, J.~A., \& West, A.~A., 2016.
\newblock {THE} {ROTATION} {AND} {GALACTIC} {KINEMATICS} {OF} {MID} {M} {DWARFS} {IN} {THE} {SOLAR} {NEIGHBORHOOD}, {\it The Astrophysical Journal\/}, {\bf 821}(2), 93, Publisher: American Astronomical Society.

\bibitem[Newton et~al.(2019)Newton, Mann, Tofflemire, Pearce, Rizzuto, Vanderburg, Martinez, Wang, Ruffio, Kraus, Johnson, Thao, Wood, Rampalli, Nielsen, Collins, Dragomir, Hellier, Anderson, Barclay, Brown, Feiden, Hart, Isopi, Kielkopf, Mallia, Nelson, Rodriguez, Stockdale, Waite, Wright, Lissauer, Ricker, Vanderspek, Latham, Seager, Winn, Jenkins, Bouma, Burke, Davies, Fausnaugh, Li, Morris, Mukai, Villaseñor, Villeneuva, De~Rosa, Macintosh, Mengel, Okumura, \& Wittenmyer]{newton_tess_2019}
Newton, E.~R., Mann, A.~W., Tofflemire, B.~M., Pearce, L., Rizzuto, A.~C., Vanderburg, A., Martinez, R.~A., Wang, J.~J., Ruffio, J.-B., Kraus, A.~L., Johnson, M.~C., Thao, P.~C., Wood, M.~L., Rampalli, R., Nielsen, E.~L., Collins, K.~A., Dragomir, D., Hellier, C., Anderson, D.~R., Barclay, T., Brown, C., Feiden, G., Hart, R., Isopi, G., Kielkopf, J.~F., Mallia, F., Nelson, P., Rodriguez, J.~E., Stockdale, C., Waite, I.~A., Wright, D.~J., Lissauer, J.~J., Ricker, G.~R., Vanderspek, R., Latham, D.~W., Seager, S., Winn, J.~N., Jenkins, J.~M., Bouma, L.~G., Burke, C.~J., Davies, M., Fausnaugh, M., Li, J., Morris, R.~L., Mukai, K., Villaseñor, J., Villeneuva, S., De~Rosa, R.~J., Macintosh, B., Mengel, M.~W., Okumura, J., \& Wittenmyer, R.~A., 2019.
\newblock {TESS} {Hunt} for {Young} and {Maturing} {Exoplanets} ({THYME}): {A} {Planet} in the 45 {Myr} {Tucana}-{Horologium} {Association}, {\it The Astrophysical Journal\/}, {\bf 880}, L17, Publisher: IOP ADS Bibcode: 2019ApJ...880L..17N.

\bibitem[Offner et~al.(2023)Offner, Moe, Kratter, Sadavoy, Jensen, \& Tobin]{offner_origin_2023}
Offner, S. S.~R., Moe, M., Kratter, K.~M., Sadavoy, S.~I., Jensen, E. L.~N., \& Tobin, J.~J., 2023.
\newblock The {Origin} and {Evolution} of {Multiple} {Star} {Systems}, {\bf 534}, 275, Conference Name: Protostars and Planets VII Place: eprint: arXiv:2203.10066 ADS Bibcode: 2023ASPC..534..275O.

\bibitem[Papaloizou \& Larwood(2000)]{papaloizou_orbital_2000}
Papaloizou, J. C.~B. \& Larwood, J.~D., 2000.
\newblock On the orbital evolution and growth of protoplanets embedded in a gaseous disc, {\it Monthly Notices of the Royal Astronomical Society\/}, {\bf 315}, 823--833, Publisher: OUP ADS Bibcode: 2000MNRAS.315..823P.

\bibitem[Park et~al.(2014)Park, Jaffe, Yuk, Chun, Pak, Kim, Pavel, Lee, Oh, Jeong, Sim, Lee, Nguyen~Le, Strubhar, Gully-Santiago, Oh, Cha, Moon, Park, Brooks, Ko, Han, Nah, Hill, Lee, Barnes, Yu, Kaplan, Mace, Kim, Lee, Hwang, \& Park]{park_design_2014}
Park, C., Jaffe, D.~T., Yuk, I.-S., Chun, M.-Y., Pak, S., Kim, K.-M., Pavel, M., Lee, H., Oh, H., Jeong, U., Sim, C.~K., Lee, H.-I., Nguyen~Le, H.~A., Strubhar, J., Gully-Santiago, M., Oh, J.~S., Cha, S.-M., Moon, B., Park, K., Brooks, C., Ko, K., Han, J.-Y., Nah, J., Hill, P.~C., Lee, S., Barnes, S., Yu, Y.~S., Kaplan, K., Mace, G., Kim, H., Lee, J.-J., Hwang, N., \& Park, B.-G., 2014.
\newblock Design and early performance of {IGRINS} ({Immersion} {Grating} {Infrared} {Spectrometer}), {\bf 9147}, 91471D, Conference Name: Ground-based and Airborne Instrumentation for Astronomy V ADS Bibcode: 2014SPIE.9147E..1DP.

\bibitem[Parviainen \& Aigrain(2015)]{parviainen_ldtk_2015}
Parviainen, H. \& Aigrain, S., 2015.
\newblock ldtk: {Limb} {Darkening} {Toolkit}, {\it Monthly Notices of the Royal Astronomical Society\/}, {\bf 453}(4), 3821--3826.

\bibitem[Petrovich et~al.(2020)Petrovich, Muñoz, Kratter, \& Malhotra]{petrovich_disk-driven_2020}
Petrovich, C., Muñoz, D.~J., Kratter, K.~M., \& Malhotra, R., 2020.
\newblock A {Disk}-driven {Resonance} as the {Origin} of {High} {Inclinations} of {Close}-in {Planets}, {\it The Astrophysical Journal\/}, {\bf 902}, L5, Publisher: IOP ADS Bibcode: 2020ApJ...902L...5P.

\bibitem[Plavchan et~al.(2020)Plavchan, Barclay, Gagné, Gao, Cale, Matzko, Dragomir, Quinn, Feliz, Stassun, Crossfield, Berardo, Latham, Tieu, Anglada-Escudé, Ricker, Vanderspek, Seager, Winn, Jenkins, Rinehart, Krishnamurthy, Dynes, Doty, Adams, Afanasev, Beichman, Bottom, Bowler, Brinkworth, Brown, Cancino, Ciardi, Clampin, Clark, Collins, Davison, Foreman-Mackey, Furlan, Gaidos, Geneser, Giddens, Gilbert, Hall, Hellier, Henry, Horner, Howard, Huang, Huber, Kane, Kenworthy, Kielkopf, Kipping, Klenke, Kruse, Latouf, Lowrance, Mennesson, Mengel, Mills, Morton, Narita, Newton, Nishimoto, Okumura, Palle, Pepper, Quintana, Roberge, Roccatagliata, Schlieder, Tanner, Teske, Tinney, Vanderburg, von Braun, Walp, Wang, Xuesong~Wang, Weigand, White, Wittenmyer, Wright, Youngblood, Zhang, \& Zilberman]{plavchan_planet_2020}
Plavchan, P., Barclay, T., Gagné, J., Gao, P., Cale, B., Matzko, W., Dragomir, D., Quinn, S., Feliz, D., Stassun, K., Crossfield, I.~J., Berardo, D.~A., Latham, D.~W., Tieu, B., Anglada-Escudé, G., Ricker, G., Vanderspek, R., Seager, S., Winn, J.~N., Jenkins, J.~M., Rinehart, S., Krishnamurthy, A., Dynes, S., Doty, J., Adams, F., Afanasev, D.~A., Beichman, C., Bottom, M., Bowler, B.~P., Brinkworth, C., Brown, C.~J., Cancino, A., Ciardi, D.~R., Clampin, M., Clark, J.~T., Collins, K., Davison, C., Foreman-Mackey, D., Furlan, E., Gaidos, E., Geneser, C., Giddens, F., Gilbert, E., Hall, R., Hellier, C., Henry, T., Horner, J., Howard, A.~W., Huang, C., Huber, J., Kane, S.~R., Kenworthy, M., Kielkopf, J., Kipping, D., Klenke, C., Kruse, E., Latouf, N., Lowrance, P., Mennesson, B., Mengel, M., Mills, S.~M., Morton, T., Narita, N., Newton, E., Nishimoto, A., Okumura, J., Palle, E., Pepper, J., Quintana, E.~V., Roberge, A., Roccatagliata, V., Schlieder, J.~E., Tanner, A., Teske, J., Tinney, C.~G., Vanderburg, A.,
  von Braun, K., Walp, B., Wang, J., Xuesong~Wang, S., Weigand, D., White, R., Wittenmyer, R.~A., Wright, D.~J., Youngblood, A., Zhang, H., \& Zilberman, P., 2020.
\newblock A planet within the debris disk around the pre-main sequence star {AU} {Mic}, {\it Nature\/}, {\bf 582}(7813), 497--500.

\bibitem[Prusti et~al.(2016)Prusti, Bruijne, Brown, Vallenari, Babusiaux, Bailer-Jones, Bastian, Biermann, Evans, Eyer, Jansen, Jordi, Klioner, Lammers, Lindegren, Luri, Mignard, Milligan, Panem, Poinsignon, Pourbaix, Randich, Sarri, Sartoretti, Siddiqui, Soubiran, Valette, Leeuwen, Walton, Aerts, Arenou, Cropper, Drimmel, Høg, Katz, Lattanzi, O’Mullane, Grebel, Holland, Huc, Passot, Bramante, Cacciari, Castañeda, Chaoul, Cheek, Angeli, Fabricius, Guerra, Hernández, Jean-Antoine-Piccolo, Masana, Messineo, Mowlavi, Nienartowicz, Ordóñez-Blanco, Panuzzo, Portell, Richards, Riello, Seabroke, Tanga, Thévenin, Torra, Els, Gracia-Abril, Comoretto, Garcia-Reinaldos, Lock, Mercier, Altmann, Andrae, Astraatmadja, Bellas-Velidis, Benson, Berthier, Blomme, Busso, Carry, Cellino, Clementini, Cowell, Creevey, Cuypers, Davidson, Ridder, Torres, Delchambre, Dell’Oro, Ducourant, Frémat, García-Torres, Gosset, Halbwachs, Hambly, Harrison, Hauser, Hestroffer, Hodgkin, Huckle, Hutton, Jasniewicz, Jordan, Kontizas,
  Korn, Lanzafame, Manteiga, Moitinho, Muinonen, Osinde, Pancino, Pauwels, Petit, Recio-Blanco, Robin, Sarro, Siopis, Smith, Smith, Sozzetti, Thuillot, Reeven, Viala, Abbas, Aramburu, Accart, Aguado, Allan, Allasia, Altavilla, Álvarez, Alves, Anderson, Andrei, Varela, Antiche, Antoja, Antón, Arcay, Atzei, Ayache, Bach, Baker, Balaguer-Núñez, Barache, Barata, Barbier, Barblan, Baroni, Navascués, Barros, Barstow, Becciani, Bellazzini, Bellei, García, Belokurov, Bendjoya, Berihuete, Bianchi, Bienaymé, Billebaud, Blagorodnova, Blanco-Cuaresma, Boch, Bombrun, Borrachero, Bouquillon, Bourda, Bouy, Bragaglia, Breddels, Brouillet, Brüsemeister, Bucciarelli, Budnik, Burgess, Burgon, Burlacu, Busonero, Buzzi, Caffau, Cambras, Campbell, Cancelliere, Cantat-Gaudin, Carlucci, Carrasco, Castellani, Charlot, Charnas, Charvet, Chassat, Chiavassa, Clotet, Cocozza, Collins, Collins, Costigan, Crifo, Cross, Crosta, Crowley, Dafonte, Damerdji, Dapergolas, David, David, Cat, Felice, Laverny, Luise, March, Martino, Souza,
  Debosscher, Pozo, Delbo, Delgado, Delgado, Marco, Matteo, Diakite, Distefano, Dolding, Anjos, Drazinos, Durán, Dzigan, Ecale, Edvardsson, Enke, Erdmann, Escolar, Espina, Evans, Bontemps, Fabre, Fabrizio, Faigler, Falcão, Casas, Faye, Federici, Fedorets, Fernández-Hernández, Fernique, Fienga, Figueras, Filippi, Findeisen, Fonti, Fouesneau, Fraile, Fraser, Fuchs, Furnell, Gai, Galleti, Galluccio, Garabato, García-Sedano, Garé, Garofalo, Garralda, Gavras, Gerssen, Geyer, Gilmore, Girona, Giuffrida, Gomes, González-Marcos, González-Núñez, González-Vidal, Granvik, Guerrier, Guillout, Guiraud, Gúrpide, Gutiérrez-Sánchez, Guy, Haigron, Hatzidimitriou, Haywood, Heiter, Helmi, Hobbs, Hofmann, Holl, Holland, Hunt, Hypki, Icardi, Irwin, Fombelle, Jofré, Jonker, Jorissen, Julbe, Karampelas, Kochoska, Kohley, Kolenberg, Kontizas, Koposov, Kordopatis, Koubsky, Kowalczyk, Krone-Martins, Kudryashova, Kull, Bachchan, Lacoste-Seris, Lanza, Lavigne, Poncin-Lafitte, Lebreton, Lebzelter, Leccia, Leclerc,
  Lecoeur-Taibi, Lemaitre, Lenhardt, Leroux, Liao, Licata, Lindstrøm, Lister, Livanou, Lobel, Löffler, López, Lopez-Lozano, Lorenz, Loureiro, MacDonald, Fernandes, Managau, Mann, Mantelet, Marchal, Marchant, Marconi, Marie, Marinoni, Marrese, Marschalkó, Marshall, Martín-Fleitas, Martino, Mary, Matijevič, Mazeh, McMillan, Messina, Mestre, Michalik, Millar, Miranda, Molina, Molinaro, Molinaro, Molnár, Moniez, Montegriffo, Monteiro, Mor, Mora, Morbidelli, Morel, Morgenthaler, Morley, Morris, Mulone, Muraveva, Musella, Narbonne, Nelemans, Nicastro, Noval, Ordénovic, Ordieres-Meré, Osborne, Pagani, Pagano, Pailler, Palacin, Palaversa, Parsons, Paulsen, Pecoraro, Pedrosa, Pentikäinen, Pereira, Pichon, Piersimoni, Pineau, Plachy, Plum, Poujoulet, Prša, Pulone, Ragaini, Rago, Rambaux, Ramos-Lerate, Ranalli, Rauw, Read, Regibo, Renk, Reylé, Ribeiro, Rimoldini, Ripepi, Riva, Rixon, Roelens, Romero-Gómez, Rowell, Royer, Rudolph, Ruiz-Dern, Sadowski, Sellés, Sahlmann, Salgado, Salguero, Sarasso, Savietto,
  Schnorhk, Schultheis, Sciacca, Segol, Segovia, Segransan, Serpell, Shih, Smareglia, Smart, Smith, Solano, Solitro, Sordo, Nieto, Souchay, Spagna, Spoto, Stampa, Steele, Steidelmüller, Stephenson, Stoev, Suess, Süveges, Surdej, Szabados, Szegedi-Elek, Tapiador, Taris, Tauran, Taylor, Teixeira, Terrett, Tingley, Trager, Turon, Ulla, Utrilla, Valentini, Elteren, Hemelryck, Leeuwen, Varadi, Vecchiato, Veljanoski, Via, Vicente, Vogt, Voss, Votruba, Voutsinas, Walmsley, Weiler, Weingrill, Werner, Wevers, Whitehead, Wyrzykowski, Yoldas, Žerjal, Zucker, Zurbach, Zwitter, Alecu, Allen, Prieto, Amorim, Anglada-Escudé, Arsenijevic, Azaz, Balm, Beck, Bernstein, Bigot, Bijaoui, Blasco, Bonfigli, Bono, Boudreault, Bressan, Brown, Brunet, Bunclark, Buonanno, Butkevich, Carret, Carrion, Chemin, Chéreau, Corcione, Darmigny, Boer, Teodoro, Zeeuw, Luche, Domingues, Dubath, Fodor, Frézouls, Fries, Fustes, Fyfe, Gallardo, Gallegos, Gardiol, Gebran, Gomboc, Gómez, Grux, Gueguen, Heyrovsky, Hoar, Iannicola, Parache,
  Janotto, Joliet, Jonckheere, Keil, Kim, Klagyivik, Klar, Knude, Kochukhov, Kolka, Kos, Kutka, Lainey, LeBouquin, Liu, Loreggia, Makarov, Marseille, Martayan, Martinez-Rubi, Massart, Meynadier, Mignot, Munari, Nguyen, Nordlander, Ocvirk, O’Flaherty, Sanz, Ortiz, Osorio, Oszkiewicz, Ouzounis, Palmer, Park, Pasquato, Peltzer, Peralta, Péturaud, Pieniluoma, Pigozzi, Poels, Prat, Prod’homme, Raison, Rebordao, Risquez, Rocca-Volmerange, Rosen, Ruiz-Fuertes, Russo, Sembay, Vizcaino, Short, Siebert, Silva, Sinachopoulos, Slezak, Soffel, Sosnowska, Straižys, Linden, Terrell, Theil, Tiede, Troisi, Tsalmantza, Tur, Vaccari, Vachier, Valles, Hamme, Veltz, Virtanen, Wallut, Wichmann, Wilkinson, Ziaeepour, \& Zschocke]{prusti_gaia_2016}
Prusti, T., Bruijne, J. H. J.~d., Brown, A. G.~A., Vallenari, A., Babusiaux, C., Bailer-Jones, C. a.~L., Bastian, U., Biermann, M., Evans, D.~W., Eyer, L., Jansen, F., Jordi, C., Klioner, S.~A., Lammers, U., Lindegren, L., Luri, X., Mignard, F., Milligan, D.~J., Panem, C., Poinsignon, V., Pourbaix, D., Randich, S., Sarri, G., Sartoretti, P., Siddiqui, H.~I., Soubiran, C., Valette, V., Leeuwen, F.~v., Walton, N.~A., Aerts, C., Arenou, F., Cropper, M., Drimmel, R., Høg, E., Katz, D., Lattanzi, M.~G., O’Mullane, W., Grebel, E.~K., Holland, A.~D., Huc, C., Passot, X., Bramante, L., Cacciari, C., Castañeda, J., Chaoul, L., Cheek, N., Angeli, F.~D., Fabricius, C., Guerra, R., Hernández, J., Jean-Antoine-Piccolo, A., Masana, E., Messineo, R., Mowlavi, N., Nienartowicz, K., Ordóñez-Blanco, D., Panuzzo, P., Portell, J., Richards, P.~J., Riello, M., Seabroke, G.~M., Tanga, P., Thévenin, F., Torra, J., Els, S.~G., Gracia-Abril, G., Comoretto, G., Garcia-Reinaldos, M., Lock, T., Mercier, E., Altmann, M., Andrae,
  R., Astraatmadja, T.~L., Bellas-Velidis, I., Benson, K., Berthier, J., Blomme, R., Busso, G., Carry, B., Cellino, A., Clementini, G., Cowell, S., Creevey, O., Cuypers, J., Davidson, M., Ridder, J.~D., Torres, A.~d., Delchambre, L., Dell’Oro, A., Ducourant, C., Frémat, Y., García-Torres, M., Gosset, E., Halbwachs, J.-L., Hambly, N.~C., Harrison, D.~L., Hauser, M., Hestroffer, D., Hodgkin, S.~T., Huckle, H.~E., Hutton, A., Jasniewicz, G., Jordan, S., Kontizas, M., Korn, A.~J., Lanzafame, A.~C., Manteiga, M., Moitinho, A., Muinonen, K., Osinde, J., Pancino, E., Pauwels, T., Petit, J.-M., Recio-Blanco, A., Robin, A.~C., Sarro, L.~M., Siopis, C., Smith, M., Smith, K.~W., Sozzetti, A., Thuillot, W., Reeven, W.~v., Viala, Y., Abbas, U., Aramburu, A.~A., Accart, S., Aguado, J.~J., Allan, P.~M., Allasia, W., Altavilla, G., Álvarez, M.~A., Alves, J., Anderson, R.~I., Andrei, A.~H., Varela, E.~A., Antiche, E., Antoja, T., Antón, S., Arcay, B., Atzei, A., Ayache, L., Bach, N., Baker, S.~G., Balaguer-Núñez, L.,
  Barache, C., Barata, C., Barbier, A., Barblan, F., Baroni, M., Navascués, D. B.~y., Barros, M., Barstow, M.~A., Becciani, U., Bellazzini, M., Bellei, G., García, A.~B., Belokurov, V., Bendjoya, P., Berihuete, A., Bianchi, L., Bienaymé, O., Billebaud, F., Blagorodnova, N., Blanco-Cuaresma, S., Boch, T., Bombrun, A., Borrachero, R., Bouquillon, S., Bourda, G., Bouy, H., Bragaglia, A., Breddels, M.~A., Brouillet, N., Brüsemeister, T., Bucciarelli, B., Budnik, F., Burgess, P., Burgon, R., Burlacu, A., Busonero, D., Buzzi, R., Caffau, E., Cambras, J., Campbell, H., Cancelliere, R., Cantat-Gaudin, T., Carlucci, T., Carrasco, J.~M., Castellani, M., Charlot, P., Charnas, J., Charvet, P., Chassat, F., Chiavassa, A., Clotet, M., Cocozza, G., Collins, R.~S., Collins, P., Costigan, G., Crifo, F., Cross, N. J.~G., Crosta, M., Crowley, C., Dafonte, C., Damerdji, Y., Dapergolas, A., David, P., David, M., Cat, P.~D., Felice, F.~d., Laverny, P.~d., Luise, F.~D., March, R.~D., Martino, D.~d., Souza, R.~d., Debosscher,
  J., Pozo, E.~d., Delbo, M., Delgado, A., Delgado, H.~E., Marco, F.~d., Matteo, P.~D., Diakite, S., Distefano, E., Dolding, C., Anjos, S.~D., Drazinos, P., Durán, J., Dzigan, Y., Ecale, E., Edvardsson, B., Enke, H., Erdmann, M., Escolar, D., Espina, M., Evans, N.~W., Bontemps, G.~E., Fabre, C., Fabrizio, M., Faigler, S., Falcão, A.~J., Casas, M.~F., Faye, F., Federici, L., Fedorets, G., Fernández-Hernández, J., Fernique, P., Fienga, A., Figueras, F., Filippi, F., Findeisen, K., Fonti, A., Fouesneau, M., Fraile, E., Fraser, M., Fuchs, J., Furnell, R., Gai, M., Galleti, S., Galluccio, L., Garabato, D., García-Sedano, F., Garé, P., Garofalo, A., Garralda, N., Gavras, P., Gerssen, J., Geyer, R., Gilmore, G., Girona, S., Giuffrida, G., Gomes, M., González-Marcos, A., González-Núñez, J., González-Vidal, J.~J., Granvik, M., Guerrier, A., Guillout, P., Guiraud, J., Gúrpide, A., Gutiérrez-Sánchez, R., Guy, L.~P., Haigron, R., Hatzidimitriou, D., Haywood, M., Heiter, U., Helmi, A., Hobbs, D., Hofmann,
  W., Holl, B., Holland, G., Hunt, J. a.~S., Hypki, A., Icardi, V., Irwin, M., Fombelle, G. J.~d., Jofré, P., Jonker, P.~G., Jorissen, A., Julbe, F., Karampelas, A., Kochoska, A., Kohley, R., Kolenberg, K., Kontizas, E., Koposov, S.~E., Kordopatis, G., Koubsky, P., Kowalczyk, A., Krone-Martins, A., Kudryashova, M., Kull, I., Bachchan, R.~K., Lacoste-Seris, F., Lanza, A.~F., Lavigne, J.-B., Poncin-Lafitte, C.~L., Lebreton, Y., Lebzelter, T., Leccia, S., Leclerc, N., Lecoeur-Taibi, I., Lemaitre, V., Lenhardt, H., Leroux, F., Liao, S., Licata, E., Lindstrøm, H. E.~P., Lister, T.~A., Livanou, E., Lobel, A., Löffler, W., López, M., Lopez-Lozano, A., Lorenz, D., Loureiro, T., MacDonald, I., Fernandes, T.~M., Managau, S., Mann, R.~G., Mantelet, G., Marchal, O., Marchant, J.~M., Marconi, M., Marie, J., Marinoni, S., Marrese, P.~M., Marschalkó, G., Marshall, D.~J., Martín-Fleitas, J.~M., Martino, M., Mary, N., Matijevič, G., Mazeh, T., McMillan, P.~J., Messina, S., Mestre, A., Michalik, D., Millar, N.~R.,
  Miranda, B. M.~H., Molina, D., Molinaro, R., Molinaro, M., Molnár, L., Moniez, M., Montegriffo, P., Monteiro, D., Mor, R., Mora, A., Morbidelli, R., Morel, T., Morgenthaler, S., Morley, T., Morris, D., Mulone, A.~F., Muraveva, T., Musella, I., Narbonne, J., Nelemans, G., Nicastro, L., Noval, L., Ordénovic, C., Ordieres-Meré, J., Osborne, P., Pagani, C., Pagano, I., Pailler, F., Palacin, H., Palaversa, L., Parsons, P., Paulsen, T., Pecoraro, M., Pedrosa, R., Pentikäinen, H., Pereira, J., Pichon, B., Piersimoni, A.~M., Pineau, F.-X., Plachy, E., Plum, G., Poujoulet, E., Prša, A., Pulone, L., Ragaini, S., Rago, S., Rambaux, N., Ramos-Lerate, M., Ranalli, P., Rauw, G., Read, A., Regibo, S., Renk, F., Reylé, C., Ribeiro, R.~A., Rimoldini, L., Ripepi, V., Riva, A., Rixon, G., Roelens, M., Romero-Gómez, M., Rowell, N., Royer, F., Rudolph, A., Ruiz-Dern, L., Sadowski, G., Sellés, T.~S., Sahlmann, J., Salgado, J., Salguero, E., Sarasso, M., Savietto, H., Schnorhk, A., Schultheis, M., Sciacca, E., Segol, M.,
  Segovia, J.~C., Segransan, D., Serpell, E., Shih, I.-C., Smareglia, R., Smart, R.~L., Smith, C., Solano, E., Solitro, F., Sordo, R., Nieto, S.~S., Souchay, J., Spagna, A., Spoto, F., Stampa, U., Steele, I.~A., Steidelmüller, H., Stephenson, C.~A., Stoev, H., Suess, F.~F., Süveges, M., Surdej, J., Szabados, L., Szegedi-Elek, E., Tapiador, D., Taris, F., Tauran, G., Taylor, M.~B., Teixeira, R., Terrett, D., Tingley, B., Trager, S.~C., Turon, C., Ulla, A., Utrilla, E., Valentini, G., Elteren, A.~v., Hemelryck, E.~V., Leeuwen, M.~v., Varadi, M., Vecchiato, A., Veljanoski, J., Via, T., Vicente, D., Vogt, S., Voss, H., Votruba, V., Voutsinas, S., Walmsley, G., Weiler, M., Weingrill, K., Werner, D., Wevers, T., Whitehead, G., Wyrzykowski, L., Yoldas, A., Žerjal, M., Zucker, S., Zurbach, C., Zwitter, T., Alecu, A., Allen, M., Prieto, C.~A., Amorim, A., Anglada-Escudé, G., Arsenijevic, V., Azaz, S., Balm, P., Beck, M., Bernstein, H.-H., Bigot, L., Bijaoui, A., Blasco, C., Bonfigli, M., Bono, G., Boudreault, S.,
  Bressan, A., Brown, S., Brunet, P.-M., Bunclark, P., Buonanno, R., Butkevich, A.~G., Carret, C., Carrion, C., Chemin, L., Chéreau, F., Corcione, L., Darmigny, E., Boer, K. S.~d., Teodoro, P.~d., Zeeuw, P. T.~d., Luche, C.~D., Domingues, C.~D., Dubath, P., Fodor, F., Frézouls, B., Fries, A., Fustes, D., Fyfe, D., Gallardo, E., Gallegos, J., Gardiol, D., Gebran, M., Gomboc, A., Gómez, A., Grux, E., Gueguen, A., Heyrovsky, A., Hoar, J., Iannicola, G., Parache, Y.~I., Janotto, A.-M., Joliet, E., Jonckheere, A., Keil, R., Kim, D.-W., Klagyivik, P., Klar, J., Knude, J., Kochukhov, O., Kolka, I., Kos, J., Kutka, A., Lainey, V., LeBouquin, D., Liu, C., Loreggia, D., Makarov, V.~V., Marseille, M.~G., Martayan, C., Martinez-Rubi, O., Massart, B., Meynadier, F., Mignot, S., Munari, U., Nguyen, A.-T., Nordlander, T., Ocvirk, P., O’Flaherty, K.~S., Sanz, A.~O., Ortiz, P., Osorio, J., Oszkiewicz, D., Ouzounis, A., Palmer, M., Park, P., Pasquato, E., Peltzer, C., Peralta, J., Péturaud, F., Pieniluoma, T., Pigozzi,
  E., Poels, J., Prat, G., Prod’homme, T., Raison, F., Rebordao, J.~M., Risquez, D., Rocca-Volmerange, B., Rosen, S., Ruiz-Fuertes, M.~I., Russo, F., Sembay, S., Vizcaino, I.~S., Short, A., Siebert, A., Silva, H., Sinachopoulos, D., Slezak, E., Soffel, M., Sosnowska, D., Straižys, V., Linden, M.~t., Terrell, D., Theil, S., Tiede, C., Troisi, L., Tsalmantza, P., Tur, D., Vaccari, M., Vachier, F., Valles, P., Hamme, W.~V., Veltz, L., Virtanen, J., Wallut, J.-M., Wichmann, R., Wilkinson, M.~I., Ziaeepour, H., \& Zschocke, S., 2016.
\newblock The {Gaia} mission, {\it Astronomy \& Astrophysics\/}, {\bf 595}, A1, Publisher: EDP Sciences.

\bibitem[Rebull et~al.(2018)Rebull, Stauffer, Cody, Hillenbrand, David, \& Pinsonneault]{rebull_rotation_2018}
Rebull, L.~M., Stauffer, J.~R., Cody, A.~M., Hillenbrand, L.~A., David, T.~J., \& Pinsonneault, M., 2018.
\newblock Rotation of {Low}-mass {Stars} in {Upper} {Scorpius} and\${\textbackslash}less\$i\${\textbackslash}greater\$\${\textbackslash}uprho\$\${\textbackslash}less\$/i\${\textbackslash}greater\${Ophiuchus} with\${\textbackslash}less\$i\${\textbackslash}greater\${K2}\${\textbackslash}less\$/i\${\textbackslash}greater\$, {\it The Astronomical Journal\/}, {\bf 155}(5), 196, Publisher: American Astronomical Society.

\bibitem[Reid et~al.(2002)Reid, Kirkpatrick, Liebert, Gizis, Dahn, \& Monet]{reid_high-resolution_2002}
Reid, I.~N., Kirkpatrick, J.~D., Liebert, J., Gizis, J.~E., Dahn, C.~C., \& Monet, D.~G., 2002.
\newblock High-{Resolution} {Spectroscopy} of {Ultracool} {M} {Dwarfs}, {\it The Astronomical Journal\/}, {\bf 124}, 519--540, Publisher: IOP ADS Bibcode: 2002AJ....124..519R.

\bibitem[Reiners et~al.(2018)Reiners, Zechmeister, Caballero, Ribas, Morales, Jeffers, Schöfer, Tal-Or, Quirrenbach, Amado, Kaminski, Seifert, Abril, Aceituno, Alonso-Floriano, Eiff, Antona, Anglada-Escudé, Anwand-Heerwart, Arroyo-Torres, Azzaro, Baroch, Barrado, Bauer, Becerril, Béjar, Benítez, Berdina\~s, Bergond, Blümcke, Brinkmöller, Burgo, Cano, Vázquez, Casal, Cifuentes, Claret, Colomé, Cortés-Contreras, Czesla, Díez-Alonso, Dreizler, Feiz, Fernández, Ferro, Fuhrmeister, Galadí-Enríquez, Garcia-Piquer, Vargas, Gesa, Galera, Hernández, González-Peinado, Grözinger, Grohnert, Guàrdia, Guenther, Guijarro, Guindos, Gutiérrez-Soto, Hagen, Hatzes, Hauschildt, Hedrosa, Helmling, Henning, Hermelo, Arabí, Castaño, Hernando, Herrero, Huber, Huke, Johnson, Juan, Kim, Klein, Klüter, Klutsch, Kürster, Lafarga, Lamert, Lampón, Lara, Laun, Lemke, Lenzen, Launhardt, Fresno, López-González, López-Puertas, Salas, López-Santiago, Luque, Madinabeitia, Mall, Mancini, Mandel, Marfil, Molina,
  Fernández, Martín, Martín-Ruiz, Marvin, Mathar, Mirabet, Montes, Moreno-Raya, Moya, Mundt, Nagel, Naranjo, Nortmann, Nowak, Ofir, Oreiro, Pallé, Panduro, Pascual, Passegger, Pavlov, Pedraz, Pérez-Calpena, Medialdea, Perger, Perryman, Pluto, Rabaza, Ramón, Rebolo, Redondo, Reffert, Reinhart, Rhode, Rix, Rodler, Rodríguez, Rodríguez-López, Trinidad, Rohloff, Rosich, Sadegi, Sánchez-Blanco, Carrasco, Sánchez-López, Sanz-Forcada, Sarkis, Sarmiento, Schäfer, Schmitt, Schiller, Schweitzer, Solano, Stahl, Strachan, Stürmer, Suárez, Tabernero, Tala, Trifonov, Tulloch, Ulbrich, Veredas, Linares, Vilardell, Wagner, Winkler, Wolthoff, Xu, Yan, \& Osorio]{reiners_carmenes_2018}
Reiners, A., Zechmeister, M., Caballero, J.~A., Ribas, I., Morales, J.~C., Jeffers, S.~V., Schöfer, P., Tal-Or, L., Quirrenbach, A., Amado, P.~J., Kaminski, A., Seifert, W., Abril, M., Aceituno, J., Alonso-Floriano, F.~J., Eiff, M. A.-v., Antona, R., Anglada-Escudé, G., Anwand-Heerwart, H., Arroyo-Torres, B., Azzaro, M., Baroch, D., Barrado, D., Bauer, F.~F., Becerril, S., Béjar, V. J.~S., Benítez, D., Berdina\~s, Z.~M., Bergond, G., Blümcke, M., Brinkmöller, M., Burgo, C.~d., Cano, J., Vázquez, M. C.~C., Casal, E., Cifuentes, C., Claret, A., Colomé, J., Cortés-Contreras, M., Czesla, S., Díez-Alonso, E., Dreizler, S., Feiz, C., Fernández, M., Ferro, I.~M., Fuhrmeister, B., Galadí-Enríquez, D., Garcia-Piquer, A., Vargas, M. L.~G., Gesa, L., Galera, V.~G., Hernández, J. I.~G., González-Peinado, R., Grözinger, U., Grohnert, S., Guàrdia, J., Guenther, E.~W., Guijarro, A., Guindos, E.~d., Gutiérrez-Soto, J., Hagen, H.-J., Hatzes, A.~P., Hauschildt, P.~H., Hedrosa, R.~P., Helmling, J., Henning,
  T., Hermelo, I., Arabí, R.~H., Castaño, L.~H., Hernando, F.~H., Herrero, E., Huber, A., Huke, P., Johnson, E.~N., Juan, E.~d., Kim, M., Klein, R., Klüter, J., Klutsch, A., Kürster, M., Lafarga, M., Lamert, A., Lampón, M., Lara, L.~M., Laun, W., Lemke, U., Lenzen, R., Launhardt, R., Fresno, M. L.~d., López-González, J., López-Puertas, M., Salas, J. F.~L., López-Santiago, J., Luque, R., Madinabeitia, H.~M., Mall, U., Mancini, L., Mandel, H., Marfil, E., Molina, J. A.~M., Fernández, D.~M., Martín, E.~L., Martín-Ruiz, S., Marvin, C.~J., Mathar, R.~J., Mirabet, E., Montes, D., Moreno-Raya, M.~E., Moya, A., Mundt, R., Nagel, E., Naranjo, V., Nortmann, L., Nowak, G., Ofir, A., Oreiro, R., Pallé, E., Panduro, J., Pascual, J., Passegger, V.~M., Pavlov, A., Pedraz, S., Pérez-Calpena, A., Medialdea, D.~P., Perger, M., Perryman, M. a.~C., Pluto, M., Rabaza, O., Ramón, A., Rebolo, R., Redondo, P., Reffert, S., Reinhart, S., Rhode, P., Rix, H.-W., Rodler, F., Rodríguez, E., Rodríguez-López, C.,
  Trinidad, A.~R., Rohloff, R.-R., Rosich, A., Sadegi, S., Sánchez-Blanco, E., Carrasco, M. A.~S., Sánchez-López, A., Sanz-Forcada, J., Sarkis, P., Sarmiento, L.~F., Schäfer, S., Schmitt, J. H. M.~M., Schiller, J., Schweitzer, A., Solano, E., Stahl, O., Strachan, J. B.~P., Stürmer, J., Suárez, J.~C., Tabernero, H.~M., Tala, M., Trifonov, T., Tulloch, S.~M., Ulbrich, R.~G., Veredas, G., Linares, J. I.~V., Vilardell, F., Wagner, K., Winkler, J., Wolthoff, V., Xu, W., Yan, F., \& Osorio, M. R.~Z., 2018.
\newblock The {CARMENES} search for exoplanets around {M} dwarfs - {High}-resolution optical and near-infrared spectroscopy of 324 survey stars, {\it Astronomy \& Astrophysics\/}, {\bf 612}, A49, Publisher: EDP Sciences.

\bibitem[Rizzuto et~al.(2016)Rizzuto, Ireland, Dupuy, \& Kraus]{rizzuto_dynamical_2016}
Rizzuto, A.~C., Ireland, M.~J., Dupuy, T.~J., \& Kraus, A.~L., 2016.
\newblock Dynamical {Masses} of {Young} {Stars}. {I}. {Discordant} {Model} {Ages} of {Upper} {Scorpius}, {\it The Astrophysical Journal\/}, {\bf 817}, 164.

\bibitem[Rizzuto et~al.(2020)Rizzuto, Newton, Mann, Tofflemire, Vanderburg, Kraus, Wood, Quinn, Zhou, Thao, Law, Ziegler, \& Briceño]{rizzuto_tess_2020}
Rizzuto, A.~C., Newton, E.~R., Mann, A.~W., Tofflemire, B.~M., Vanderburg, A., Kraus, A.~L., Wood, M.~L., Quinn, S.~N., Zhou, G., Thao, P.~C., Law, N.~M., Ziegler, C., \& Briceño, C., 2020.
\newblock {TESS} {Hunt} for {Young} and {Maturing} {Exoplanets} ({THYME}). {II}. {A} 17 {Myr} {Old} {Transiting} {Hot} {Jupiter} in the {Sco}-{Cen} {Association}, {\it The Astronomical Journal\/}, {\bf 160}(1), 33, Publisher: American Astronomical Society.

\bibitem[Rogers et~al.(2012)Rogers, Lin, \& Lau]{rogers_internal_2012}
Rogers, T.~M., Lin, D. N.~C., \& Lau, H. H.~B., 2012.
\newblock Internal {Gravity} {Waves} {Modulate} the {Apparent} {Misalignment} of {Exoplanets} around {Hot} {Stars}, {\it The Astrophysical Journal\/}, {\bf 758}, L6, Publisher: IOP ADS Bibcode: 2012ApJ...758L...6R.

\bibitem[Rosotti \& Clarke(2018)]{rosotti_evolution_2018}
Rosotti, G.~P. \& Clarke, C.~J., 2018.
\newblock The evolution of photoevaporating viscous discs in binaries, {\it Monthly Notices of the Royal Astronomical Society\/}, {\bf 473}(4), 5630--5640.

\bibitem[Schlieder et~al.(2012)Schlieder, Lépine, \& Simon]{schlieder_cool_2012}
Schlieder, J.~E., Lépine, S., \& Simon, M., 2012.
\newblock Cool {Young} {Stars} in the {Northern} {Hemisphere}: $\beta$ {Pictoris} and {AB} {Doradus} {Moving} {Group} {Candidates}, {\it The Astronomical Journal\/}, {\bf 143}, 80, Publisher: IOP ADS Bibcode: 2012AJ....143...80S.

\bibitem[Scholz et~al.(2007)Scholz, Coffey, Brandeker, \& Jayawardhana]{scholz_rotation_2007}
Scholz, A., Coffey, J., Brandeker, A., \& Jayawardhana, R., 2007.
\newblock Rotation and {Activity} of {Pre}-{Main}-{Sequence} {Stars}, {\it The Astrophysical Journal\/}, {\bf 662}, 1254--1267, Publisher: IOP ADS Bibcode: 2007ApJ...662.1254S.

\bibitem[Seager \& Mallén-Ornelas(2003)]{seager_unique_2003}
Seager, S. \& Mallén-Ornelas, G., 2003.
\newblock A {Unique} {Solution} of {Planet} and {Star} {Parameters} from an {Extrasolar} {Planet} {Transit} {Light} {Curve}, {\it The Astrophysical Journal\/}, {\bf 585}(2), 1038, Publisher: IOP Publishing.

\bibitem[Shkolnik et~al.(2017)Shkolnik, Allers, Kraus, Liu, \& Flagg]{shkolnik_all-sky_2017}
Shkolnik, E.~L., Allers, K.~N., Kraus, A.~L., Liu, M.~C., \& Flagg, L., 2017.
\newblock All-sky {Co}-moving {Recovery} {Of} {Nearby} {Young} {Members} ({ACRONYM}). {II}. {The} $\beta$ {Pictoris} {Moving} {Group}, {\it The Astronomical Journal\/}, {\bf 154}, 69, Publisher: IOP ADS Bibcode: 2017AJ....154...69S.

\bibitem[Skrutskie et~al.(2006)Skrutskie, Cutri, Stiening, Weinberg, Schneider, Carpenter, Beichman, Capps, Chester, Elias, Huchra, Liebert, Lonsdale, Monet, Price, Seitzer, Jarrett, Kirkpatrick, Gizis, Howard, Evans, Fowler, Fullmer, Hurt, Light, Kopan, Marsh, McCallon, Tam, Dyk, \& Wheelock]{skrutskie_two_2006}
Skrutskie, M.~F., Cutri, R.~M., Stiening, R., Weinberg, M.~D., Schneider, S., Carpenter, J.~M., Beichman, C., Capps, R., Chester, T., Elias, J., Huchra, J., Liebert, J., Lonsdale, C., Monet, D.~G., Price, S., Seitzer, P., Jarrett, T., Kirkpatrick, J.~D., Gizis, J.~E., Howard, E., Evans, T., Fowler, J., Fullmer, L., Hurt, R., Light, R., Kopan, E.~L., Marsh, K.~A., McCallon, H.~L., Tam, R., Dyk, S.~V., \& Wheelock, S., 2006.
\newblock The {Two} {Micron} {All} {Sky} {Survey} ({2MASS}), {\it The Astronomical Journal\/}, {\bf 131}(2), 1163, Publisher: IOP Publishing.

\bibitem[Somers et~al.(2020)Somers, Cao, \& Pinsonneault]{somers_spots_2020}
Somers, G., Cao, L., \& Pinsonneault, M.~H., 2020.
\newblock The {SPOTS} {Models}: {A} {Grid} of {Theoretical} {Stellar} {Evolution} {Tracks} and {Isochrones} for {Testing} the {Effects} of {Starspots} on {Structure} and {Colors}, {\it The Astrophysical Journal\/}, {\bf 891}, 29, Publisher: IOP ADS Bibcode: 2020ApJ...891...29S.

\bibitem[Sperauskas et~al.(2019)Sperauskas, Deveikis, \& Tokovinin]{sperauskas_spectroscopic_2019}
Sperauskas, J., Deveikis, V., \& Tokovinin, A., 2019.
\newblock Spectroscopic orbits of nearby stars, {\it Astronomy \& Astrophysics\/}, {\bf 626}, A31, Publisher: EDP Sciences.

\bibitem[Spina et~al.(2017)Spina, Randich, Magrini, Jeffries, Friel, Sacco, Pancino, Bonito, Bravi, Franciosini, Klutsch, Montes, Gilmore, Vallenari, Bensby, Bragaglia, Flaccomio, Koposov, Korn, Lanzafame, Smiljanic, Bayo, Carraro, Casey, Costado, Damiani, Donati, Frasca, Hourihane, Jofré, Lewis, Lind, Monaco, Morbidelli, Prisinzano, Sousa, Worley, \& Zaggia]{spina_gaia-eso_2017}
Spina, L., Randich, S., Magrini, L., Jeffries, R.~D., Friel, E.~D., Sacco, G.~G., Pancino, E., Bonito, R., Bravi, L., Franciosini, E., Klutsch, A., Montes, D., Gilmore, G., Vallenari, A., Bensby, T., Bragaglia, A., Flaccomio, E., Koposov, S.~E., Korn, A.~J., Lanzafame, A.~C., Smiljanic, R., Bayo, A., Carraro, G., Casey, A.~R., Costado, M.~T., Damiani, F., Donati, P., Frasca, A., Hourihane, A., Jofré, P., Lewis, J., Lind, K., Monaco, L., Morbidelli, L., Prisinzano, L., Sousa, S.~G., Worley, C.~C., \& Zaggia, S., 2017.
\newblock The {Gaia}-{ESO} {Survey}: the present-day radial metallicity distribution of the {Galactic} disc probed by pre-main-sequence clusters, {\it Astronomy and Astrophysics\/}, {\bf 601}, A70, ADS Bibcode: 2017A\&A...601A..70S.

\bibitem[Stassun \& Torres(2021)]{stassun_parallax_2021}
Stassun, K.~G. \& Torres, G., 2021.
\newblock Parallax {Systematics} and {Photocenter} {Motions} of {Benchmark} {Eclipsing} {Binaries} in {Gaia} {EDR3}, {\it The Astrophysical Journal Letters\/}, {\bf 907}(2), L33, Publisher: The American Astronomical Society.

\bibitem[Stassun et~al.(2019)Stassun, Oelkers, Paegert, Torres, Pepper, De~Lee, Collins, Latham, Muirhead, Chittidi, Rojas-Ayala, Fleming, Rose, Tenenbaum, Ting, Kane, Barclay, Bean, Brassuer, Charbonneau, Ge, Lissauer, Mann, McLean, Mullally, Narita, Plavchan, Ricker, Sasselov, Seager, Sharma, Shiao, Sozzetti, Stello, Vanderspek, Wallace, \& Winn]{stassun_revised_2019}
Stassun, K.~G., Oelkers, R.~J., Paegert, M., Torres, G., Pepper, J., De~Lee, N., Collins, K., Latham, D.~W., Muirhead, P.~S., Chittidi, J., Rojas-Ayala, B., Fleming, S.~W., Rose, M.~E., Tenenbaum, P., Ting, E.~B., Kane, S.~R., Barclay, T., Bean, J.~L., Brassuer, C.~E., Charbonneau, D., Ge, J., Lissauer, J.~J., Mann, A.~W., McLean, B., Mullally, S., Narita, N., Plavchan, P., Ricker, G.~R., Sasselov, D., Seager, S., Sharma, S., Shiao, B., Sozzetti, A., Stello, D., Vanderspek, R., Wallace, G., \& Winn, J.~N., 2019.
\newblock The {Revised} {TESS} {Input} {Catalog} and {Candidate} {Target} {List}, {\it The Astronomical Journal\/}, {\bf 158}, 138, ADS Bibcode: 2019AJ....158..138S.

\bibitem[{STScI Development Team}(2018)]{stsci_development_team_synphot_2018}
{STScI Development Team}, 2018.
\newblock synphot: {Synthetic} photometry using {Astropy}, {\it Astrophysics Source Code Library\/}, p. ascl:1811.001, ADS Bibcode: 2018ascl.soft11001S.

\bibitem[Takaishi et~al.(2020)Takaishi, Tsukamoto, \& Suto]{takaishi_stardisc_2020}
Takaishi, D., Tsukamoto, Y., \& Suto, Y., 2020.
\newblock Star–disc alignment in the protoplanetary discs: {SPH} simulation of the collapse of turbulent molecular cloud cores, {\it Monthly Notices of the Royal Astronomical Society\/}, {\bf 492}(4), 5641--5654.

\bibitem[Tayar et~al.(2022)Tayar, Claytor, Huber, \& van Saders]{tayar_guide_2022}
Tayar, J., Claytor, Z.~R., Huber, D., \& van Saders, J., 2022.
\newblock A {Guide} to {Realistic} {Uncertainties} on the {Fundamental} {Properties} of {Solar}-type {Exoplanet} {Host} {Stars}, {\it The Astrophysical Journal\/}, {\bf 927}, 31, Publisher: IOP ADS Bibcode: 2022ApJ...927...31T.

\bibitem[Tazzari et~al.(2017)Tazzari, Testi, Natta, Ansdell, Carpenter, Guidi, Hogerheijde, Manara, Miotello, Marel, Dishoeck, \& Williams]{tazzari_physical_2017}
Tazzari, M., Testi, L., Natta, A., Ansdell, M., Carpenter, J., Guidi, G., Hogerheijde, M., Manara, C.~F., Miotello, A., Marel, N. v.~d., Dishoeck, E. F.~v., \& Williams, J.~P., 2017.
\newblock Physical properties of dusty protoplanetary disks in {Lupus}: evidence for viscous evolution?, {\it Astronomy \& Astrophysics\/}, {\bf 606}, A88, Publisher: EDP Sciences.

\bibitem[Thao et~al.(2024)Thao, Mann, Feinstein, Gao, Thorngren, Rotman, Welbanks, Brown, Duvvuri, France, Longo, Sandoval, Schneider, Wilson, Youngblood, Vanderburg, Barber, Wood, Batalha, Kraus, Murray, Newton, Rizzuto, Tofflemire, Tsai, Bean, Berta-Thompson, Evans-Soma, Froning, Kempton, Miguel, \& Pineda]{thao_featherweight_2024}
Thao, P.~C., Mann, A.~W., Feinstein, A.~D., Gao, P., Thorngren, D., Rotman, Y., Welbanks, L., Brown, A., Duvvuri, G.~M., France, K., Longo, I., Sandoval, A., Schneider, P.~C., Wilson, D.~J., Youngblood, A., Vanderburg, A., Barber, M.~G., Wood, M.~L., Batalha, N.~E., Kraus, A.~L., Murray, C.~A., Newton, E.~R., Rizzuto, A., Tofflemire, B.~M., Tsai, S.-M., Bean, J.~L., Berta-Thompson, Z.~K., Evans-Soma, T.~M., Froning, C.~S., Kempton, E. M.~R., Miguel, Y., \& Pineda, J.~S., 2024.
\newblock The {Featherweight} {Giant}: {Unraveling} the {Atmosphere} of a 17 {Myr} {Planet} with {JWST}, Publication Title: arXiv e-prints ADS Bibcode: 2024arXiv240916355T.

\bibitem[Torres et~al.(2006)Torres, Quast, Silva, Reza, Melo, \& Sterzik]{torres_search_2006}
Torres, C. a.~O., Quast, G.~R., Silva, L.~d., Reza, R. d.~l., Melo, C. H.~F., \& Sterzik, M., 2006.
\newblock Search for associations containing young stars ({SACY}) - {I}. {Sample} and searching method, {\it Astronomy \& Astrophysics\/}, {\bf 460}(3), 695--708, Number: 3 Publisher: EDP Sciences.

\bibitem[Tripathi et~al.(2017)Tripathi, Andrews, Birnstiel, \& Wilner]{tripathi_millimeter_2017}
Tripathi, A., Andrews, S.~M., Birnstiel, T., \& Wilner, D.~J., 2017.
\newblock A millimeter {Continuum} {Size}-{Luminosity} {Relationship} for {Protoplanetary} {Disks}, {\it The Astrophysical Journal\/}, {\bf 845}, 44, Publisher: IOP ADS Bibcode: 2017ApJ...845...44T.

\bibitem[Vallenari et~al.(2023)Vallenari, Brown, Prusti, Bruijne, Arenou, Babusiaux, Biermann, Creevey, Ducourant, Evans, Eyer, Guerra, Hutton, Jordi, Klioner, Lammers, Lindegren, Luri, Mignard, Panem, Pourbaix, Randich, Sartoretti, Soubiran, Tanga, Walton, Bailer-Jones, Bastian, Drimmel, Jansen, Katz, Lattanzi, Leeuwen, Bakker, Cacciari, Castañeda, Angeli, Fabricius, Fouesneau, Frémat, Galluccio, Guerrier, Heiter, Masana, Messineo, Mowlavi, Nicolas, Nienartowicz, Pailler, Panuzzo, Riclet, Roux, Seabroke, Sordo, Thévenin, Gracia-Abril, Portell, Teyssier, Altmann, Andrae, Audard, Bellas-Velidis, Benson, Berthier, Blomme, Burgess, Busonero, Busso, Cánovas, Carry, Cellino, Cheek, Clementini, Damerdji, Davidson, Teodoro, Campos, Delchambre, Dell’Oro, Esquej, Fernández-Hernández, Fraile, Garabato, García-Lario, Gosset, Haigron, Halbwachs, Hambly, Harrison, Hernández, Hestroffer, Hodgkin, Holl, Janßen, Fombelle, Jordan, Krone-Martins, Lanzafame, Löffler, Marchal, Marrese, Moitinho, Muinonen, Osborne,
  Pancino, Pauwels, Recio-Blanco, Reylé, Riello, Rimoldini, Roegiers, Rybizki, Sarro, Siopis, Smith, Sozzetti, Utrilla, Leeuwen, Abbas, Ábrahám, Aramburu, Aerts, Aguado, Ajaj, Aldea-Montero, Altavilla, Álvarez, Alves, Anders, Anderson, Varela, Antoja, Baines, Baker, Balaguer-Núñez, Balbinot, Balog, Barache, Barbato, Barros, Barstow, Bartolomé, Bassilana, Bauchet, Becciani, Bellazzini, Berihuete, Bernet, Bertone, Bianchi, Binnenfeld, Blanco-Cuaresma, Blazere, Boch, Bombrun, Bossini, Bouquillon, Bragaglia, Bramante, Breedt, Bressan, Brouillet, Brugaletta, Bucciarelli, Burlacu, Butkevich, Buzzi, Caffau, Cancelliere, Cantat-Gaudin, Carballo, Carlucci, Carnerero, Carrasco, Casamiquela, Castellani, Castro-Ginard, Chaoul, Charlot, Chemin, Chiaramida, Chiavassa, Chornay, Comoretto, Contursi, Cooper, Cornez, Cowell, Crifo, Cropper, Crosta, Crowley, Dafonte, Dapergolas, David, David, Laverny, Luise, March, Ridder, Souza, Torres, Peloso, Pozo, Delbo, Delgado, Delisle, Demouchy, Dharmawardena, Matteo, Diakite,
  Diener, Distefano, Dolding, Edvardsson, Enke, Fabre, Fabrizio, Faigler, Fedorets, Fernique, Fienga, Figueras, Fournier, Fouron, Fragkoudi, Gai, Garcia-Gutierrez, Garcia-Reinaldos, García-Torres, Garofalo, Gavel, Gavras, Gerlach, Geyer, Giacobbe, Gilmore, Girona, Giuffrida, Gomel, Gomez, González-Núñez, González-Santamaría, González-Vidal, Granvik, Guillout, Guiraud, Gutiérrez-Sánchez, Guy, Hatzidimitriou, Hauser, Haywood, Helmer, Helmi, Sarmiento, Hidalgo, Hilger, Hładczuk, Hobbs, Holland, Huckle, Jardine, Jasniewicz, Piccolo, Jiménez-Arranz, Jorissen, Campillo, Julbe, Karbevska, Kervella, Khanna, Kontizas, Kordopatis, Korn, Kóspál, Kostrzewa-Rutkowska, Kruszyńska, Kun, Laizeau, Lambert, Lanza, Lasne, Campion, Lebreton, Lebzelter, Leccia, Leclerc, Lecoeur-Taibi, Liao, Licata, Lindstrøm, Lister, Livanou, Lobel, Lorca, Loup, Pardo, Romeo, Managau, Mann, Manteiga, Marchant, Marconi, Marcos, Santos, Pina, Marinoni, Marocco, Marshall, Polo, Martín-Fleitas, Marton, Mary, Masip, Massari,
  Mastrobuono-Battisti, Mazeh, McMillan, Messina, Michalik, Millar, Mints, Molina, Molinaro, Molnár, Monari, Monguió, Montegriffo, Montero, Mor, Mora, Morbidelli, Morel, Morris, Muraveva, Murphy, Musella, Nagy, Noval, Ocaña, Ogden, Ordenovic, Osinde, Pagani, Pagano, Palaversa, Palicio, Pallas-Quintela, Panahi, Payne-Wardenaar, Esteller, Penttilä, Pichon, Piersimoni, Pineau, Plachy, Plum, Poggio, Prša, Pulone, Racero, Ragaini, Rainer, Raiteri, Rambaux, Ramos, Ramos-Lerate, Fiorentin, Regibo, Richards, Diaz, Ripepi, Riva, Rix, Rixon, Robichon, Robin, Robin, Roelens, Rogues, Rohrbasser, Romero-Gómez, Rowell, Royer, Mieres, Rybicki, Sadowski, Núñez, Sellés, Sahlmann, Salguero, Samaras, Gimenez, Sanna, Santoveña, Sarasso, Schultheis, Sciacca, Segol, Segovia, Ségransan, Semeux, Shahaf, Siddiqui, Siebert, Siltala, Silvelo, Slezak, Slezak, Smart, Snaith, Solano, Solitro, Souami, Souchay, Spagna, Spina, Spoto, Steele, Steidelmüller, Stephenson, Süveges, Surdej, Szabados, Szegedi-Elek, Taris, Taylor,
  Teixeira, Tolomei, Tonello, Torra, Torra, Elipe, Trabucchi, Tsounis, Turon, Ulla, Unger, Vaillant, Dillen, Reeven, Vanel, Vecchiato, Viala, Vicente, Voutsinas, Weiler, Wevers, Wyrzykowski, Yoldas, Yvard, Zhao, Zorec, Zucker, \& Zwitter]{vallenari_gaia_2023}
Vallenari, A., Brown, A. G.~A., Prusti, T., Bruijne, J. H. J.~d., Arenou, F., Babusiaux, C., Biermann, M., Creevey, O.~L., Ducourant, C., Evans, D.~W., Eyer, L., Guerra, R., Hutton, A., Jordi, C., Klioner, S.~A., Lammers, U.~L., Lindegren, L., Luri, X., Mignard, F., Panem, C., Pourbaix, D., Randich, S., Sartoretti, P., Soubiran, C., Tanga, P., Walton, N.~A., Bailer-Jones, C. a.~L., Bastian, U., Drimmel, R., Jansen, F., Katz, D., Lattanzi, M.~G., Leeuwen, F.~v., Bakker, J., Cacciari, C., Castañeda, J., Angeli, F.~D., Fabricius, C., Fouesneau, M., Frémat, Y., Galluccio, L., Guerrier, A., Heiter, U., Masana, E., Messineo, R., Mowlavi, N., Nicolas, C., Nienartowicz, K., Pailler, F., Panuzzo, P., Riclet, F., Roux, W., Seabroke, G.~M., Sordo, R., Thévenin, F., Gracia-Abril, G., Portell, J., Teyssier, D., Altmann, M., Andrae, R., Audard, M., Bellas-Velidis, I., Benson, K., Berthier, J., Blomme, R., Burgess, P.~W., Busonero, D., Busso, G., Cánovas, H., Carry, B., Cellino, A., Cheek, N., Clementini, G., Damerdji,
  Y., Davidson, M., Teodoro, P.~d., Campos, M.~N., Delchambre, L., Dell’Oro, A., Esquej, P., Fernández-Hernández, J., Fraile, E., Garabato, D., García-Lario, P., Gosset, E., Haigron, R., Halbwachs, J.-L., Hambly, N.~C., Harrison, D.~L., Hernández, J., Hestroffer, D., Hodgkin, S.~T., Holl, B., Janßen, K., Fombelle, G. J.~d., Jordan, S., Krone-Martins, A., Lanzafame, A.~C., Löffler, W., Marchal, O., Marrese, P.~M., Moitinho, A., Muinonen, K., Osborne, P., Pancino, E., Pauwels, T., Recio-Blanco, A., Reylé, C., Riello, M., Rimoldini, L., Roegiers, T., Rybizki, J., Sarro, L.~M., Siopis, C., Smith, M., Sozzetti, A., Utrilla, E., Leeuwen, M.~v., Abbas, U., Ábrahám, P., Aramburu, A.~A., Aerts, C., Aguado, J.~J., Ajaj, M., Aldea-Montero, F., Altavilla, G., Álvarez, M.~A., Alves, J., Anders, F., Anderson, R.~I., Varela, E.~A., Antoja, T., Baines, D., Baker, S.~G., Balaguer-Núñez, L., Balbinot, E., Balog, Z., Barache, C., Barbato, D., Barros, M., Barstow, M.~A., Bartolomé, S., Bassilana, J.-L., Bauchet,
  N., Becciani, U., Bellazzini, M., Berihuete, A., Bernet, M., Bertone, S., Bianchi, L., Binnenfeld, A., Blanco-Cuaresma, S., Blazere, A., Boch, T., Bombrun, A., Bossini, D., Bouquillon, S., Bragaglia, A., Bramante, L., Breedt, E., Bressan, A., Brouillet, N., Brugaletta, E., Bucciarelli, B., Burlacu, A., Butkevich, A.~G., Buzzi, R., Caffau, E., Cancelliere, R., Cantat-Gaudin, T., Carballo, R., Carlucci, T., Carnerero, M.~I., Carrasco, J.~M., Casamiquela, L., Castellani, M., Castro-Ginard, A., Chaoul, L., Charlot, P., Chemin, L., Chiaramida, V., Chiavassa, A., Chornay, N., Comoretto, G., Contursi, G., Cooper, W.~J., Cornez, T., Cowell, S., Crifo, F., Cropper, M., Crosta, M., Crowley, C., Dafonte, C., Dapergolas, A., David, M., David, P., Laverny, P.~d., Luise, F.~D., March, R.~D., Ridder, J.~D., Souza, R.~d., Torres, A.~d., Peloso, E. F.~d., Pozo, E.~d., Delbo, M., Delgado, A., Delisle, J.-B., Demouchy, C., Dharmawardena, T.~E., Matteo, P.~D., Diakite, S., Diener, C., Distefano, E., Dolding, C., Edvardsson,
  B., Enke, H., Fabre, C., Fabrizio, M., Faigler, S., Fedorets, G., Fernique, P., Fienga, A., Figueras, F., Fournier, Y., Fouron, C., Fragkoudi, F., Gai, M., Garcia-Gutierrez, A., Garcia-Reinaldos, M., García-Torres, M., Garofalo, A., Gavel, A., Gavras, P., Gerlach, E., Geyer, R., Giacobbe, P., Gilmore, G., Girona, S., Giuffrida, G., Gomel, R., Gomez, A., González-Núñez, J., González-Santamaría, I., González-Vidal, J.~J., Granvik, M., Guillout, P., Guiraud, J., Gutiérrez-Sánchez, R., Guy, L.~P., Hatzidimitriou, D., Hauser, M., Haywood, M., Helmer, A., Helmi, A., Sarmiento, M.~H., Hidalgo, S.~L., Hilger, T., Hładczuk, N., Hobbs, D., Holland, G., Huckle, H.~E., Jardine, K., Jasniewicz, G., Piccolo, A. J.-A., Jiménez-Arranz, O., Jorissen, A., Campillo, J.~J., Julbe, F., Karbevska, L., Kervella, P., Khanna, S., Kontizas, M., Kordopatis, G., Korn, A.~J., Kóspál, A., Kostrzewa-Rutkowska, Z., Kruszyńska, K., Kun, M., Laizeau, P., Lambert, S., Lanza, A.~F., Lasne, Y., Campion, J.-F.~L., Lebreton, Y.,
  Lebzelter, T., Leccia, S., Leclerc, N., Lecoeur-Taibi, I., Liao, S., Licata, E.~L., Lindstrøm, H. E.~P., Lister, T.~A., Livanou, E., Lobel, A., Lorca, A., Loup, C., Pardo, P.~M., Romeo, A.~M., Managau, S., Mann, R.~G., Manteiga, M., Marchant, J.~M., Marconi, M., Marcos, J., Santos, M. M. S.~M., Pina, D.~M., Marinoni, S., Marocco, F., Marshall, D.~J., Polo, L.~M., Martín-Fleitas, J.~M., Marton, G., Mary, N., Masip, A., Massari, D., Mastrobuono-Battisti, A., Mazeh, T., McMillan, P.~J., Messina, S., Michalik, D., Millar, N.~R., Mints, A., Molina, D., Molinaro, R., Molnár, L., Monari, G., Monguió, M., Montegriffo, P., Montero, A., Mor, R., Mora, A., Morbidelli, R., Morel, T., Morris, D., Muraveva, T., Murphy, C.~P., Musella, I., Nagy, Z., Noval, L., Ocaña, F., Ogden, A., Ordenovic, C., Osinde, J.~O., Pagani, C., Pagano, I., Palaversa, L., Palicio, P.~A., Pallas-Quintela, L., Panahi, A., Payne-Wardenaar, S., Esteller, X.~P., Penttilä, A., Pichon, B., Piersimoni, A.~M., Pineau, F.-X., Plachy, E., Plum, G.,
  Poggio, E., Prša, A., Pulone, L., Racero, E., Ragaini, S., Rainer, M., Raiteri, C.~M., Rambaux, N., Ramos, P., Ramos-Lerate, M., Fiorentin, P.~R., Regibo, S., Richards, P.~J., Diaz, C.~R., Ripepi, V., Riva, A., Rix, H.-W., Rixon, G., Robichon, N., Robin, A.~C., Robin, C., Roelens, M., Rogues, H. R.~O., Rohrbasser, L., Romero-Gómez, M., Rowell, N., Royer, F., Mieres, D.~R., Rybicki, K.~A., Sadowski, G., Núñez, A.~S., Sellés, A.~S., Sahlmann, J., Salguero, E., Samaras, N., Gimenez, V.~S., Sanna, N., Santoveña, R., Sarasso, M., Schultheis, M., Sciacca, E., Segol, M., Segovia, J.~C., Ségransan, D., Semeux, D., Shahaf, S., Siddiqui, H.~I., Siebert, A., Siltala, L., Silvelo, A., Slezak, E., Slezak, I., Smart, R.~L., Snaith, O.~N., Solano, E., Solitro, F., Souami, D., Souchay, J., Spagna, A., Spina, L., Spoto, F., Steele, I.~A., Steidelmüller, H., Stephenson, C.~A., Süveges, M., Surdej, J., Szabados, L., Szegedi-Elek, E., Taris, F., Taylor, M.~B., Teixeira, R., Tolomei, L., Tonello, N., Torra, F., Torra,
  J., Elipe, G.~T., Trabucchi, M., Tsounis, A.~T., Turon, C., Ulla, A., Unger, N., Vaillant, M.~V., Dillen, E.~v., Reeven, W.~v., Vanel, O., Vecchiato, A., Viala, Y., Vicente, D., Voutsinas, S., Weiler, M., Wevers, T., Wyrzykowski, L., Yoldas, A., Yvard, P., Zhao, H., Zorec, J., Zucker, S., \& Zwitter, T., 2023.
\newblock Gaia {Data} {Release} 3 - {Summary} of the content and survey properties, {\it Astronomy \& Astrophysics\/}, {\bf 674}, A1, Publisher: EDP Sciences.

\bibitem[Van~Eylen \& Albrecht(2015)]{van_eylen_eccentricity_2015}
Van~Eylen, V. \& Albrecht, S., 2015.
\newblock {ECCENTRICITY} {FROM} {TRANSIT} {PHOTOMETRY}: {SMALL} {PLANETS} {IN} {KEPLER} {MULTI}-{PLANET} {SYSTEMS} {HAVE} {LOW} {ECCENTRICITIES}, {\it The Astrophysical Journal\/}, {\bf 808}(2), 126, Publisher: The American Astronomical Society.

\bibitem[Vogt et~al.(1983)Vogt, Soderblom, \& Penrod]{vogt_rotational_1983}
Vogt, S.~S., Soderblom, D.~R., \& Penrod, G.~D., 1983.
\newblock Rotational studies of late-type stars. {III}. {Rotation} among {BY} {Draconis} stars., {\it The Astrophysical Journal\/}, {\bf 269}, 250--252, Publisher: IOP ADS Bibcode: 1983ApJ...269..250V.

\bibitem[Wan et~al.(2018)Wan, Kafle, Lewis, Mackey, Sharma, \& Ibata]{wan_galactic_2018}
Wan, Z., Kafle, P.~R., Lewis, G.~F., Mackey, D., Sharma, S., \& Ibata, R.~A., 2018.
\newblock Galactic cartography with {SkyMapper} - {I}. {Population} substructure and the stellar number density of the inner halo, {\it Monthly Notices of the Royal Astronomical Society\/}, {\bf 480}, 1218--1228, Publisher: OUP ADS Bibcode: 2018MNRAS.480.1218W.

\bibitem[Watson et~al.(2011)Watson, Littlefair, Diamond, Collier~Cameron, Fitzsimmons, Simpson, Moulds, \& Pollacco]{2011MNRAS.413L..71W}
Watson, C.~A., Littlefair, S.~P., Diamond, C., Collier~Cameron, A., Fitzsimmons, A., Simpson, E., Moulds, V., \& Pollacco, D., 2011.
\newblock On the alignment of debris discs and their host stars' rotation axis - implications for spin-orbit misalignment in exoplanetary systems, {\it Monthly Notices of the Royal Astronomical Society\/}, {\bf 413}, L71--L75, ADS Bibcode: 2011MNRAS.413L..71W.

\bibitem[Weise et~al.(2010)Weise, Launhardt, Setiawan, \& Henning]{weise_rotational_2010}
Weise, P., Launhardt, R., Setiawan, J., \& Henning, T., 2010.
\newblock Rotational velocities of nearby young stars, {\it Astronomy and Astrophysics\/}, {\bf 517}, A88, ADS Bibcode: 2010A\&A...517A..88W.

\bibitem[West \& Basri(2009)]{west_first_2009}
West, A.~A. \& Basri, G., 2009.
\newblock A {FIRST} {LOOK} {AT} {ROTATION} {IN} {INACTIVE} {LATE}-{TYPE} {M} {DWARFS}, {\it The Astrophysical Journal\/}, {\bf 693}(2), 1283--1289, Publisher: American Astronomical Society.

\bibitem[Williams \& Cieza(2011)]{williams_protoplanetary_2011}
Williams, J.~P. \& Cieza, L.~A., 2011.
\newblock Protoplanetary {Disks} and {Their} {Evolution}, {\it Annual Review of Astronomy and Astrophysics\/}, {\bf 49}(1), 67--117, \_eprint: https://doi.org/10.1146/annurev-astro-081710-102548.

\bibitem[Winn et~al.(2010)Winn, Fabrycky, Albrecht, \& Johnson]{winn_hot_2010}
Winn, J.~N., Fabrycky, D., Albrecht, S., \& Johnson, J.~A., 2010.
\newblock {HOT} {STARS} {WITH} {HOT} {JUPITERS} {HAVE} {HIGH} {OBLIQUITIES}, {\it The Astrophysical Journal\/}, {\bf 718}(2), L145--L149, Publisher: American Astronomical Society.

\bibitem[Wood et~al.(2021)Wood, Mann, \& Kraus]{wood_characterizing_2021}
Wood, M.~L., Mann, A.~W., \& Kraus, A.~L., 2021.
\newblock Characterizing {Undetected} {Stellar} {Companions} with {Combined} {Data} {Sets}, {\it The Astronomical Journal\/}, {\bf 162}(4), 128, Publisher: The American Astronomical Society.

\bibitem[Wood et~al.(2023{\natexlab{a}})Wood, Mann, Barber, Bush, Kraus, Tofflemire, Vanderburg, Newton, Feiden, Zhou, Bouma, Quinn, Armstrong, Osborn, Adibekyan, Mena, Sousa, Gagné, Fields, Milburn, Thao, Schmidt, Gnilka, Howell, Law, Ziegler, Briceño, Ricker, Vanderspek, Latham, Seager, Winn, Jenkins, Schlieder, Osborn, Twicken, Ciardi, \& Huang]{wood_tess_2023}
Wood, M.~L., Mann, A.~W., Barber, M.~G., Bush, J.~L., Kraus, A.~L., Tofflemire, B.~M., Vanderburg, A., Newton, E.~R., Feiden, G.~A., Zhou, G., Bouma, L.~G., Quinn, S.~N., Armstrong, D.~J., Osborn, A., Adibekyan, V., Mena, E.~D., Sousa, S.~G., Gagné, J., Fields, M.~J., Milburn, R.~P., Thao, P.~C., Schmidt, S.~P., Gnilka, C.~L., Howell, S.~B., Law, N.~M., Ziegler, C., Briceño, C., Ricker, G.~R., Vanderspek, R., Latham, D.~W., Seager, S., Winn, J.~N., Jenkins, J.~M., Schlieder, J.~E., Osborn, H.~P., Twicken, J.~D., Ciardi, D.~R., \& Huang, C.~X., 2023{\natexlab{a}}.
\newblock {TESS} {Hunt} for {Young} and {Maturing} {Exoplanets} ({THYME}). {IX}. {A} 27 {Myr} {Extended} {Population} of {Lower} {Centaurus} {Crux} with a {Transiting} {Two}-planet {System}, {\it The Astronomical Journal\/}, {\bf 165}(3), 85, Publisher: The American Astronomical Society.

\bibitem[Wood et~al.(2023{\natexlab{b}})Wood, Mann, Barber, Bush, Milburn, Thao, Schmidt, Tofflemire, \& Kraus]{wood_lithium_2023}
Wood, M.~L., Mann, A.~W., Barber, M.~G., Bush, J.~L., Milburn, R.~P., Thao, P.~C., Schmidt, S.~P., Tofflemire, B.~M., \& Kraus, A.~L., 2023{\natexlab{b}}.
\newblock A {Lithium} {Depletion} {Age} for the {Carina} {Association}, {\it The Astronomical Journal\/}, {\bf 166}, 247, Publisher: IOP ADS Bibcode: 2023AJ....166..247W.

\bibitem[Wright et~al.(2010)Wright, Eisenhardt, Mainzer, Ressler, Cutri, Jarrett, Kirkpatrick, Padgett, McMillan, Skrutskie, Stanford, Cohen, Walker, Mather, Leisawitz, Gautier, McLean, Benford, Lonsdale, Blain, Mendez, Irace, Duval, Liu, Royer, Heinrichsen, Howard, Shannon, Kendall, Walsh, Larsen, Cardon, Schick, Schwalm, Abid, Fabinsky, Naes, \& Tsai]{wright_wide-field_2010}
Wright, E.~L., Eisenhardt, P. R.~M., Mainzer, A.~K., Ressler, M.~E., Cutri, R.~M., Jarrett, T., Kirkpatrick, J.~D., Padgett, D., McMillan, R.~S., Skrutskie, M., Stanford, S.~A., Cohen, M., Walker, R.~G., Mather, J.~C., Leisawitz, D., Gautier, III, T.~N., McLean, I., Benford, D., Lonsdale, C.~J., Blain, A., Mendez, B., Irace, W.~R., Duval, V., Liu, F., Royer, D., Heinrichsen, I., Howard, J., Shannon, M., Kendall, M., Walsh, A.~L., Larsen, M., Cardon, J.~G., Schick, S., Schwalm, M., Abid, M., Fabinsky, B., Naes, L., \& Tsai, C.-W., 2010.
\newblock The {Wide}-field {Infrared} {Survey} {Explorer} ({WISE}): {Mission} {Description} and {Initial} {On}-orbit {Performance}, {\it The Astronomical Journal\/}, {\bf 140}, 1868--1881, Publisher: IOP ADS Bibcode: 2010AJ....140.1868W.

\bibitem[Wu \& Murray(2003)]{wu_planet_2003}
Wu, Y. \& Murray, N., 2003.
\newblock Planet {Migration} and {Binary} {Companions}: {The} {Case} of {HD} 80606b, {\it The Astrophysical Journal\/}, {\bf 589}(1), 605, Publisher: IOP Publishing.

\bibitem[Zúñiga-Fernández et~al.(2021)Zúñiga-Fernández, Bayo, Elliott, Zamora, Corvalán, Haubois, Corral-Santana, Olofsson, Huélamo, Sterzik, Torres, Quast, \& Melo]{zuniga-fernandez_search_2021}
Zúñiga-Fernández, S., Bayo, A., Elliott, P., Zamora, C., Corvalán, G., Haubois, X., Corral-Santana, J.~M., Olofsson, J., Huélamo, N., Sterzik, M.~F., Torres, C. a.~O., Quast, G.~R., \& Melo, C. H.~F., 2021.
\newblock Search for associations containing young stars ({SACY}) - {VIII}. {An} updated census of spectroscopic binary systems exhibiting hints of non-universal multiplicity among their associations, {\it Astronomy \& Astrophysics\/}, {\bf 645}, A30, Publisher: EDP Sciences.

\end{thebibliography}

\appendix

\section{\lowercase{\rstarcode{}} Computational Cost Considerations} \label{apdx:rstar-speed}

Following Section~\ref{sec:phot_model}, the pure-\texttt{synphot} extinction calculation can be computationally expensive. To work around this, we implemented a method which uses \texttt{synphot} to create the blackbody spectrum and the extinction model, but all of the relevant calculations are performed `by hand' with \texttt{numpy} \citep{harris_array_2020}. The pared-down \texttt{numpy} approach already makes a noticeable speed improvement, but the extinction calculation could be performed tens of times for every iteration of the simulation, so it is important that our methods are as computationally inexpensive as possible. 

To this end, \rstarcode{} employs \texttt{numba} just-in-time (JIT) compilation \citep{lam_numba_2015} to as many aspects of the extinction procedure as possible. JIT is a method which converts Python code to optimized machine code when first compiled. The first function call can oftentimes be more computationally expensive than the original (non-JIT) function, but every subsequent call will show vast speed improvements. Thus, it is a much more advantageous method to use when a function is called multiple times, as is the case with the extinction calculation. From the pure-\texttt{synphot} method to the combined \texttt{numpy} and \texttt{numba} method, we estimate a 10--15x increase in compilation speed depending on the number of bandpass filters used in the calculation.\footnote{Speed improvements were estimated on a machine with an 8-core Intel~Xeon~W processor and 32\,GB of RAM.}

\section{\lowercase{\rstarcode{}} Extinction Calculation} \label{apdx:rstar-calc}

Following Section~\ref{sec:phot_model}, the `by-hand' \texttt{numpy}-based extinction calculation procedure is designed to reproduce the values derived by the pure-\texttt{synphot} method. The extinction value in a particular bandpass filter (\Alambda{}) is calculated by
\begin{equation}
    \label{eqn:extinction}
    A_{\lambda} = \mathcal{S}_{\mathrm{eff}}^{\prime} - \mathcal{S}_{\mathrm{eff}}
\end{equation}
where $\mathcal{S}_{\mathrm{eff}}^{\prime}$ and $\mathcal{S}_{\mathrm{eff}}$ are the extincted and pure blackbody spectra, respectively. Effective stimulus is the flux density an observer would measure given a certain amount of flux generated from the synthetic (blackbody or extincted) source spectrum through a particular bandpass filter (in \flam{} units; \flamunits{}). Effective stimulus is calculated by
\begin{equation}
\label{eqn:effstim}
    \mathcal{S}_{\mathrm{eff}} = \frac{\int F(\lambda) \, e(\lambda) \, \lambda \, d\lambda}{\int e(\lambda) \, \lambda \, d\lambda}
\end{equation}
with flux \flamalt{} in \flam{} units, wavelength $\lambda$ in \AA, and the bandpass
filter response function $e(\lambda)$ (also called `bandpass transmission function') in dimensionless
fractions between 0 (no transmission) and 1 (full transmission).

Equations~\ref{eqn:extinction} and~\ref{eqn:effstim} give their results in \flam{} units. If \Alambda{} needs to be presented in different units (e.g., magnitudes, as is the case by default within \rstarcode{}), both $\mathcal{S}_{eff}^{\prime}$ and $\mathcal{S}_{eff}$ need to be converted individually before being used in Equation~\ref{eqn:extinction}. When converting to AB magnitudes, for example, \flam{} must be converted to \fnu{} via
\begin{equation}
    \label{eqn:fnu}
    F_{\nu} = \frac{\lambda_{\textnormal{piv}}^2}{c} \, F_{\lambda}
\end{equation}
where $c$ is the speed of light and \lambdapiv{} is the pivot wavelength for a particular bandpass filter. CCDs follow the `equal-energy convention' for \lambdapiv{} given as
\begin{equation}
    \label{eqn:equal_energy}
    \lambda_{\textnormal{piv}} = \sqrt{\frac{\int e(\lambda) d\lambda}{\int e(\lambda) \lambda^{-2} d\lambda}}
\end{equation}
whereas the \texttt{synphot} method follows the `quantum-efficiency convention' given as
\begin{equation}
    \lambda_{\textnormal{piv}} = \sqrt{\frac{\int e(\lambda) \lambda \, d\lambda}{\int e(\lambda) \lambda^{-1} d\lambda}}
\end{equation}
Both pivot wavelengths lead to values of extinction that differ to $\ll$1\% in magnitudes.

\bsp	
\label{lastpage}
\end{document}